\documentclass[aps,prd,twocolumn,floatfix,
letterpaper,superscriptaddress]{revtex4}
\usepackage{graphicx}\usepackage{natbib}\usepackage{amsmath}
\usepackage{times}
\newcommand{\kms}{\ensuremath{{\rm km\,s}^{-1}}}
\newcommand{\msun}{\ensuremath{{\rm M}_{\odot}}}

\newcommand{\yr}{\ensuremath{\rm yr}}

\newcommand{\sigmab}{\ensuremath{\sigma_{\log B}}}
\newcommand{\ergs}{\ensuremath{\rm erg}\,{\rm s}^{-1}}
\newcommand{\ergcms}{\ensuremath{\rm erg}\,{\rm cm}^{-2}\,{\rm s}^{-1}}

\newcommand{\note}[1]{}
\begin{document}

\title{Young and Millisecond Pulsar GeV Gamma-ray Fluxes from the Galactic Center and Beyond}

\author{Ryan M.\ O'Leary}
\affiliation{JILA, 440 UCB, University of Colorado, Boulder, CO 80309-0440, USA}
\author{Matthew D.\ Kistler}
\affiliation{Kavli Institute for Particle Astrophysics and Cosmology, Stanford University,
SLAC National Accelerator Laboratory, Menlo Park, CA 94025, USA}
\author{Matthew Kerr}
\affiliation{CSIRO Astronomy and Space Science, Australia Telescope National Facility, PO Box 76, Epping NSW 171, Australia}
\author{Jason Dexter}
\affiliation{Max Planck Institute for Extraterrestrial Physics, Giessenbachstr. 1, 85748 Garching, Germany}

\date{\today}

\begin{abstract}
Gamma-ray observations have shown pulsars to be efficient converters of rotational energy into GeV photons and it is of wide-ranging interest to determine their contribution to the gamma-ray background.  We arrive at flux predictions from both the young ($\lesssim\,$Myr) and millisecond ($\sim\,$Gyr) Galactic pulsar populations.  We find that unresolved pulsars can yield both a significant fraction of the excess GeV gamma rays near the Galactic Center and an inverse Compton flux in the inner kpc similar to that inferred by {\it Fermi}. 
We compare models of the young pulsar population and millisecond pulsar population to constraints from gamma-ray and radio observations.   Overall, we find that the young pulsars should outnumber millisecond pulsars as unassociated gamma-ray point sources in this region.  The number of young radio pulsars discovered near the Galactic Center is in agreement with our model of the young pulsar population.  Deeper radio observations at higher latitudes can constrain the total gamma-ray emission from both young and millisecond pulsars from the inner galaxy. 
 While this is a step towards better understanding of pulsars, cosmic rays in the Milky Way, and searches for dark matter, we also discuss a few interesting puzzles that arise from the underlying physics of pulsar emission and evolution.
\end{abstract}

\maketitle

%
\section{Introduction}
%

\begin{figure*}[t!]
\includegraphics[width=\textwidth,clip=true]{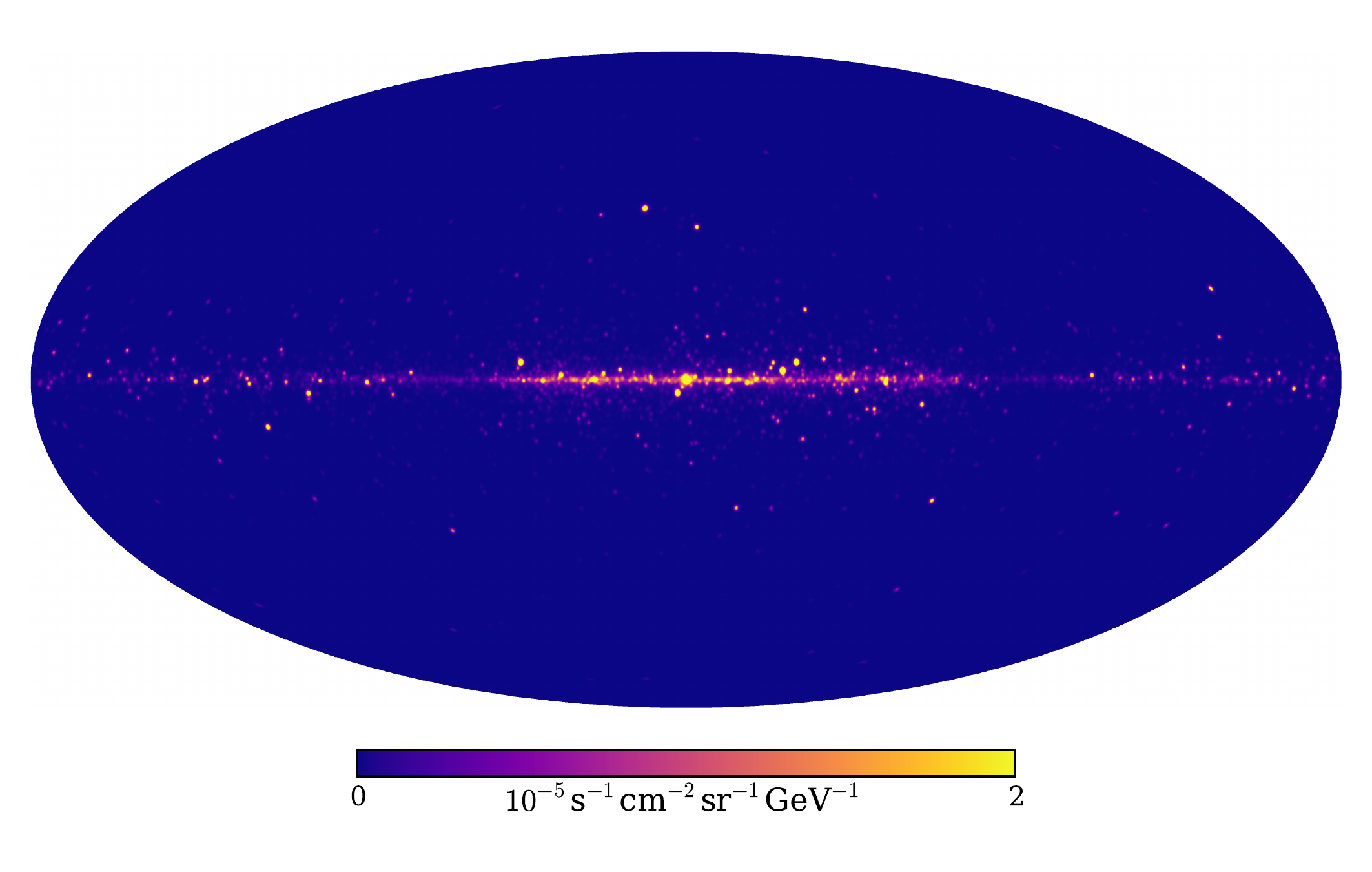}\vspace{-0.0cm}
\caption{\label{fig:allsky} All-sky model of the gamma-ray flux ($dN/dE$) from young pulsars in the Milky Way (MW) at 2\,GeV, as discussed in \S~\ref{sec:gammaray}.  The bright peak in the center results from young pulsars arising in the Central Molecular Zone (CMZ) of the Galactic Center (GC). We have set the maximum flux at $2\times 10^{-5}$s$^{-1}$cm$^{-2}$sr$^{-1}$GeV$^{-1}$ in order to enhance the visibility of the diffuse plane emission.}
\end{figure*}

In recent years, the {\em Fermi} Large Area Telescope \cite{2009ApJ...697.1071A} has revitalized the gamma-ray pulsar field \cite{2010ApJS..187..460A} with detections approaching 200 in total, split among young radio-loud (YRL) pulsars, young radio-quiet (YRQ) pulsars, and old, recycled millisecond pulsars (MSPs) \cite{2013pulsarcatalog,2015arXiv150203251L}.  This has allowed studies specifically based on gamma-ray pulsars alone.  The brightest gamma-ray pulsars tend to be young, $\tau \lesssim 10^6\,\yr$, with high spin-down power, $\dot{E} > 10^{35}\,{\rm erg\,s}^{-1}$, so such studies have focused on the tens of {\em local} pulsars, e.g., \cite{2011ApJ...727..123W,2012A&A...545A..42P,2013ApJ...776...61P,2014ApJ...796...14C,2015A&A...575A...3P}.

However, these nearby detections imply that there should be far more gamma-ray pulsars throughout the galaxy that are yet below the threshold for individual discovery by {\em Fermi}.  That these should lead to a diffuse background has long been expected \cite[e.g.,][]{1976ApJ...208L.107H,1977MNRAS.179P..69S,1978A&A....67..149B,1981Natur.290..316H,1981ApJ...247..639H,1991JApA...12...17B,1992ApJ...391..659B,1994ApJ...420L..75D,1995ApJ...449..211Y,1996ApJ...461..872S,1997ApJ...476..238B,1997ApJ...491..159P,1998MNRAS.301..841Z,2000ApJ...538..818M,2010JCAP...01..005F}, though the flux level depends on a variety of parameters related to pulsars and the Milky Way (MW) that need to be carefully examined.

Here, we elaborate on the model developed in \citet{2015PF}, where we demonstrated that that young pulsars (YPs) resulting from supernovae (SNe) near the Galactic Center (GC) and Galactic disk lead to a gamma-ray flux and angular distribution closely resembling the GC GeV gamma-ray Excess \cite{2009arXiv0910.2998G,2011JCAP...03..010A,2011PhLB..697..412H,2011PhRvD..84l3005H,2012PhRvD..86h3511A,2013PDU.....2..118H,2013PhRvD..88h3521G,2014PhRvD..89f3515M,2014PhRvD..90b3526A,2014daylanetal,2014arXiv1406.6948Z,2014arXiv1409.0042C,2014caloreetal,FermiGC} seen in {\it Fermi} data.  Though in general this follows directly from SN pulsar kicks, e.g., young pulsars escaping the GC should roughly follow a $\rho \sim r^{-2}$ number density profile (smoothed by their distribution of velocities) while getting fainter with time so that the flux density further steepens, the details remain to be precisely resolved.

For example, in Fig.~\ref{fig:allsky} we show the all-sky map of the contribution of young pulsars to the gamma-ray sky in our Fiducial Model \cite{2015PF}.  Only a small fraction of the visible point sources are bright enough to be blindly detected as pulsars, and a few hundred more are above the {\em Fermi} point source detection threshold.  Thousands of faint pulsars blend together to produce a diffuse background that peaks at the Galactic Center with emission all along the Galactic plane.

In this work, we present a comprehensive analysis of the contribution of young pulsars and MSPs to the gamma-ray sky.  For the young pulsars we follow the approach of \citet{2015PF}, but explore the expansive parameter space of pulsar properties, with a particular focus on three different models for for gamma-ray emission.  In addition, we present a new (forward) model for the population of MSPs within the Galaxy.  With these models, we address the young pulsar and MSP contributions of diffuse GeV gamma rays, the number of unassociated {\em Fermi} point sources, cosmic-ray electron production within the inner kpc of the MW, as well as the underlying distribution of unresolved gamma-ray point sources.  This latter case is particular relevant to the recent evidence that the excess gamma-ray emission in the Galactic Center is not smooth \cite{2015arXiv150605124L,2015arXiv150605104B}.  We also estimate the number of radio-loud pulsars in the Galactic Center from both YPs and MSPs, and compare our estimates to current and future radio observations.  

\begin{table}
\caption{Acronyms and Abbreviations}
\label{tab:ac}
\begin{ruledtabular}
\begin{tabular}{ll}
  AGN & Active Galactic Nucleus \\
  CC & Core Collapse \\
  CMZ & Central Molecular Zone\\
  {\em Fermi} & {\em Fermi} Large Area Telescope\\
  GC & Galactic Center \\
  IMF & Initial Mass Function \\
  MSP & Millisecond (recycled) pulsar\\
MW & Milky Way \\
SFR & Star Formation Rate\\
SN & Supernova\\
SNR & Supernova Rate\\
YP & Young Pulsar\\
YRL & Young Radio Loud Pulsar\\
YRQ & Young Radio Quiet Pulsar
\end{tabular}
\end{ruledtabular}
\end{table}

In \S~\ref{sec:pulsars}, we describe how pulsars and their associated gamma-ray properties evolve with age.  We also describe the YRL and YRQ populations of pulsars.
In \S~\ref{sec:pop}, we describe our method for modeling the Galactic populations of young and millisecond pulsars, including their birth rate and birth sites.
In \S~\ref{sec:models}, we summarize the three young pulsar emission models and the one MSP Model that we focus on throughout the paper.
In \S~\ref{sec:gammaray}, we present the Galactic gamma-ray emission arising from our four pulsar models, addressing their contribution to unassociated point sources discovered by {\em Fermi} (\S~\ref{sec:ps}) and diffuse emission near the GC (\S~\ref{sec:diffuse}).
In \S~\ref{sec:extended}, we address contributions to inverse-Compton emission from around the GC.
In \S~\ref{sec:radio} we predict the distribution of radio-loud young and millisecond pulsars in the GC region and compare this to current constraints.
In \S~\ref{sec:unresolved} we estimate the resolved and unresolved flux distribution of the pulsars and compare our results to flux distribution derived from one-point photon statistics near the GC.
Finally, in \S~\ref{sec:disc} we  summarize our main results and contributions and discuss a variety of additional implications of our models.
In the Appendix, we detail many of the underlying uncertainties in our pulsar models and how they impact expected contributions to gamma-ray excesses.  Throughout this work, we consider MSPs as pulsars that have been recycled via mass accretion from a binary companion, and all other pulsars as young pulsars, independent of age.  Table~\ref{tab:ac} collects much of the nomenclature we use in this work for easy reference.

\section{Pulsar Properties and Evolution}
\label{sec:pulsars}
In this section, we  give a general overview of pulsar properties and how they evolve in time.  We detail the birth properties of pulsars in our Fiducial Model and our empirical model for the gamma-ray properties of pulsars.  Finally we describe the properties of both YRQ and YRL  pulsars.

\subsection{Pulsar Evolution}
\label{sec:ev}
The spin-down power of a pulsar is inferred from the observations of the pulsar period, $P$, and the change of the pulse period, $\dot{P}$, as
\begin{equation}
  \label{eq:edot}
\dot{E} = 4 \pi^2 I \frac{\dot{P}}{P^3},
\end{equation}
where $I$ is the neutron star moment of inertia.  For a rotating dipole, the spin-down rate is related to the surface field of the pulsar, $B$, by
\begin{equation}
\label{eq:pdot}
   \dot{P} = \frac{8 \pi^2 R^6 B^2}{3 I c^3 P} \sin^2{\alpha}\,,
 \end{equation}
where $R$ is the neutron star radius and $c$ is the speed of light \cite{1969ApJ...157.1395O}.  While the dipole spin-down power depends on the angle between the spin axis of the pulsar and the magnetic field axis, $\alpha$, we do not include this effect in our work, and typical of population synthesis models of pulsars \cite[e.g.,][]{1969ApJ...157.1395O,fauchergiguerekaspi2006}, set $\sin{\alpha} \equiv 1$.

For an isolated pulsar, observations of only $P$ and $\dot{P}$ cannot uniquely determine the surface field, mass, and moment of inertia (or radius).  Observations of binary systems have revealed a wide range of pulsar masses, from $1.2-2.0\,\msun$ \cite{2011A&A...527A..83Z,2010Natur.467.1081D,2013Sci...340..448A}.  In this work we will use the canonical values of $R \equiv 10^6\,$cm, $M \equiv 1.4\,\msun$, and $I \equiv 1\times 10^{45}\,$g$\,$cm$^2$. The overall uncertainty in $I$ is about a factor of two.  Indeed, other works \citep[e.g.,][]{2015A&A...575A...3P}, have used a higher value of $I$.  A survey of recent neutron star mass-radius studies suggests that one could justify either choice of $I$ \cite[e.g.,][]{2015arXiv150505155O,2015arXiv151007515S}.  As such, most population studies set the moment of inertia, radius, and mass of each pulsar to be fixed, with all of the uncertainties incorporated into the underlying distribution of $B$ and the initial spin period $P_0$. We do the same in this work. Knowing $B$ and $P_0$, we can directly calculate the time evolution of the properties of each pulsar. 

The underlying distribution of surface magnetic fields is one of the most important uncertainties in determining the contribution of pulsars to the gamma-ray sky.  The $B$ distribution has been  modeled by a log-normal distribution in both population synthesis models of radio pulsars and gamma-ray pulsars, with a mean $\left< \log_{10}{B}\right> = 12.65$. However, these two independent population synthesis methods have found different preferred spreads in the distributions, $\sigma_{\log B}$. In particular, the $B$ distribution used by \citet{2011ApJ...727..123W} ($\sigma_{\log B} = 0.3$) for modeling early results of the {\em Fermi} pulsar catalog  is significantly narrower than the spread preferred from radio surveys \citep{fauchergiguerekaspi2006} ($\sigma_{\log B} = 0.55$).  Since \citet{2011ApJ...727..123W}, {\em Fermi} has discovered many more pulsars, especially at lower ${\dot E}$ and older ages.

In this work we use $\sigma_{\log B}  = 0.45$, roughly between the preferred value from previous studies in gamma rays and radio.  We find that this is most consistent with the entire population of known pulsars, even if it isn't the preferred value for either the radio discovered or gamma-ray discovered populations.  In Appendix~\ref{sec:B}, we describe the differences between the radio and gamma-ray selected pulsar populations and directly compare our models with  $\sigma_{\log B}  = 0.3$ and $\sigma_{\log B} = 0.55$ in detail.   

In order to describe the complete evolution of a pulsar, we also need to know the initial birth period distribution. We follow \citet{2011ApJ...727..123W}, and use a truncated normal distribution, with the spread and mean equal to $50\,$ms.  Likewise, we truncate the distribution below $10\,$ms.  Similar to the $B$ distribution, the preferred birth period distribution of the radio pulsars is in disagreement with the gamma-ray population. In \citet{fauchergiguerekaspi2006}, the authors found that the preferred mean was closer to $300\,$ms, significantly longer than used here. \citet{2011ApJ...727..123W} proposed that this was likely biased by the radio luminosity function of \citet{fauchergiguerekaspi2006}.  Using the luminosity function of \citet{1987A&A...171..152S}, \citet{2011ApJ...727..123W} found the radio data was consistent with the lower initial spin periods used in their own work. Nevertheless, here we do not include newly born pulsars with ages $\leq 10^4\,\yr$ in our analysis, so few pulsars have spin periods comparable to their value at birth since we use such small initial spin periods.  

For each pulsar, we use the initial $B$ and $P_0$ to evolve the observable properties of the pulsar to its present position and age (see \S~\ref{sec:ev} for more details). As the pulsar ages, its rotation slows and its period increases as
\begin{equation}
  P(t) = \sqrt{P_0^2 + \frac{16 \pi^2 R^6 B^2 t}{3 I c^3}},
\end{equation}
where we take $R \equiv 10^6\,$cm and $I \equiv 10^{45}\,$g$\,$cm$^2$, and $t$ is the pulsar's age.

\begin{figure}[t!]
\hspace*{-0.7cm}
\includegraphics[width=1.1\columnwidth,clip=true]{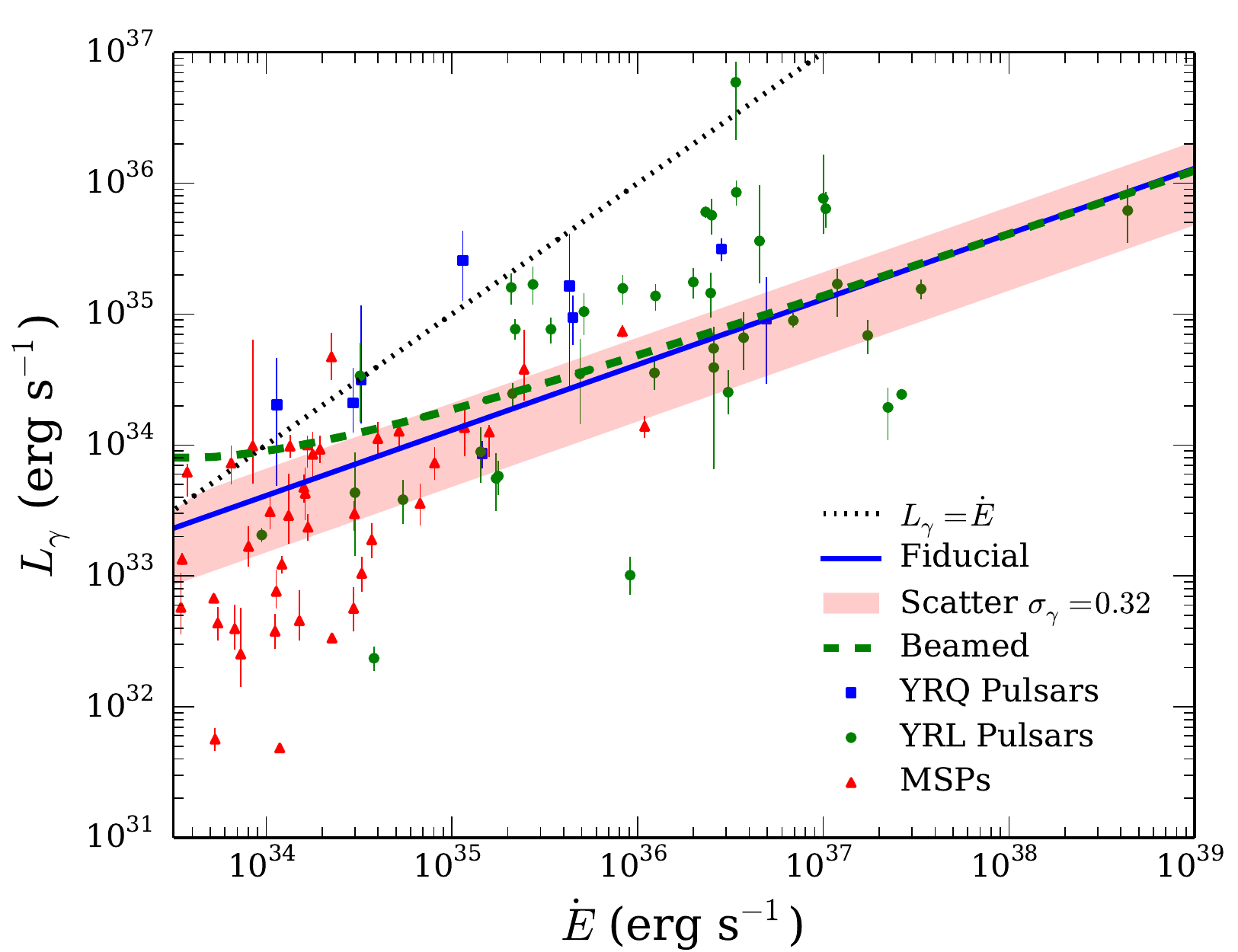}
\caption{Pulsar gamma-ray luminosity, $L_\gamma$, versus spin-down power, $\dot{E}$.
Individual markers denote YRQ pulsars (blue squares), YRL pulsars (green circles), and MSPs (red triangles) from 2PC \cite{2013pulsarcatalog}.  Error bars mostly correspond to pulsar distance uncertainties (no error bars are shown for $\dot{E}$, which is only relevant for a few MSPs).  Note that a number of YRQ pulsars lack distance estimates and are not shown.
The curves correspond to the four emission models detailed in \S~\ref{sec:models}: our young pulsar Fiducial Model with $C = 1.3$ ({\it solid blue}); 
the Scatter Model with $\sigma_\gamma = 0.32$ ($1\sigma$ range is {\it red shaded band}); the observed luminosity of {\em visible} pulsars in our YRQ Beamed Model ({\it dashed green}).  The MSP Model tracks the Fiducial curve ({\it solid blue}).    \label{fig:lumfunc}}
\end{figure}

\begin{figure}[t!]
\includegraphics[width=\columnwidth,clip=true]{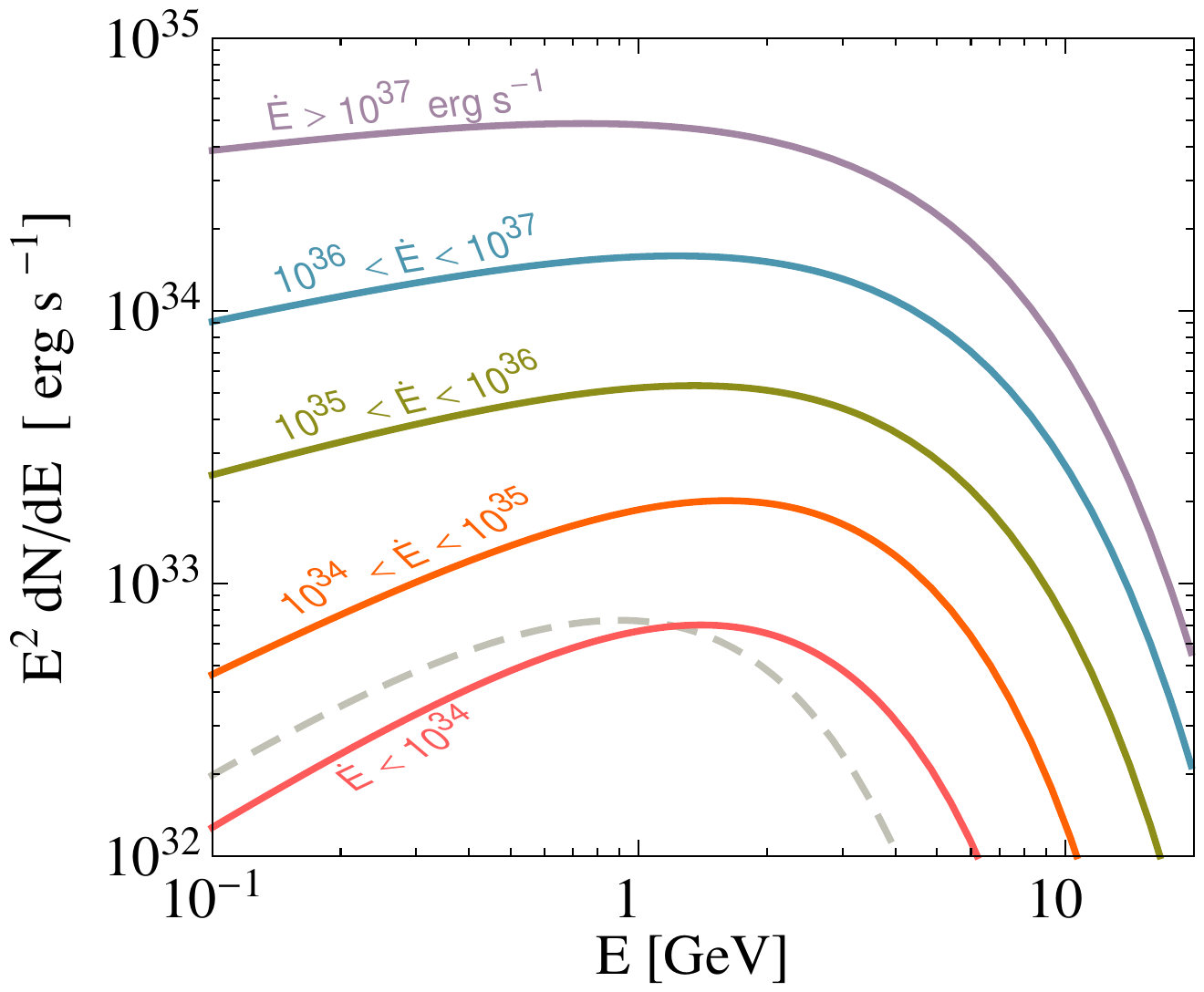}\vspace{-0.0cm}
\caption{Pulsar gamma-ray spectral fits for various $\dot{E}$ ranges (shown for the log middle energy in each band), which display a hardening at low energies with decreasing $\dot{E}$.  \label{fig:spectra}}
\end{figure}

\subsection{Gamma-ray Properties and Spectra}
\label{sec:spectra}
The explosion in gamma-ray pulsar detections by {\it Fermi} has led to a great empirical improvement in understanding of the emission properties.  As the data in Fig.~\ref{fig:lumfunc} suggest, the gamma-ray luminosity is described fairly well by the relation
\begin{equation}
\label{eq:lumfunc}
L_\gamma =  C \times \left(\frac{\dot{E}}{10^{33}\,{\rm erg\,s}^{-1}}\right)^{1/2} \times 10^{33}\,{\rm erg\,s}^{-1},
\end{equation}
with $C \approx 1$ \cite{2011ApJ...727..123W,2013pulsarcatalog}, where $\dot{E}$ is the total spin-down power of the pulsar inferred via Eq.~(\ref{eq:edot}).  However, the amount of scatter observed in the luminosity function of pulsars exceeds what is expected from uncertainties in the distance to pulsars alone.

The detailed light curves measured by {\it Fermi} can in most cases be well fit by assuming gamma-ray emission in the outer pulsar magnetosphere; however, no particular model has a clear advantage in all cases \cite{2015A&A...575A...3P}.  Not having access yet to the true underlying electrodynamical model, we make general empirically-based assumptions and discuss possible deviations. For our Fiducial Model we set $C=1.3$, which is closer to reproducing the mean integrated flux of the known pulsar population ($\left< C\right> = 1.6$), but doesn't over produce the number of unknown point sources (see \S~\ref{sec:ps}).

In the {\it Fermi} data, there is a rough correlation of the fit pulsar spectrum power-law index to the spin-down power, $a \propto \dot{E}^{0.2}$ (Eq.~\ref{eq:dnde}).  For a gamma-ray pulsar population dominated by Geminga-like ($\dot{E} \approx 10^{34.5}\,$erg~s$^{-1}$) pulsars, the average spectrum of the total emission is harder below 1~GeV than from younger pulsars.

We obtain a description of the gamma-ray spectra versus $\dot{E}$ by taking spectral fits from the Third {\em Fermi} Source Catalog (3FGL) \cite{2015arXiv150102003T} and cross correlating with the list of detected pulsars.  This is basic, but attempts to take advantage of pulsars characterized since the Second {\it Fermi} Pulsar Catalog (2PC) \cite{2013pulsarcatalog}.  This method also results in the phase averaged spectrum for each pulsar, which is required to characterize a population of unresolved sources.

We exclude pulsars without a counterpart in 3FGL and those whose spectra are fit only with a power law, as these obviously bias the spectral shape.  Generally, all pulsars are fit with cutoff spectra, but there is some lag between discovering a pulsar and updating the model in the catalog.  There are also objects, like the Crab, that do not get special treatment in the catalog process and have a strong nebular component.

This leaves 73 young, non-recycled pulsars, which we divide into five $\dot{E}$ groups.  The top ($>10^{37}$~erg~s$^{-1}$) and bottom ($<10^{34}$~erg~s$^{-1}$) groups each contain only 5 pulsars, while the others contain $\sim20$, so there is likely some small number effects in these bands.

We normalize each spectrum to its total energy flux and take the median spectrum in the band to obtain the spectra in Fig.~\ref{fig:spectra}.  There is relatively little evolution in the high-energy ($>$10 GeV) behavior until $\dot{E}$ drops below $10^{34}$~erg~s$^{-1}$.  This displays evolution at low energies toward harder spectra with lower cutoffs at low $\dot{E}$. These are fit by
\begin{equation}
\label{eq:dnde}
\frac{dN}{dE} \propto E^a e^{-E/E_c},
\end{equation}
as given in Table~\ref{tab:spectra}.  At energies below the {\it Fermi} range, the spectrum is generally expected to steepen for the pulsars of greatest interest here (e.g., \cite{1996ApJ...470..469R}).

\begin{table}[t!]
\caption{Pulsar spectrum $dN/dE \propto E^a \exp(-E/E_c)$ used in $\dot{E}$ ranges.}
\label{tab:spectra}
\begin{ruledtabular}
\begin{tabular}{lcc}
					&	$a$		& $E_c$ [GeV] \\ \hline
$\dot{E}>10^{37}\,$erg~s$^{-1}$& $-1.8$   & 3.7 \\
$10^{36}<\dot{E}<10^{37}$   	& $-1.65$   & 3.5 \\
$10^{35}<\dot{E}<10^{36}$   	& $-1.55$   & 3.0 \\
$10^{34}<\dot{E}<10^{35}$   	& $-1.2$   & 2.0 \\
$\dot{E}<10^{34}$		   	& $-1.0$   & 1.4 \\
\end{tabular}
\end{ruledtabular}
\end{table}

We note that some uncertainty exists in the bottom range set, which contains only 5 pulsars.  Our method using this small number yields $a = -1.0$ and $E_c = 0.9$~GeV (shown as the dashed line in Fig.~\ref{fig:spectra}).  At $\dot{E}<10^{34}$, we have instead used the empirical spectral relations from \citet{2015A&A...575A...3P} to obtain $a = -1.0$ and $E_c = 1.4$~GeV.  As we discuss later, this range arguably stands to benefit the most from improved gamma-ray pulsar data.

The $>10^{37}$~erg~s$^{-1}$ range contains the same small number in the sample set.  Such very young ($\lesssim 10$~kyr) pulsars tend to be abnormal spectrally.  The Crab is the canonical example, although it is not alone in this respect (e.g., PSR~B1509).  Their gamma-ray efficiencies are very low, with an energy spectrum that can peak in the MeV range.  Spin-down index estimates are only available from such very young pulsars, with $n<3$ displayed by some (unfortunately, measuring the index of the $\gtrsim 10^4$~yr sample is difficult to impossible).  However, such rogue very young pulsars would be rare in our simulations and the most subject to Poisson fluctuations, so we do not attempt to refine the spectrum of these sources any further.

{\it Fermi} data indicate that the spectra tend to have more emission at higher energies than Eq.~(\ref{eq:dnde}) alone suggests \cite{2014arXiv1407.5583C}.  For the brightest pulsars with the most spectral information, such as Geminga, residual high-energy emission is often seen beyond a simple exponential cutoff.  This work focuses on the region near $2\,$GeV, the spectral peak for such pulsars and the GC Excess, and so we do not attempt to account for such surplus emission in our analysis.

\subsection{Beaming and Radio Quiet Geminga-like Pulsars}
\label{sec:beam}
The population of young gamma-ray pulsars can be split into two classes: pulsars that have observed radio emission (radio loud; YRL hereafter) and pulsars that have little or no observed radio emission (radio quiet; YRQ hereafter).  Geminga is the prototypical example of a YRQ pulsar -- indeed radio-quiet pulsars are often described as ``Geminga-like'' -- with decades of searches finding no clear detection of a pulsed radio signal \citep[cf.][]{1997AstL...23..283K}. 

For a given pulsar, the observable signal depends upon the magnetic field geometry and the angle to the observer.
YRQ pulsars are caused by the misalignment of the radio signal (which typically has a small opening angle for young pulsars) from that of the gamma-ray emission.  The blind detection of pulsed gamma rays is then the best way to determine if a gamma-ray point source is a gamma-ray pulsar.  Conversely, the known population of radio pulsars do not all produce gamma-ray emission visible to us.  Approximately half of the young gamma-ray pulsar population is YRQ (not correcting for any observational selection effects).

Beaming of both types of emission plays  an important role in determining the fractions of YRL and YRQ pulsars.    It is expected that the width of the gamma-ray beam decreases as the efficiency of the emission increases.  Likewise, as pulsars spin down, the radio beams also shrink.  With smaller opening angles for both beams, it is expected that the YRQ pulsar fraction should increase.  
As pulsars age and ${\dot E}$ decreases, it is also thought that the magnetic field may approach alignment with the rotation axis \citep{2010MNRAS.402.1317Y}, resulting in a further increase in the YRQ fraction \cite{2011ApJ...727..123W}. Overall, recent {\em Fermi} results support both scenarios \cite{2015CRPhy..16..641G}.  

We consider for beaming the One Pole Caustic model curves in \citet[][Fig.~14]{2012A&A...545A..42P}, approximating the beaming fraction of the YRQ, Geminga-like population with an average beaming fraction described as
\begin{equation}
  \label{eq:beamrq}
  \left<f_{\rm beam, RQ}\right> = 0.85 \left(1 - \left[\frac{0.7\times 10^{33} \ergs}{{\dot E}}\right]^{0.2}\right).
\end{equation}
Similarly, for the YRL gamma-ray pulsars
\begin{equation}
  \label{eq:beamrl}
  \left<f_{\rm beam, RL}\right> = 1 - \left(\frac{0.5\times 10^{33} \ergs}{{\dot E}}\right)^{0.3}.
\end{equation}
For each beaming model, we randomly assign a fraction, $f_{\rm beam}$, of the pulsars to be beamed towards the observer, and set the flux of all other pulsars to zero.  We increase the total luminosity of the visible population to account for the beaming of the pulsars
\begin{equation}
L_\gamma =  \frac{C}{f_{\rm beam}} \times \left(\frac{\dot{E}}{10^{33}\,{\rm erg\,s}^{-1}}\right)^{1/2} \times 10^{33}\,\ergs.
\end{equation}
The total luminosity of the entire pulsar population is thus the same as if we included no beaming (cf., Eq.~\ref{eq:lumfunc}). Our Beamed Model, as we outline in \S~\ref{sec:models} uses only the YRQ beaming fraction, with $C = 1.0$.  
We defer to Appendix~\ref{app:beam} discussion on relative differences between these beaming models.

\begin{figure*}[t!]
  \includegraphics[width=.97\columnwidth,clip=true]{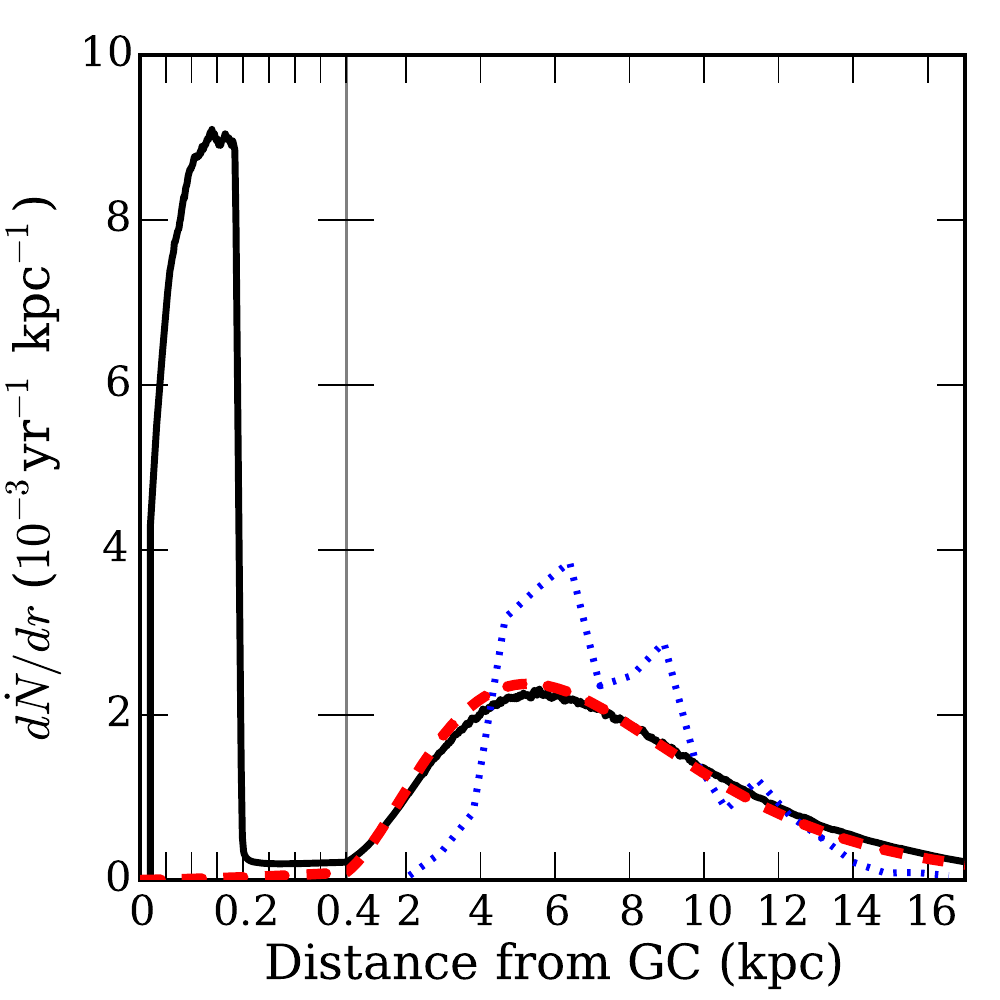}
  \includegraphics[width=.97\columnwidth,clip=true]{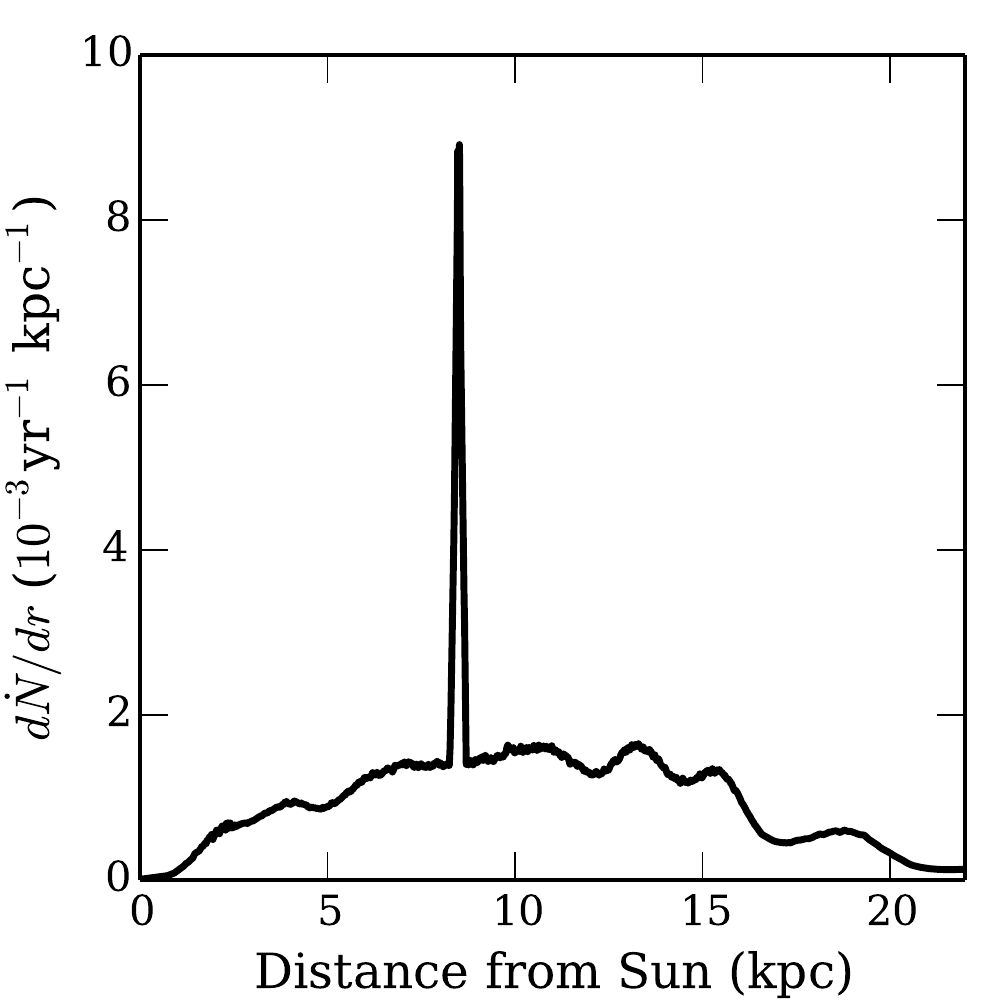}
\caption{\label{fig:dsnr}  Distribution of Galactic SNe.
{\it Left:} The differential SN rate as a function of distance from the GC in our young pulsar models, where 7\% of the rate is from the CMZ ({\it black line}).  For comparison, we show two distributions used by {\it Fermi} \cite{FermiGC} to model cosmic-ray injection:  ``Pulsars'' \cite{2004A&A...422..545Y} ({\it red dashed}) and ``OB stars'' \cite{2000A&A...358..521B} ({\it blue dotted}). All lines have the same total normalization.  Here we use two linear scales for the distance, changing at $0.4\,$kpc, twice the size of the CMZ.
{\it Right:}  The differential SN rate versus distance from the Sun.  
} 
\end{figure*}

 \begin{figure*}[]
   \includegraphics[width=.8888\columnwidth,clip=true]{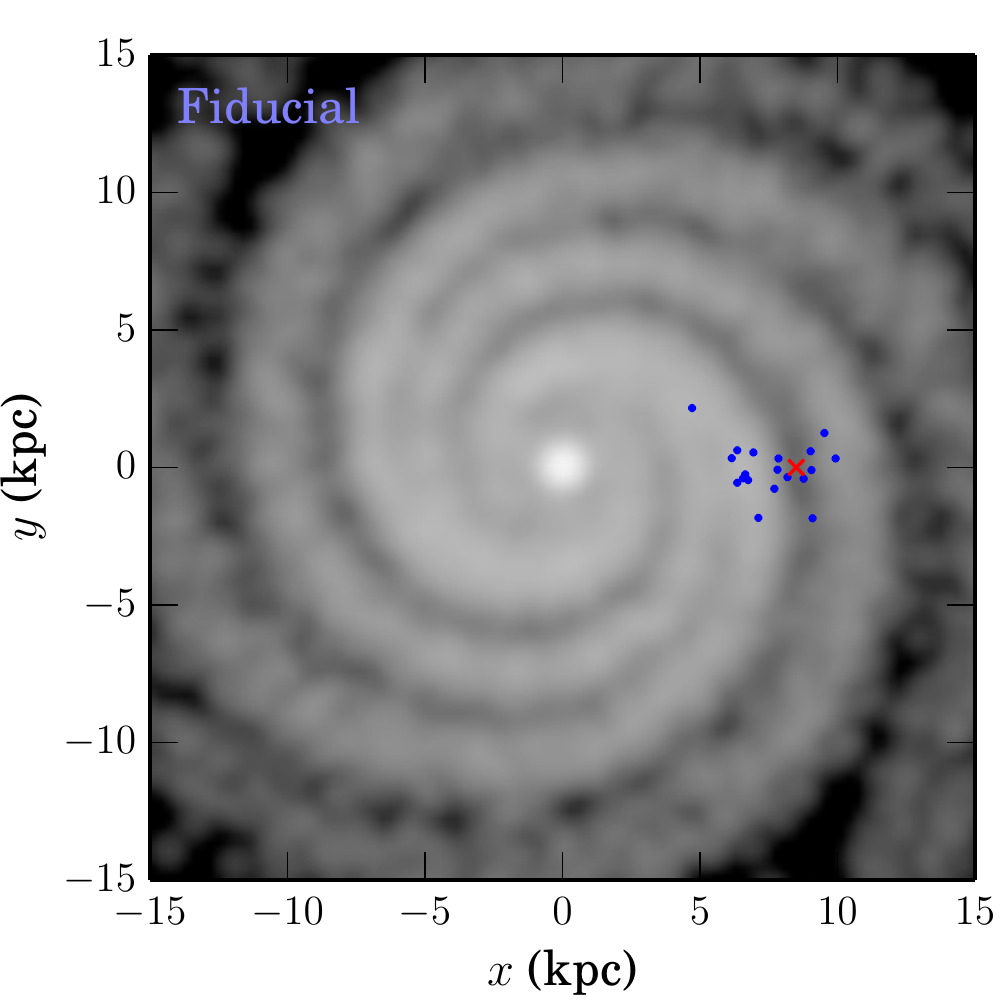}
   \includegraphics[width=1.1111\columnwidth,clip=true]{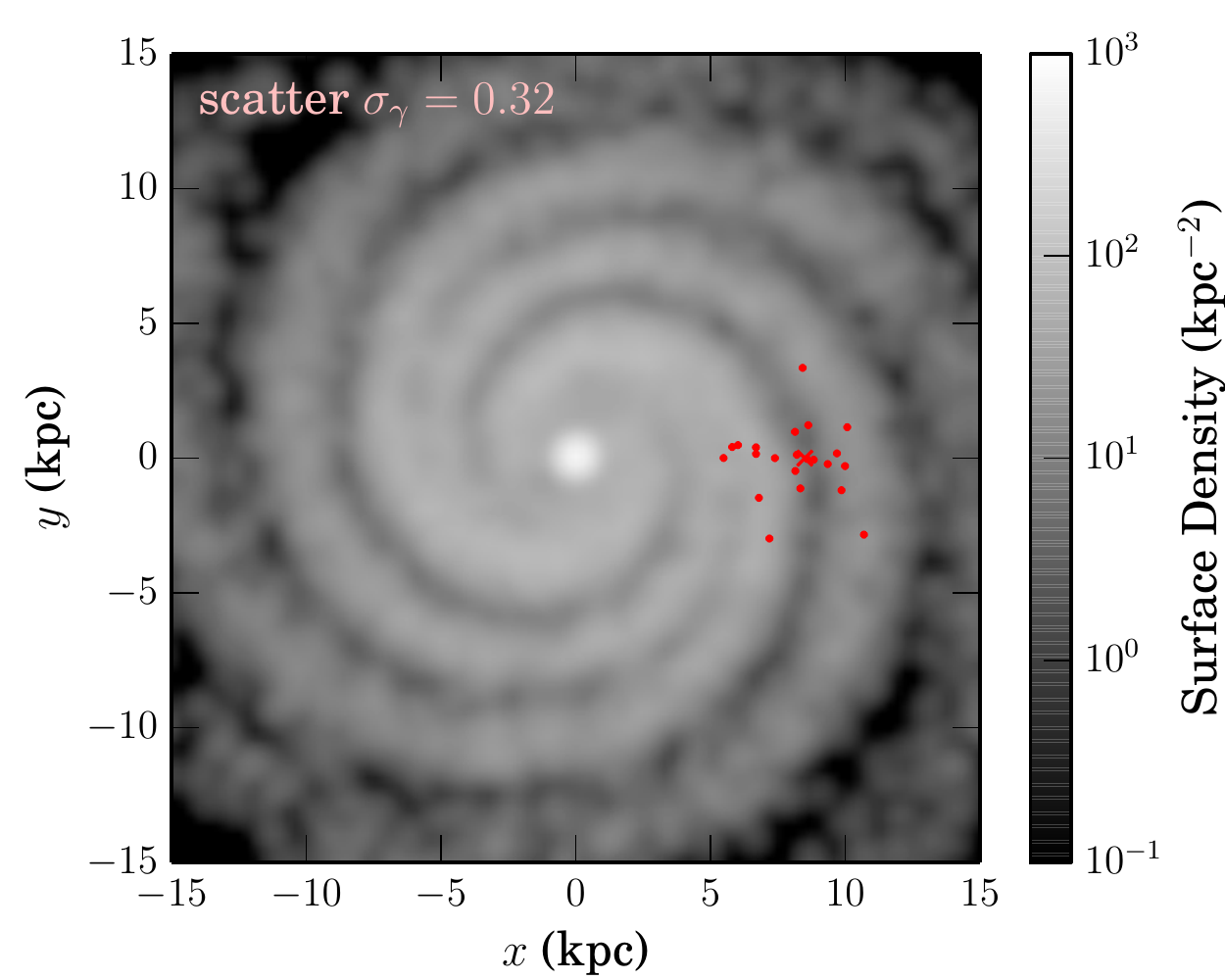}\\
   \includegraphics[width=.8888\columnwidth,clip=true]{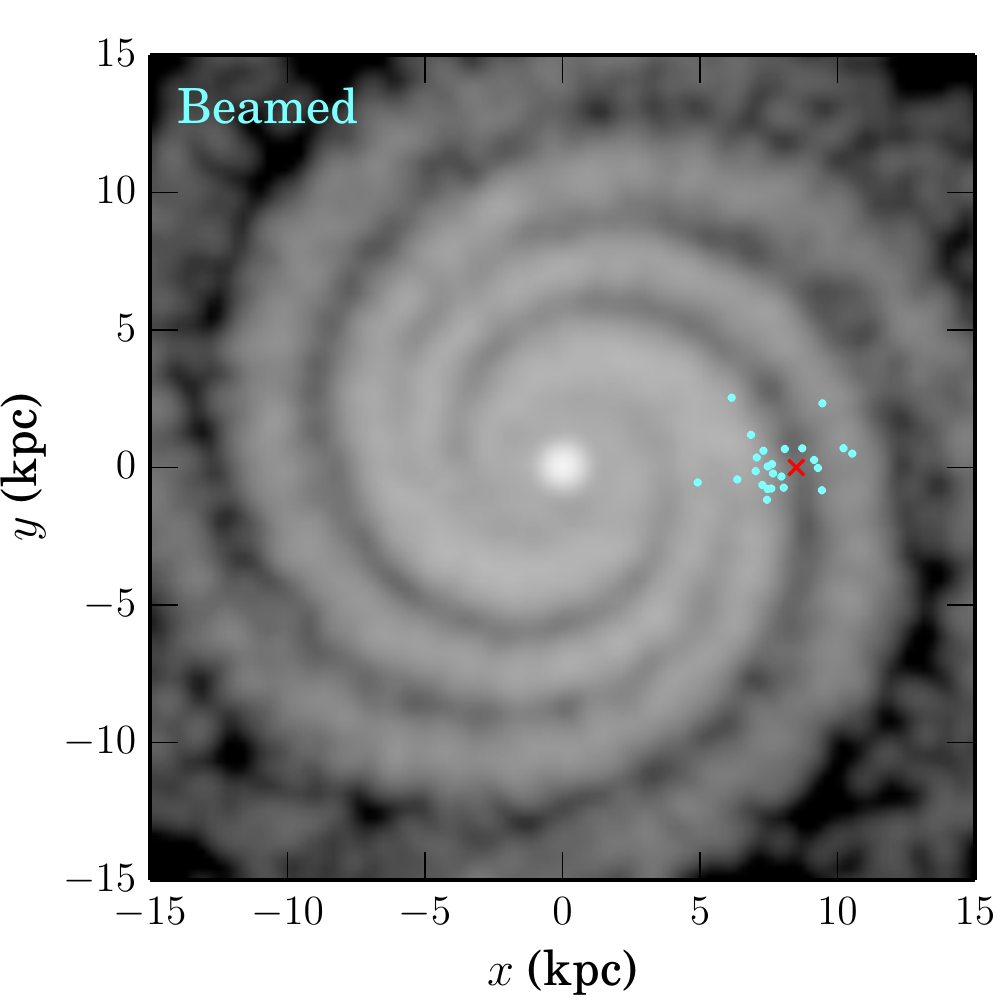}
   \includegraphics[width=1.1111\columnwidth,clip=true]{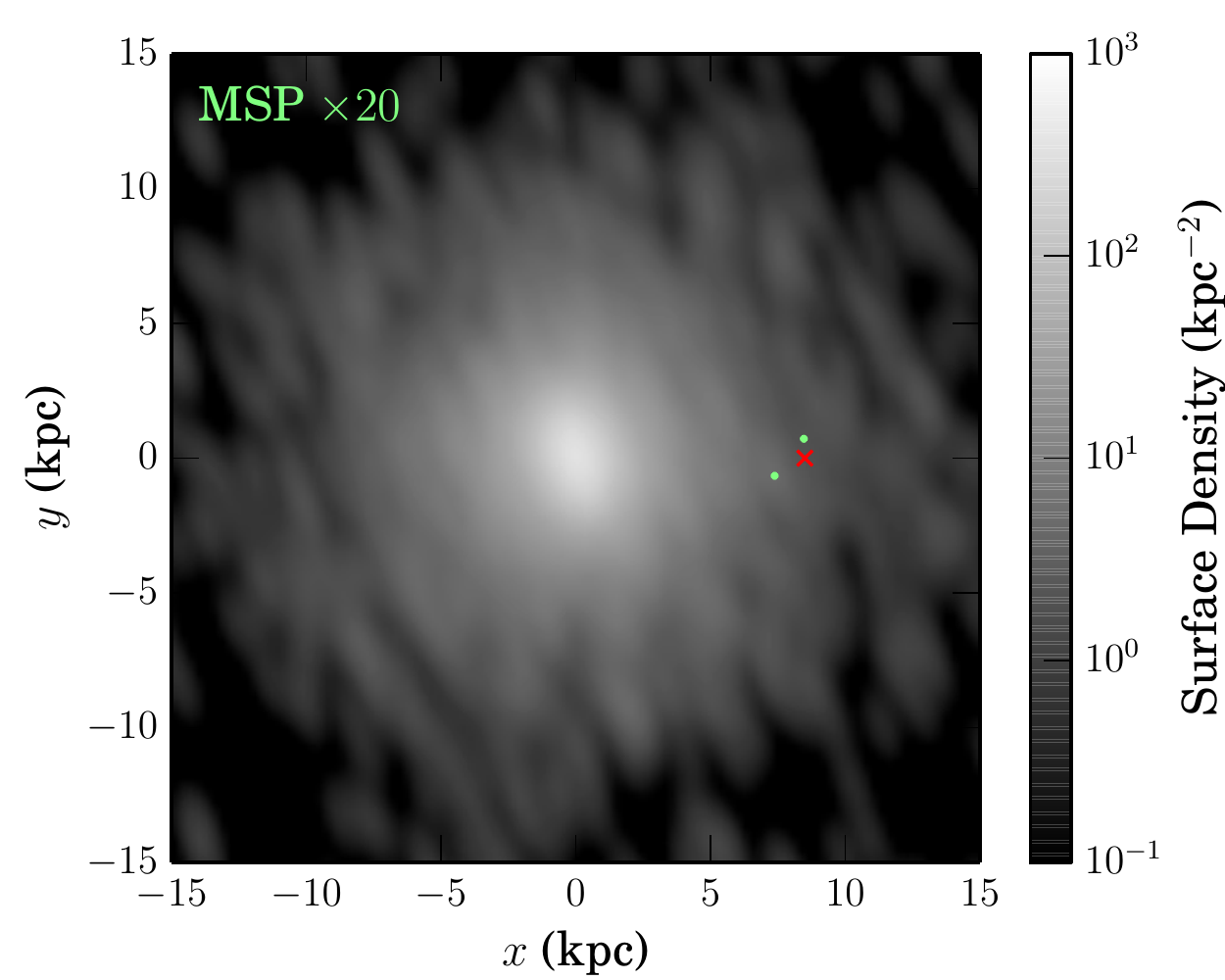}  
\caption{\label{fig:mwdist} Surface density of Galactic gamma-ray pulsars with $\dot{E} > 10^{33.5}\,$erg$\,$s$^{-1}$ in one realization from each of our four models. The panels show, from the top-left, our Fiducial Model, our Scatter Model, our Beamed Model, and the $20\times$ enhanced  millisecond pulsar (MSP) model. The points are pulsars with fluxes greater than $10^{-10}\,\ergcms$. The red cross shows the location of the Sun.  }
 \end{figure*}

\section{Galactic Pulsar Population}
\label{sec:pop}
In this section, we discuss in greater detail the birth rate and birth sites of pulsars in the Milky Way incorporated into the population synthesis model developed in \cite{2015PF}, which tracks pulsar trajectories through the MW from birth to the present.

\subsection{Birth Rates}
\label{sec:birth}
The rate of pulsar birth in the MW should generally follow from the rate of formation of stars sufficiently massive to yield a core-collapse (CC) SN, $M_* \gtrsim 8\,M_\odot$.  This is not instantaneous though, since even massive stars take $\sim\,$several--tens of Myr to evolve to explosion \cite{2013A&A...558A.131G}, so that the SN rate is based on the star formation rate (SFR) smoothed over time.

Galactic SFR estimators \cite{chomiukpovich2011} or estimates of the SN rate \cite{2011MNRAS.412.1473L} point toward a CC SN rate of $\sim\,2$ per century.  Estimates based on attempts to match the observed population of radio pulsars in the Galactic Disk typically imply higher rates, e.g., $\sim 2.8 \!\times\! 10^{-2}\,$yr$^{-1}$ in \cite{fauchergiguerekaspi2006} and $\sim 3.7 \!\times\! 10^{-2}\,$yr$^{-1}$ in \cite{2012A&A...545A..42P}.  Based on early {\it Fermi} pulsar data, \cite{2011ApJ...727..123W} estimated a MW disk gamma-ray pulsar formation rate of $\sim 1.7 \!\times\! 10^{-2}\,$yr$^{-1}$.  

Typically, the stellar evolution time delay will also smooth over SN locations in the Galaxy, so that fine structure becomes less important for the pulsar population in total, e.g., due to particular stellar clusters in spiral arms.  An exception to this would arise for star formation occurring in a location fixed in position over a period of tens of Myr.  The GC satisfies this condition, which has important implications.

While the prominent Arches and Quintuplet clusters are both within $\lesssim 50$~pc of the GC, there are young stars observed throughout the inner $\sim 200$~pc Central Molecular Zone (CMZ) region \cite{2010ApJ...725..188M}.  There is also a sizable, sustained SFR within the central few parsecs \cite{genzeletal2010}.

For young gamma-ray pulsars, we are interested in the SN rate over the past few~Myr.  While the Galactic disk rate should remain nearly proportional to the current average SFR, this does not necessarily hold in the GC.  A number of indicators suggest that the CMZ SFR may be episodic on $\sim 10\,$Myr periods \cite{2014MNRAS.440.3370K}, and could have been much higher in the past \cite{2011MNRAS.413..763C,2014MNRAS.444L..39L,2015arXiv150507111K}.  One way to circumvent this difficultly is to count the number of massive stars that are close to explosion (effectively integrating over the past SFR convolved with stellar evolution).  Overall, these considerations make estimating the present SN rate in the GC less direct than simply measuring and converting from the present SFR.

At high masses, where evolution is fastest, \citet{2015MNRAS.447.2322R,2015MNRAS.449.2436R} estimate the number of Wolf-Rayet (W-R) stars (arising from stars with initial mass $\gtrsim\! 25\,M_\odot$) in the CMZ ($\approx\! 250$) and the entire Galaxy ($\approx\! 1200 \pm 200$), finding a $\sim\! 20\%$ CMZ fraction.  For $\approx\! 250$ W-R stars with an estimated W-R lifetime of $\sim\! 0.25$~Myr \cite{2015MNRAS.447.2322R,2015MNRAS.449.2436R}, the implied CMZ SN rate from W-R stars alone is $\sim\! 10^{-3}\, {\rm yr}^{-1}$, without even considering any $M_i \lesssim 25\,M_\odot$ lower-mass stars or stars that traverse different evolutionary channels.

\citet{figer2004} used top heavy initial mass functions (IMFs) to arrive at SFRs $\sim1-2\%$ of the total Galactic rate.  We are interested only in massive stars.  They estimate $\sim$~150 O-type stars in Arches, about 3 times this in all the clusters, and 6 times this in the entire region, or $\sim$~900 O stars.  Taking a normal IMF gets us to $\sim$~(8/20)$^{-1.35} \times 900 \sim 3100$ stars above SN threshold in $\sim 3 \times 10^6$~yr of star formation, or $\sim 10^{-3}$~SN~yr$^{-1}$.
This is consistent with $\sim 10\%$ of all O / W-R stars being born in the GC.

On the present GC SFR, there are young stellar object (YSO) counts both within and outside of clusters throughout the CMZ.  These rates depend on the region under consideration and depend on the amount of time objects are in the YSO phase and estimations of contamination from older stars.  Galactic SFR estimates from YSOs tend to be on the low side of the various measures \cite{chomiukpovich2011}, despite concerns for contamination.  A SFR of 0.06$\,M_\odot\,$yr$^{-1}$ within $b < 1^\circ$ and $l<1^\circ$ is derived from free-free emission in \cite{2013MNRAS.429..987L}, though this type of measure also tends to be on the low side in the MW \cite{chomiukpovich2011}.

Going from SFR to SNR also contains a dependence on the IMF  and binary fraction of massive stars.  A number of studies have found evidence for a top-heavy IMF in the central parsec \cite{bartkoetal2010}.  Previous indications \cite{1999ApJ...525..750F} of a top-heavy IMF in the Arches and Quintuplet clusters have since been revised \cite[e.g.,][]{2009A&A...501..563E}, suggesting that the IMF in this environment is consistent with the typical \citet{1955ApJ...121..161S} IMF at the high mass end.  In all cases in the GC low mass main sequence stars have yet to be observed, making definitive estimates of the IMF challenging.  

One could bypass stellar arguments by counting supernova remnants, with a substantial number observed \citep{2015MNRAS.453..172P}.  Complications in converting to a SN rate include uncertainty in the duration of their visible lifetimes, Poisson variations in the very-recent rate, and clustering of multiple nearby SNe resulting in only a single superbubble.  For example, the XMM superbubbles in \cite{2015MNRAS.453..172P} may have resulted from multiple SNe and there are likely more yet unobserved SN remnants.

We assume for simplicity that the pulsar formation rate is constant in time and that the GC CMZ has a pulsar formation rate $7\,\%$ that of the Galactic rate, $1.5\times 10^{-3}\,{\rm yr}^{-1}$.  Considering the mass-dependent delay time between stellar birth and explosion \cite{2013A&A...558A.131G}, along with a declining IMF, the SN rate typically evolves in a much smoother manner than an episodic SFR.  This is roughly the average of the SFR over the prior $\sim30$~Myr, varying at the tens of percent level over Myr timescales.  If there were one or more CMZ star formation outbursts over that period, the SN rate today could easily be a factor of a few higher than inferred from the present SFR (depending on the outbursts intensities and durations).  Stars formed in clusters may become dispersed after disruption on a $\sim\,$few~Myr timescale \cite{2014MNRAS.440.3370K}.  We note that rates do not include exotic outcomes such as magnetar or black hole formation.

\subsection{Young Pulsar Birth Sites and Population Synthesis}
We generate the Galactic population of young gamma-ray pulsars by using a two component model to account for pulsars born in the Galactic disk and the CMZ, as in  \cite{2015PF}, with Fig.~\ref{fig:dsnr} showing the distribution of Galactic SNe in our model.  As detailed in \S~\ref{sec:birth}, we assume that pulsars are born in the disk uniformly through time, with a rate of $2.1\times10^{-2}\,$yr$^{-1}$  \cite{fauchergiguerekaspi2006,2011MNRAS.412.1473L}.  The pulsars start on circular orbits within the spiral arms of the MW using the parameters in \citet{fauchergiguerekaspi2006}, who successfully modeled the radio population of young pulsars.  Pulsars are formed following an exponential disk with scale radius of $3\,$kpc and scale height of $50\,$pc.  Each pulsar is placed in one of the four spiral arms, modeled as
\begin{equation}
  \theta(r) = k \ln(r/r_0) + \theta_0 \,,
\end{equation}
using the parameters derived from \cite{1992ApJS...83..111W}.  We assume that stars form equally across the four spiral arms. The distribution of known radio pulsars suggest that the sprial arms are not equally important.  We note that we use a slightly different coordinate system than \citet{fauchergiguerekaspi2006}, rotated by $90^\circ$.

For our Fiducial Model, we assume that the birth rate of pulsars in the CMZ is also constant in time, with a rate of $1.5\times 10^{-3}\,$yr$^{-1}$, which corresponds to approximately $7\,\%$ of the Galactic disk rate.  This is still significantly smaller than the observed fraction of Wolf-Rayet stars in the GC ($\approx 20\,\%$), suggesting that the total birth rate may be substantially higher.  We will explore variations of the CMZ SN rate in Appendix~\ref{sec:cmz}.   We assume that the CMZ pulsars are born in the Galactic plane in the CMZ between 20 and 200$\,$pc with scale height of $50\,$pc.
We initialize the orbits with random in plane motion, and follow a surface density profile $\propto r^{-1}$.

We assign each pulsar a kick randomly from a three-dimensional normal distribution with a $\approx 408\,\kms$ mean velocity \cite{2005MNRAS.360..974H}. The mean kick velocity is significantly larger than the three-dimensional velocity dispersion of GC stars $\approx 95\,\kms$ \cite{genzeletal1996}, which redistributes the CMZ pulsars throughout the Galactic Bulge.  We integrate each pulsar's orbit in the Galactic potential over its lifetime using  {\sc galpy} \footnote{Available at  {\protect\url{https://github.com/jobovy/galpy}}} with the \texttt{MWPotential2014} model \citep{bovy2015}.
We run our simulations for up to $10^8\,$yr, significantly longer than the lifetime of most gamma-ray pulsars, to ensure that we include the entire population.

In Fig.~\ref{fig:mwdist}, we show the resulting surface density profile of gamma-ray pulsars in the MW for the four models that we describe in the next section. The individual points for each model show the bright end of the pulsar population, each with a flux greater than $10^{-10}\,\ergcms$ (we will call pulsars brighter than this flux limit 'bright' throughout the text). The location of the Sun is shown by the red cross.  Since the typical young gamma-ray pulsar has a lifetime that is short compared to its orbital time, the pulsars largely map their birthplaces in the spiral arms of the MW.

In our model as in \citet{fauchergiguerekaspi2006}, the Sun is located between the Carina-Sagittarius and Perseus spiral arms, 8.5\,kpc from the GC.  Unlike \citet{2011ApJ...727..123W}, we do not include the local Orion spur in our analysis, which may systematically alter the number of bright pulsars in our simulations.   As we are mostly interested in the distant and faint pulsars that contribute to the unresolved background, our results remain mostly unaffected.

\subsection{Millisecond Pulsars}
\label{sec:msp}
The existence of MSPs as strong gamma-ray emitters was one of the surprises of the {\it Fermi} mission.  It was thought previously that they might emit through a ``pair starved'' process, tapping most of the volume above the polar cap for acceleration and emission and thus producing gamma-ray beams with shapes similar to radio beams.  Instead, it is clear they have outer magnetosphere emission with plenty of pairs to keep the gaps thin -- thus sharp caustic-like peaks.  It's not yet understood why the pair formation is so efficient -- it's possible that the compact MSP magnetosphere allows multipolar fields to contribute, and/or surface X-ray emission.

In stark contrast to the young pulsar population, MSPs are created long after the SN that produced the neutron star via the spinning-up of formerly-young pulsars by stellar material accreted from a binary companion.  Due to their weaker magnetic fields MSPs can remain gamma-ray bright for $\sim 10^9-10^{10}\,\yr$.
The contribution of unresolved MSPs to the gamma-ray background has been discussed for both the entire sky \citep{2010JCAP...01..005F}, as well as the GC \citep{2011JCAP...03..010A,2012PhRvD..86h3511A,2013MNRAS.436.2461M,  2013PhRvD..88h3009H,2014JHEAp...3....1Y,2014ApJ...796...14C}. Here, we forward model the contribution of MSPs to the GC Excess by using a similar process as we have used for young pulsars.

We assume instead that the re-birth places of MSPs follow the present-day MW stellar distribution.  In particular, we consider two components.  We assume that MSPs form continuously over time in either the Galactic Disk or the Galactic Bulge.  To generate the disk population of MSPs, we assume that the MSPs form on circular orbits in a smooth exponential disk. For each MSP, we assign the surface density of MSPs from an exponential distribution with a scale of $2.7\,$kpc, and a scale height of 50\,pc. To generate the Bulge population, we assume that the pulsars formed on isotropic orbits, from a $\Gamma$ distribution with an inner profile of $k = 1.2$, and a characteristic scale of $1.9\,$kpc.  In our Fiducial MSP Model, MSPs are randomly assigned to the disk or bulge populations in proportion to their mass ratio, $68:5$ respectively \citep{bovy2015}.

Although MSPs may not receive a substantial kick at birth, dynamical processes heat up the population over time through repeated gravitational scatterings. In our model, each MSP receives a random kick using a three-dimensional normal distribution with $\sigma_{\rm 1D} = 80$~km\,s$^{-1}$ \cite{1998MNRAS.295..743L}.  Some radio observations  suggest that the kick velocities of MSPs may be slightly smaller, with $\sigma_{\rm 1D} \approx 48$ \cite{1997ApJ...482..971C}. We have modeled the MSPs with this smaller velocity and found consistent results. We model the initial magnetic field distribution similarly to the ordinary pulsars, with a lognormal distribution, with mean of $\left<\log_{10}{B}\right> = {8.3}$ where $B$ has units of Gauss.  The dispersion is 0.3 dex.  The period distribution at re-birth follows from earlier, with a scale of $5\,$ms and a minimum birth period of 1$\,$ms.   This type of distribution, while following a common methodology used for the young population, differs from the canonical population synthesis that has been used in the past for MSPs  \cite{1997ApJ...482..971C}.  Since we wanted to forward model the MSP population, we found this to be a more natural method.  Nevertheless, we find this reproduces the observed distribution  of the brightest gamma-ray MSPs. In particular it reproduces the observed pulse period distribution and the latitude distribution of the {\em Fermi} pulsars.

In contrast to the young pulsars, the re-birth rate of MSPs is known only indirectly.  We therefore generate the Galactic MSP population by continuously populating the galaxy until there are five MSPs with a gamma-ray flux greater than $4\times10^{-10}\,\ergcms$.  Above this flux, we assume that the MSP population is completely known from gamma-ray observations.  We find that the re-birth rate of MSPs in the MW disk is $\approx 8 \pm 3\times10^{-7}\,$yr$^{-1}$.  This is smaller than that derived in \cite{1997ApJ...482..971C}, $3.0^{+1.8}_{-1.2}\times 10^{-6}\,$yr$^{-1}$, at less than the $2\sigma$ level.

\section{Our Four Pulsar Emission Models}
\label{sec:models}
In this work we focus on four different pulsar models with parameters that we outline here.  The first three models are for young pulsars, focusing on the range of uncertainties in their emission properties:  our Fiducial Model, Scatter Model, and Beamed Model.  These three models all have the same birth properties that we describe in \S~\ref{sec:pop}.  The fourth model that we show throughout the work is for MSPs.  

{\bf \emph{Fiducial Model}}: Our Fiducial Model is the model presented in 
\citet{2015PF}.  We assume that there a one-to-one relation between $L_{\gamma}$ and $\dot{E}^{1/2}$ as in Eq.~(\ref{eq:lumfunc}) with $C = 1.3$.  We impose a firm cutoff in the pulsar luminosity with $L_\gamma = 0$ for $\dot{E} < 10^{33.5}\,\ergs$.  Systematic variations of the model parameters are described in the Appendix. 

{\bf \emph{Scatter Model}}: The amount of scatter observed in the luminosity distribution of pulsars exceeds what is expected from the uncertainties in the distances to the pulsars alone.  For this reason, we assume that the luminosity of the pulsars follows the same relation as our fudicial model, but includes an intrinsic log-normal scatter. This is similar to studies of the radio luminosity functions of pulsars.  For each pulsar we randomly select the luminosity from a log-normal distribution with $C= 1.0$ for the mean of the log luminosity and a scatter, $\sigma_\gamma =  0.32\,$dex. If the gamma-ray luminosity is greater than the spin-down luminosity, we then set $L_\gamma = {\dot E}$.  The mean luminosity, under these conditions, is then equivalent to $\left< C\right >=1.3$.  This is still smaller than the mean luminosity of the observed pulsars, which have $\left< C\right >=1.6$. 

{\bf \emph{Beamed Model}}: A number of theoretical models for pulsar gamma-ray emission suggest that as pulsars age, their emission becomes strongly beamed (see \S~\ref{sec:beam}).  As such, we have included a model to approximate beaming using the relations we presented in \S~\ref{sec:beam}, for the YRQ beamed emission. In this model the average gamma-ray luminosity of all pulsars is equal to setting $C=1.0$, but a fraction $f_{\rm beam}$ of pulsars are beamed towards the observer, with a luminosity that is $f_{\rm beam}^{-1}$ greater than the sky average luminosity.

{\bf \emph{MSP $\times$20 Model}}:  The possible contribution of MSPs as an important constituent of the GC Excess has been discussed extensively (e.g., \cite{2011JCAP...03..010A,2012PhRvD..86h3511A,2013MNRAS.436.2461M,2013PhRvD..88h3009H,2014JHEAp...3....1Y,2015ApJ...812...15B}). In this model, we do not include the contribution of young pulsars, even though we argue young pulsars must contribute significantly in this region. As we will soon see, the assumption just discussed in \S\ref{sec:msp} of a direct correspondence between MSPs and bulge-to-disk mass ratio results in a low diffuse MSP GeV flux.

To see if the Excess might be explained by an overabundance of MSPs near the GC, we use the method described in \S\ref{sec:msp} with an enhancement of MSPs in the Galactic Bulge.  Such a boost might result if a fraction of the Bulge formed from the inspiral and disruption of globular clusters \citep[e.g., F.~Antonini (private comm.);][]{2015ApJ...812...15B}.  We find that to explain the full Excess would require a specific MSP formation rate
\begin{equation}
    20 \left(\frac{M_{\rm B}}{6\times 10^9\,\msun}\right)^{-1} \frac{M_{\rm D}}{6.8\times10^{10}\,\msun}
\end{equation}
times that expected from point source counts of the local disk MSP population, where $M_{\rm B}$ and $M_{\rm D}$ are the mass of the bulge and disk respectively.  The enhancement factor is degenerate with the assumed bulge mass, e.g., $M_{\rm B} \approx 2\times 10^{10}\,\msun$ \citep{2015arXiv151007425V} reduces the required boost to a factor of $6$.  Overall, the Bulge would need be composed of a few hundred massive globular clusters similar to 47~Tuc, a globular cluster known for its rich MSP population.  
We include this $\sim 20\times$-enhanced MSP model throughout our analysis, with the MSP luminosity function chosen to be the same as for young pulsars, with $C = 1.3$ \cite{2013pulsarcatalog}.

In Fig.~\ref{fig:lumfunc}, we plot the three different emission models we use for the young pulsars.  We also show the observed gamma-ray luminosities of pulsars, $L_\gamma$,  from the 2PC as a function of $\dot{E}$.  A number of the YRQ pulsars have no known distances, especially those with $\dot{E} \lesssim 10^{33.5}\ergs$, and so are not shown.  The MSP emission model is identical to our Fiducial Model with $C=1.3$, as shown by the solid blue line.

\section{Gamma rays from Galactic Pulsars}
\label{sec:gammaray}
%
\subsection{{\em Fermi} Point Sources}
\label{sec:ps}

 \begin{figure*}[]
 \includegraphics[width=0.28\textwidth,clip=true]{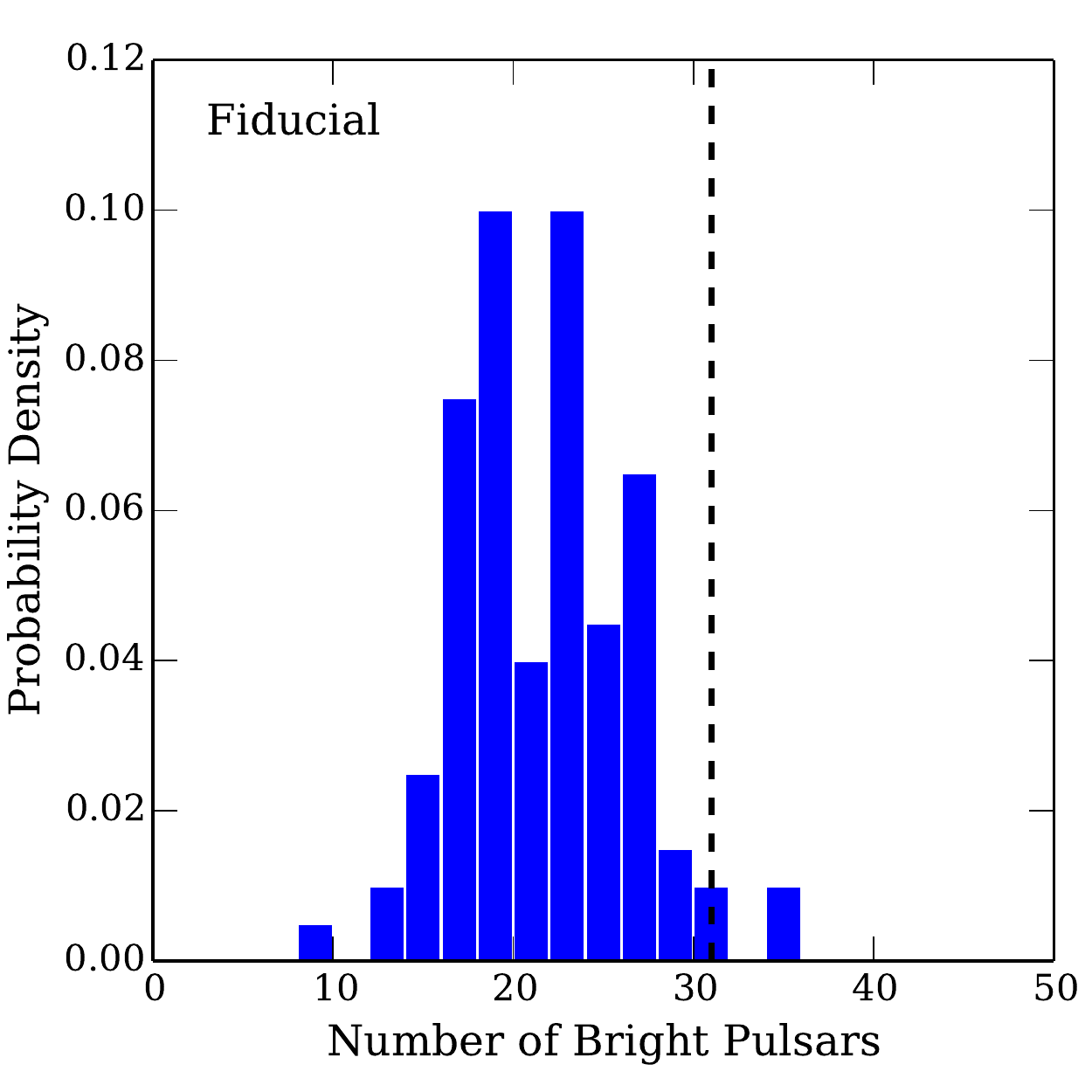}
 \includegraphics[width=0.28\textwidth,clip=true]{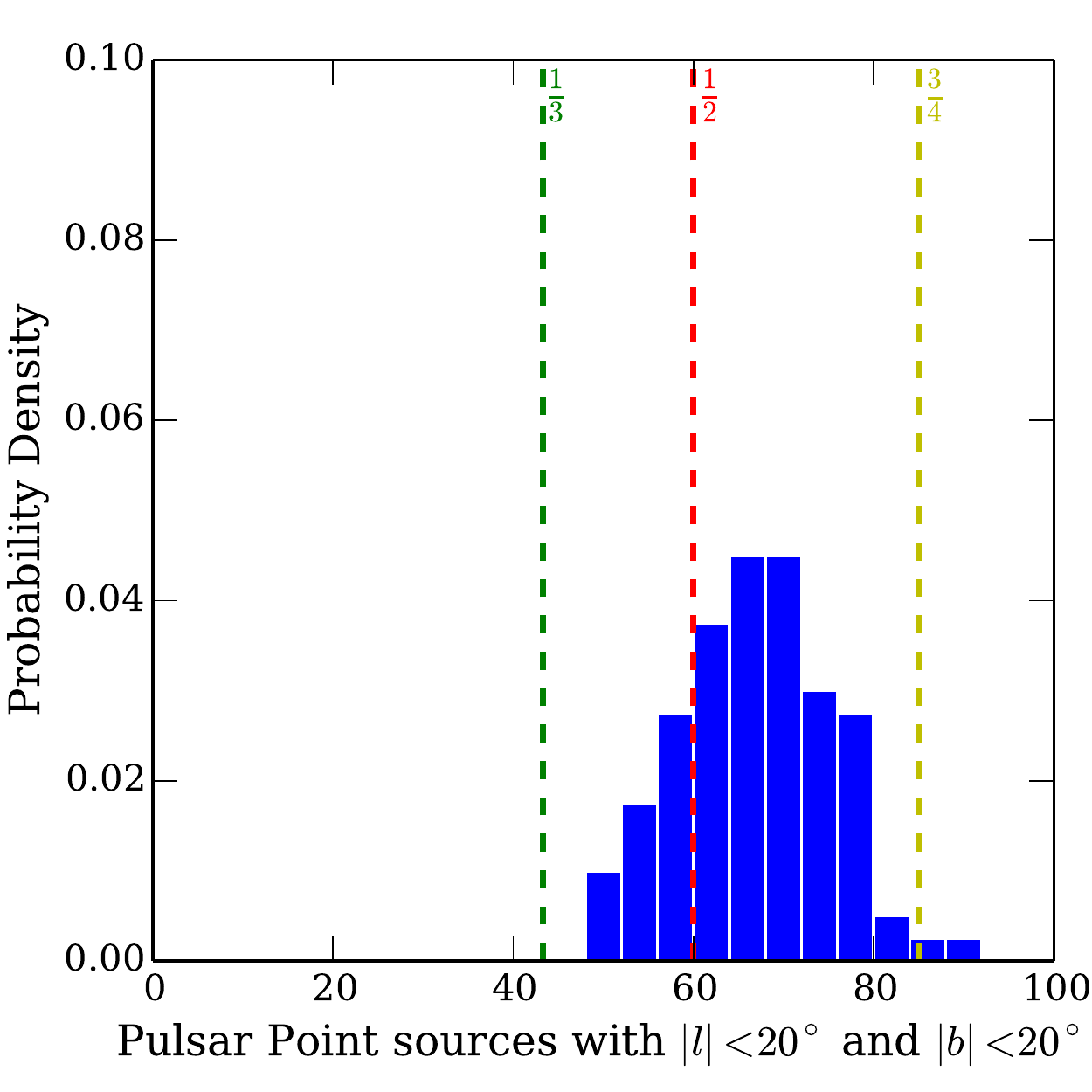}
 \includegraphics[width=0.28\textwidth,clip=true]{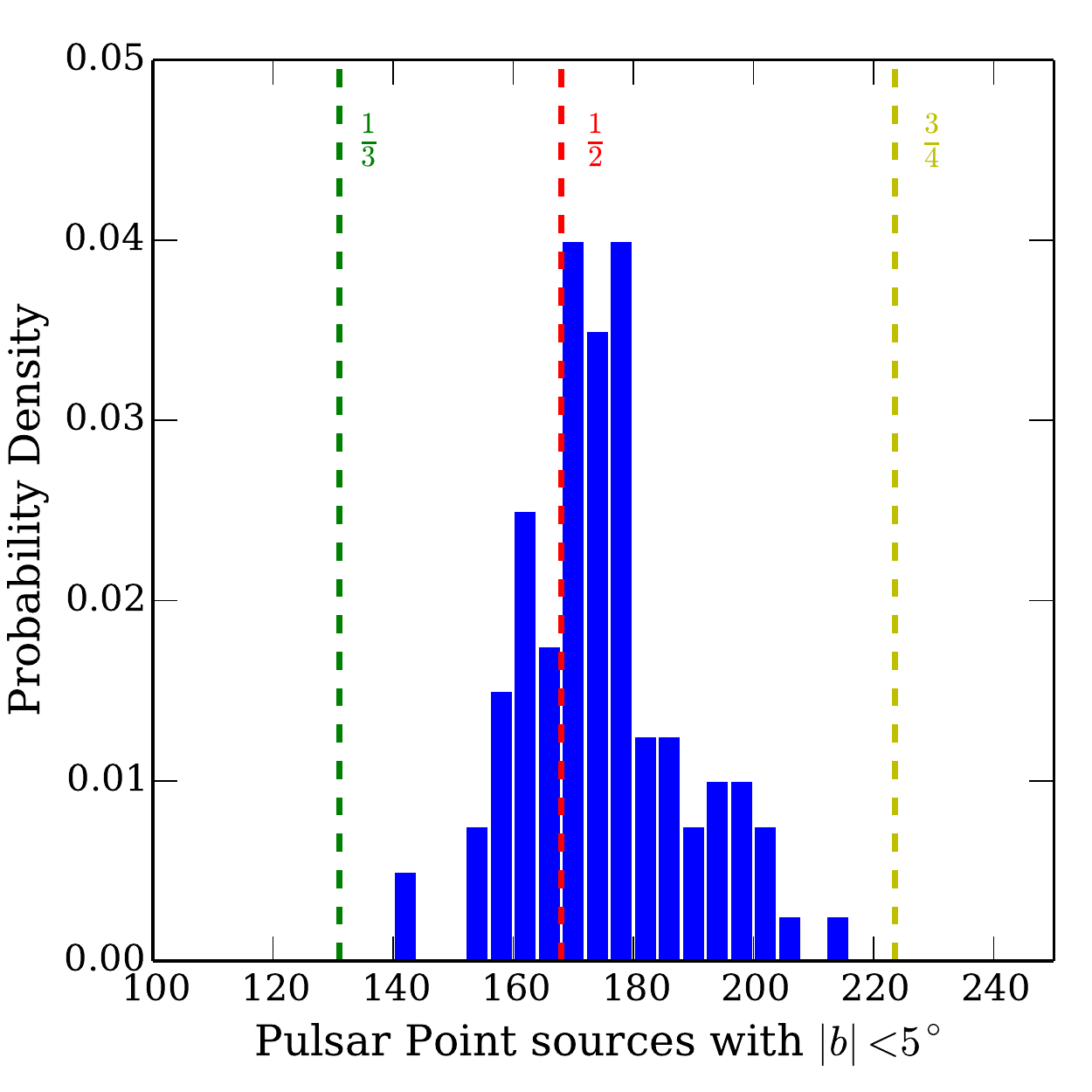}\\
 \includegraphics[width=0.28\textwidth,clip=true]{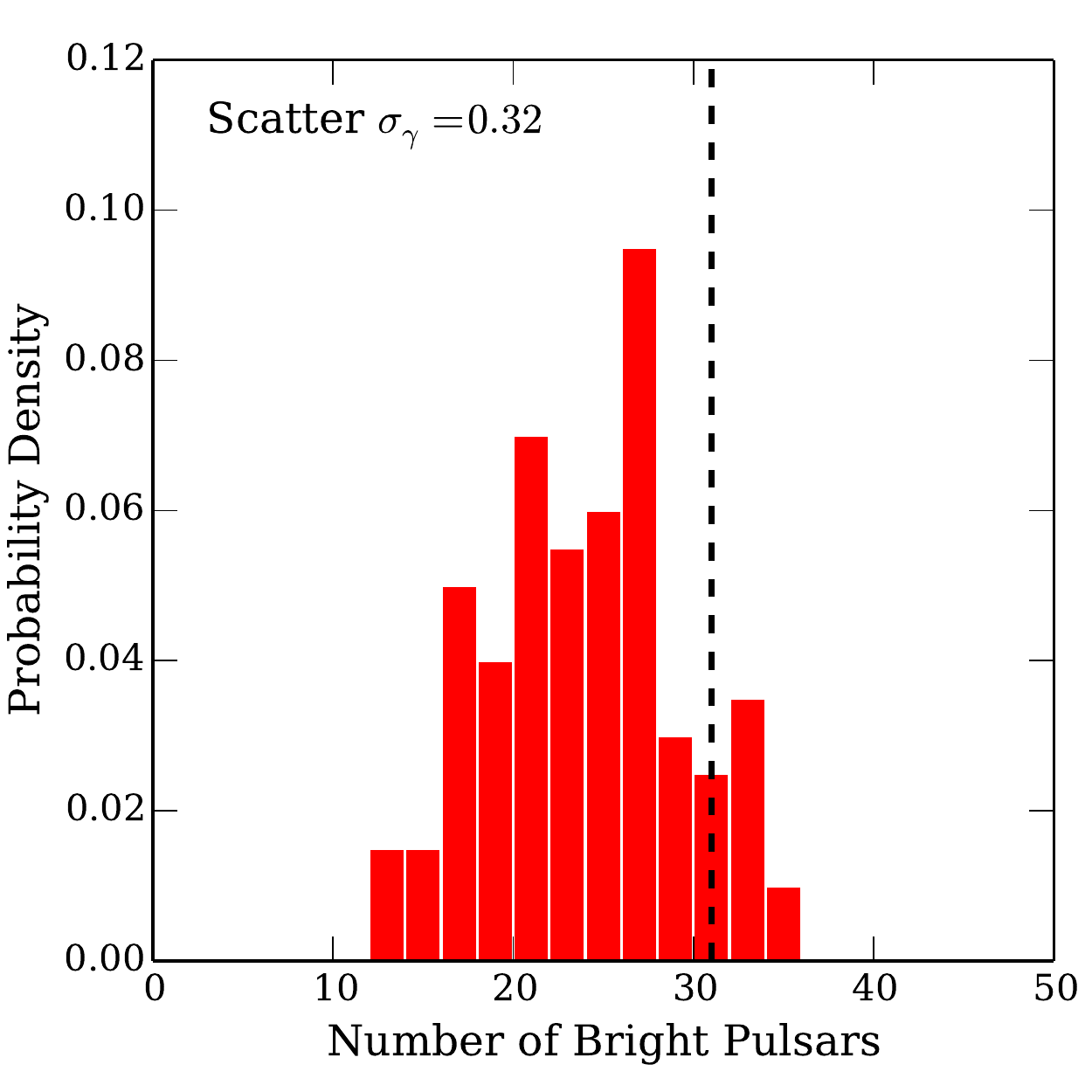}
 \includegraphics[width=0.28\textwidth,clip=true]{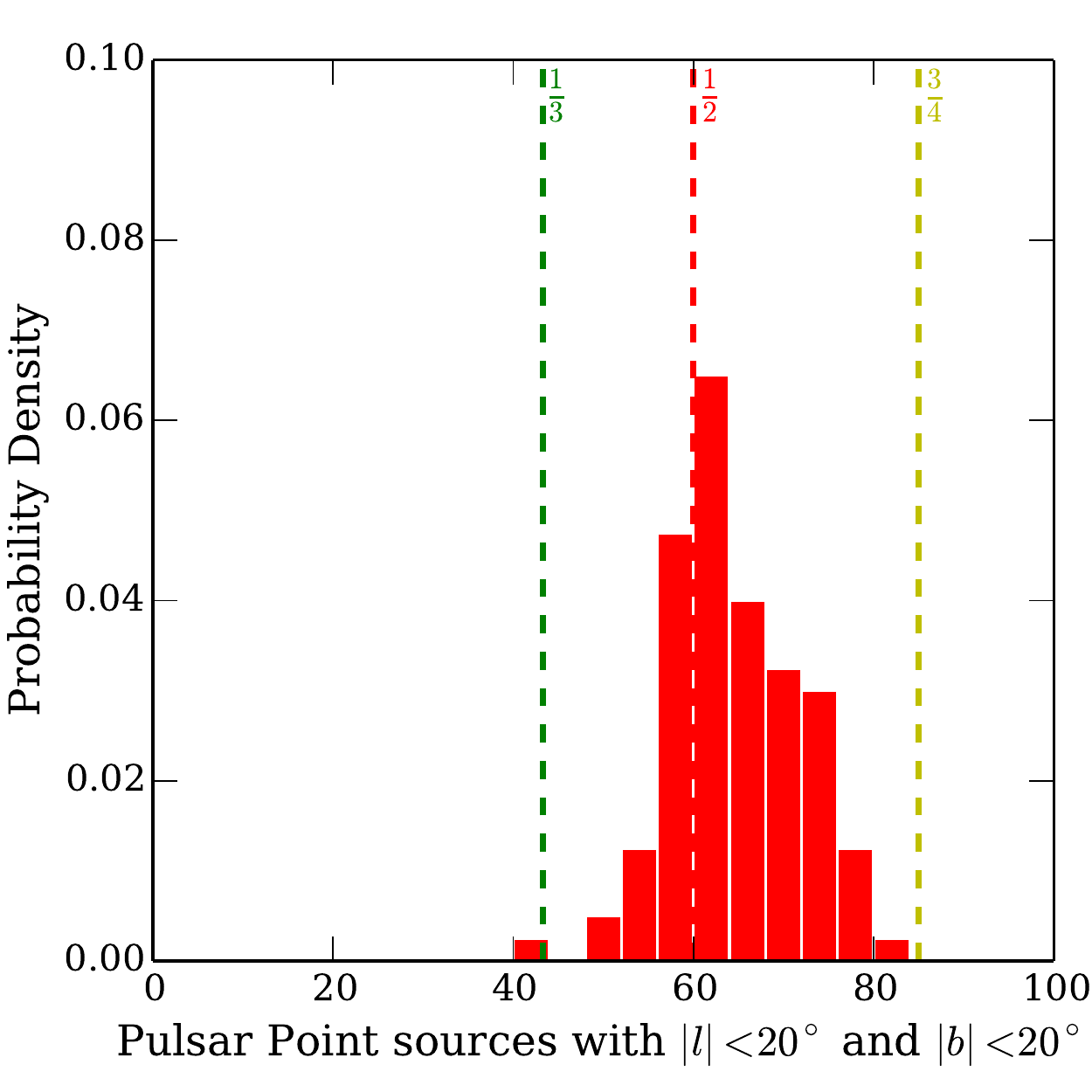}
 \includegraphics[width=0.28\textwidth,clip=true]{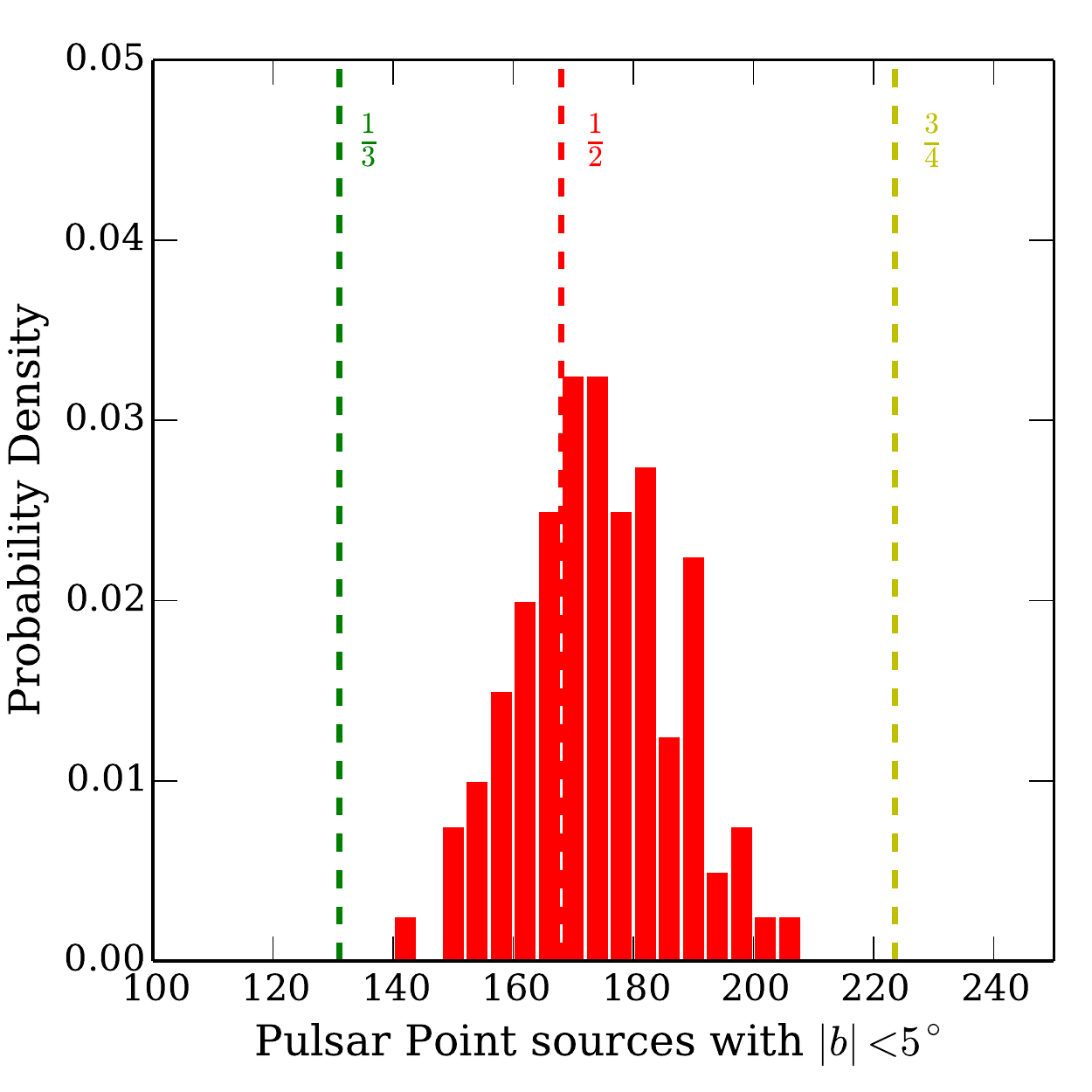}\\
 \includegraphics[width=0.28\textwidth,clip=true]{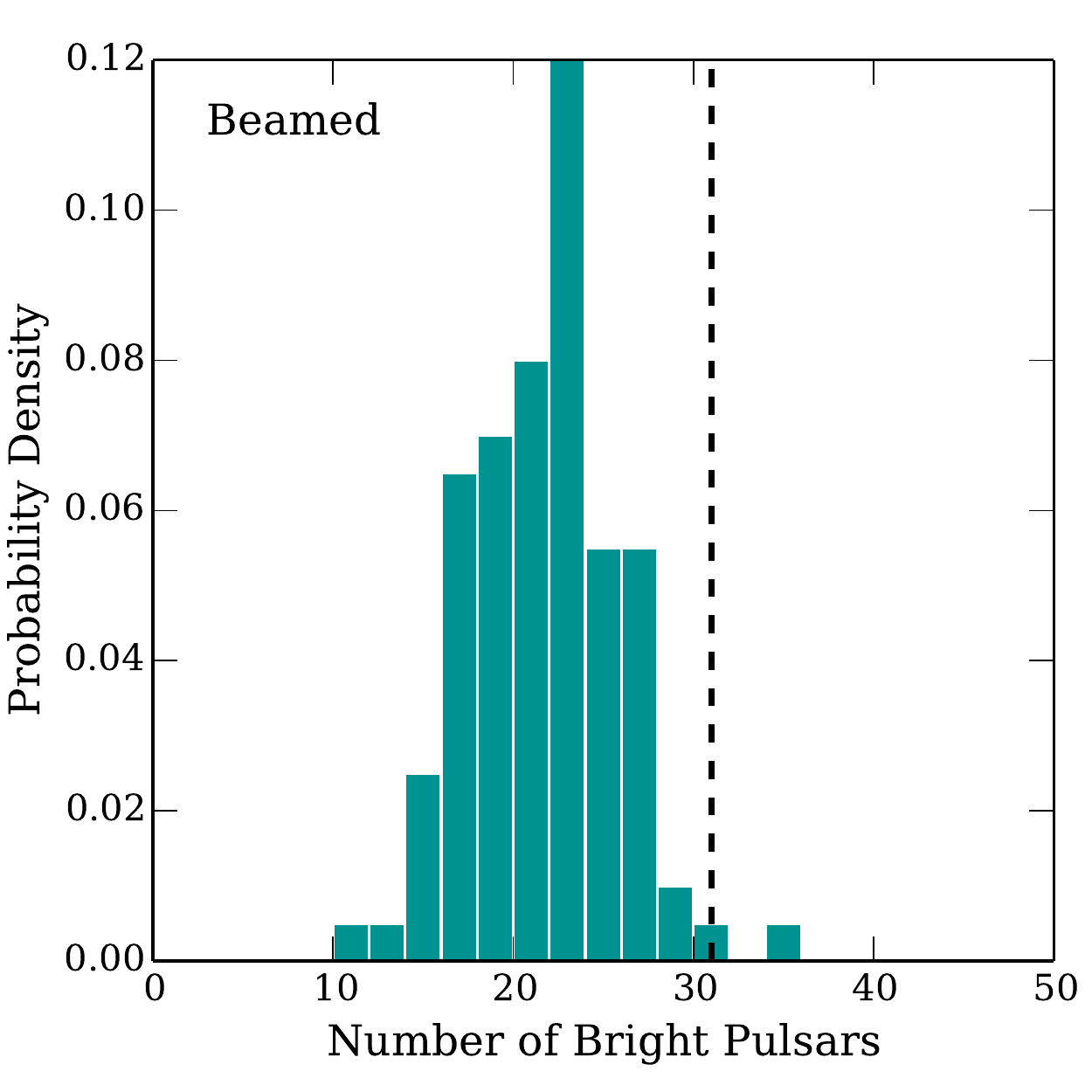}
 \includegraphics[width=0.28\textwidth,clip=true]{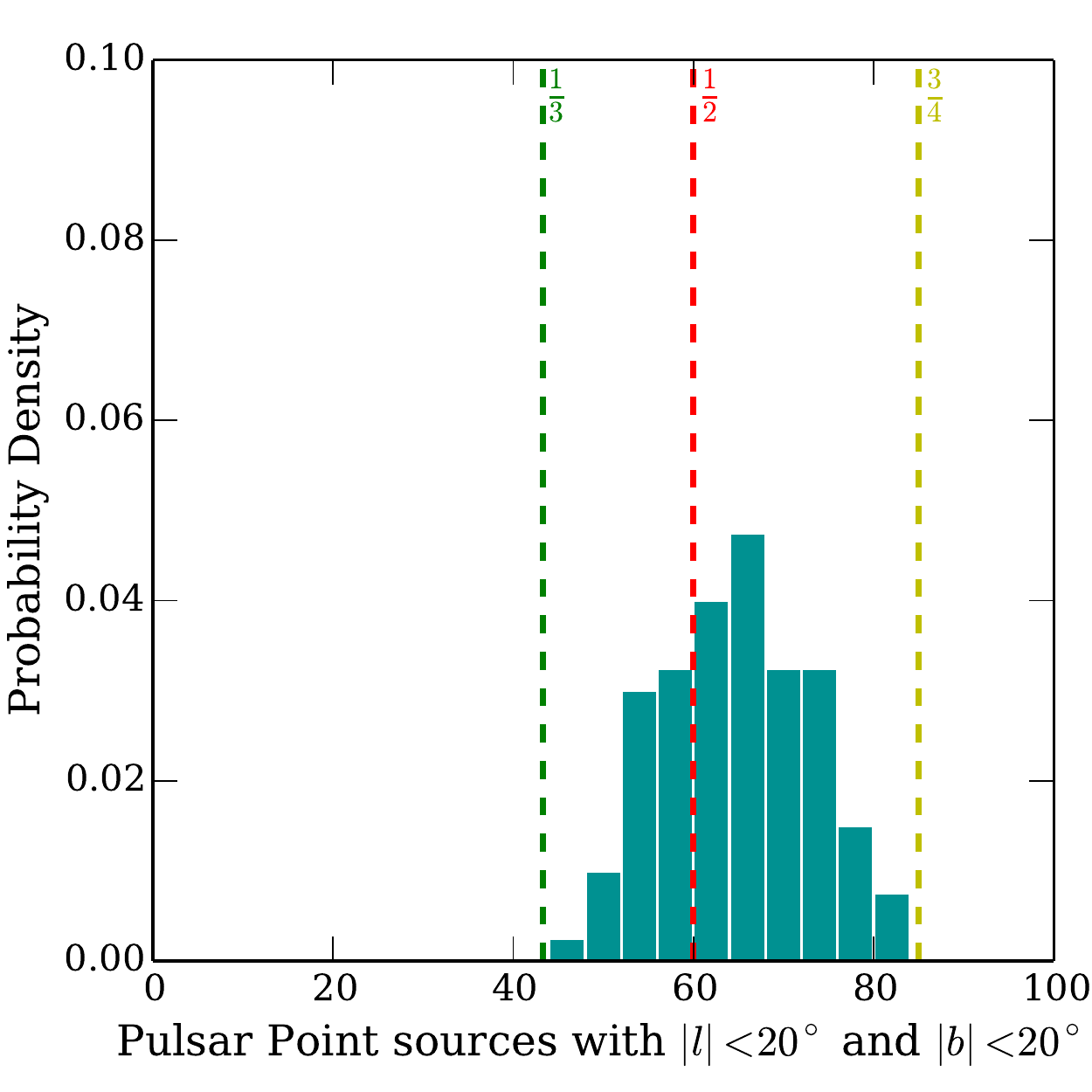}
 \includegraphics[width=0.28\textwidth,clip=true]{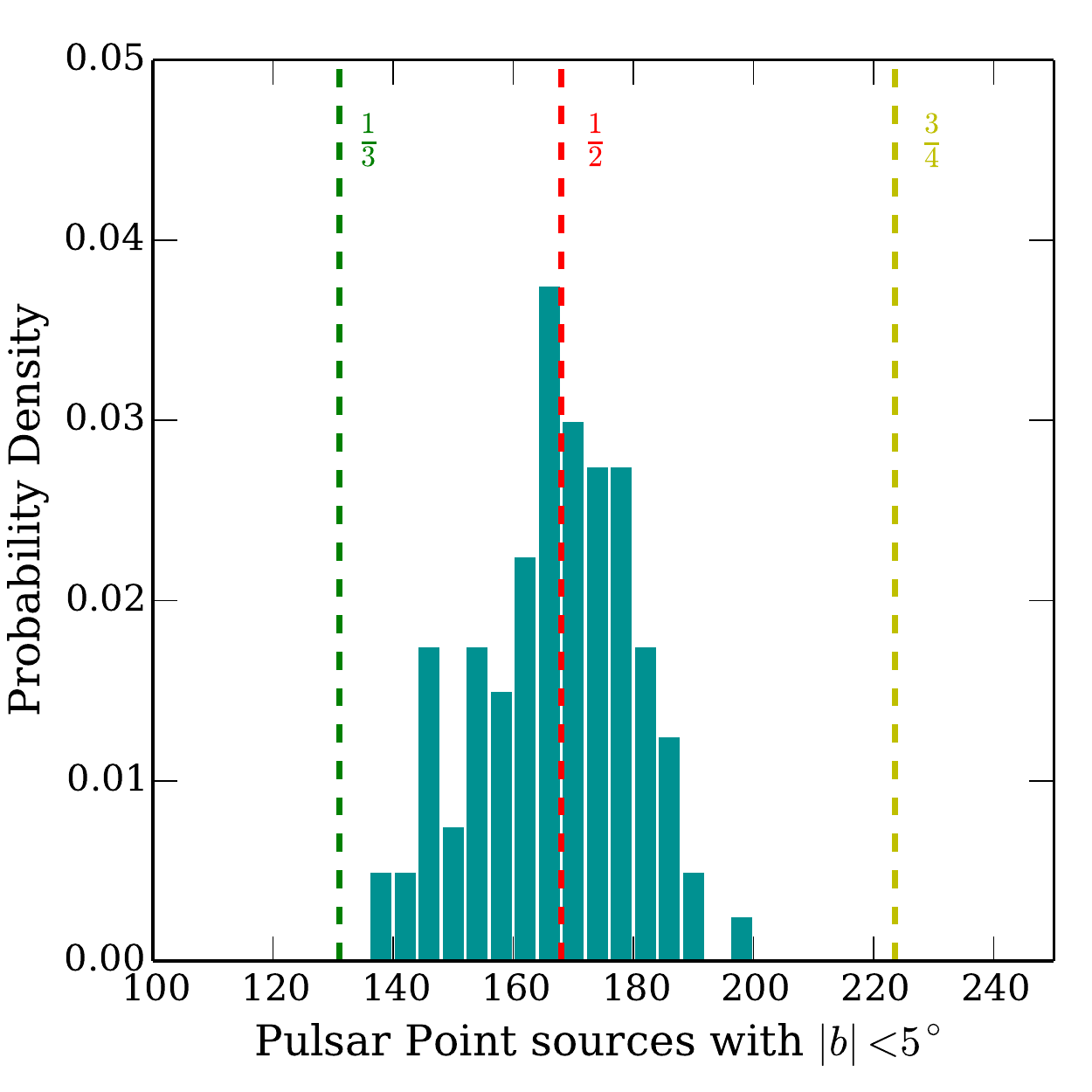}\\
 \includegraphics[width=0.28\textwidth,clip=true]{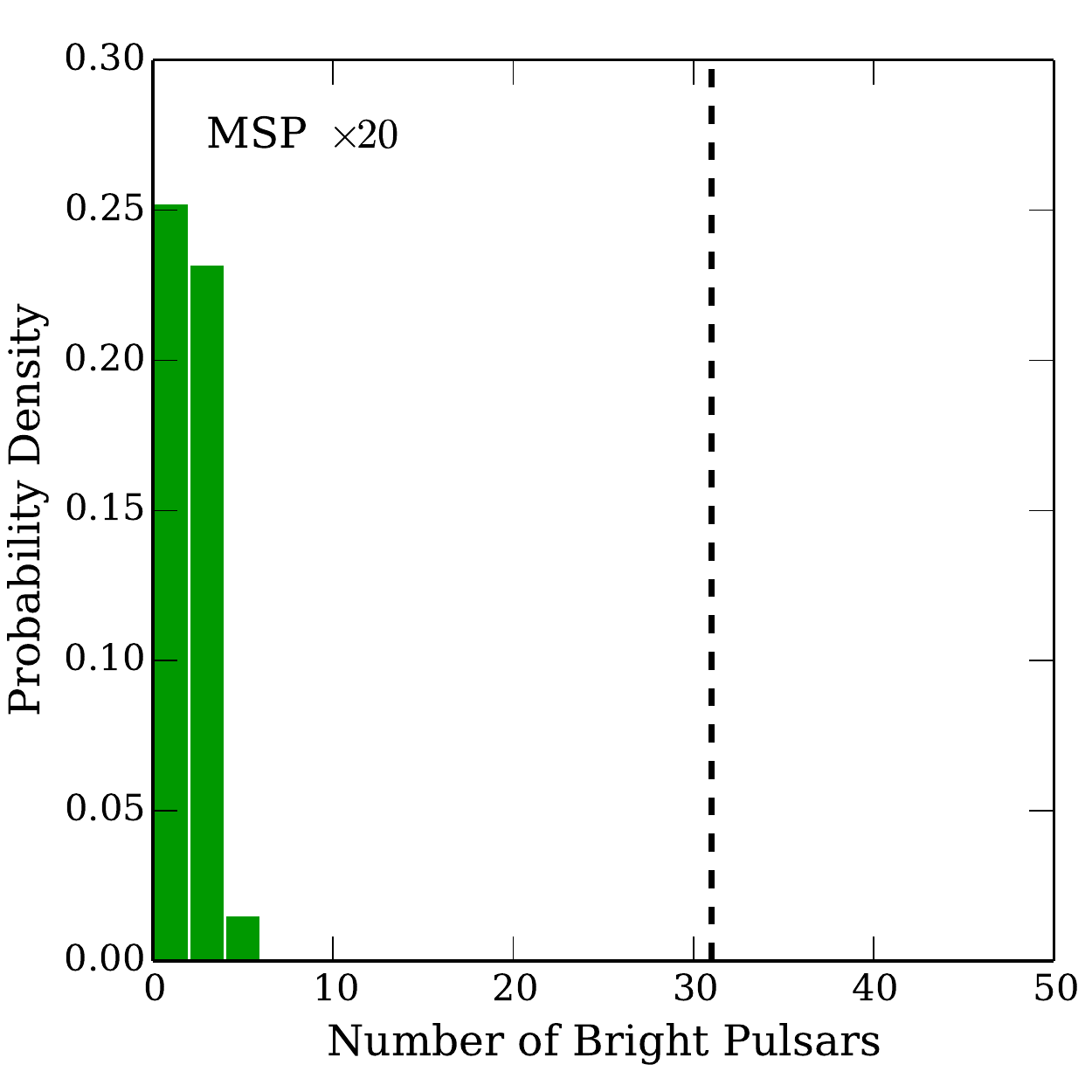}
 \includegraphics[width=0.28\textwidth,clip=true]{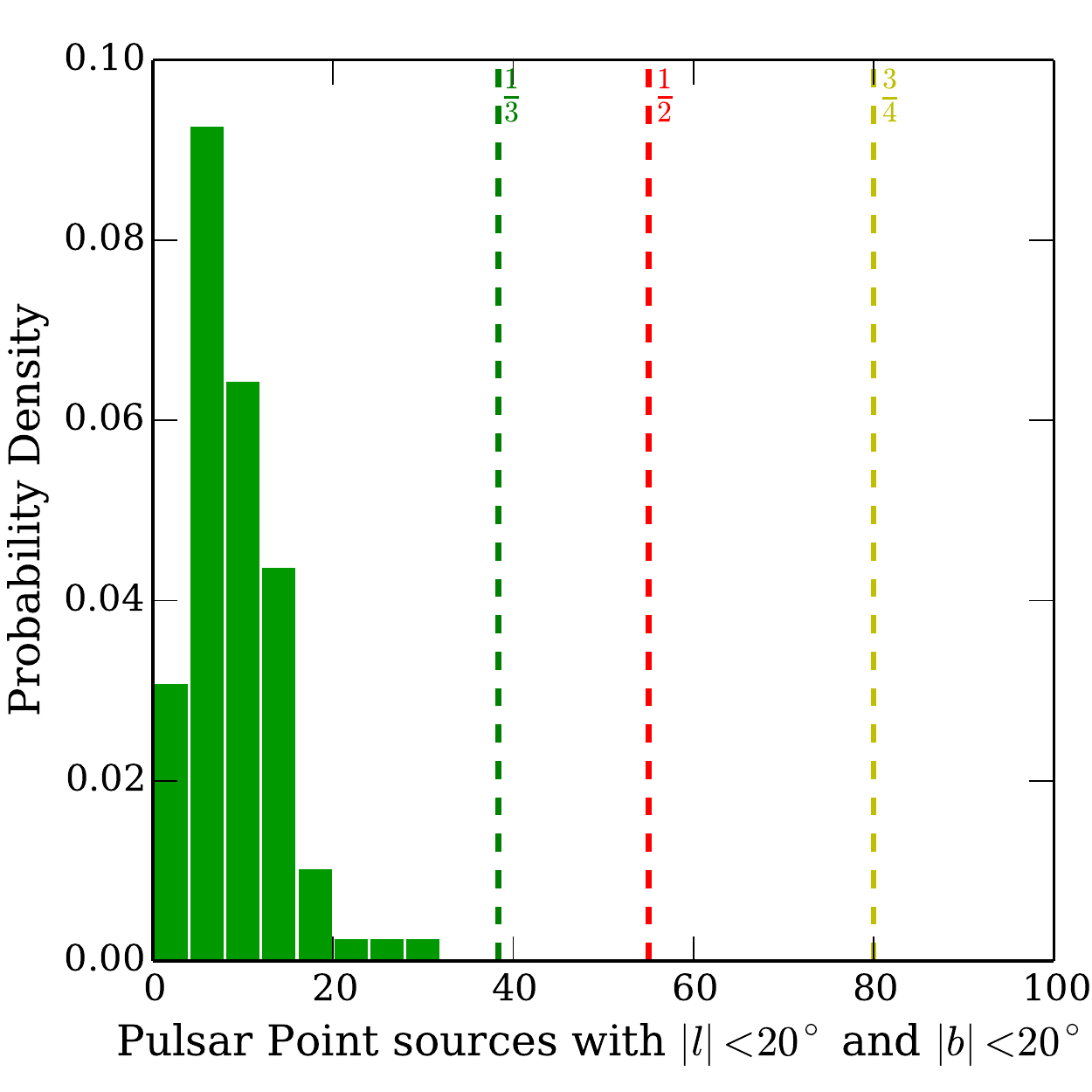}
 \includegraphics[width=0.28\textwidth,clip=true]{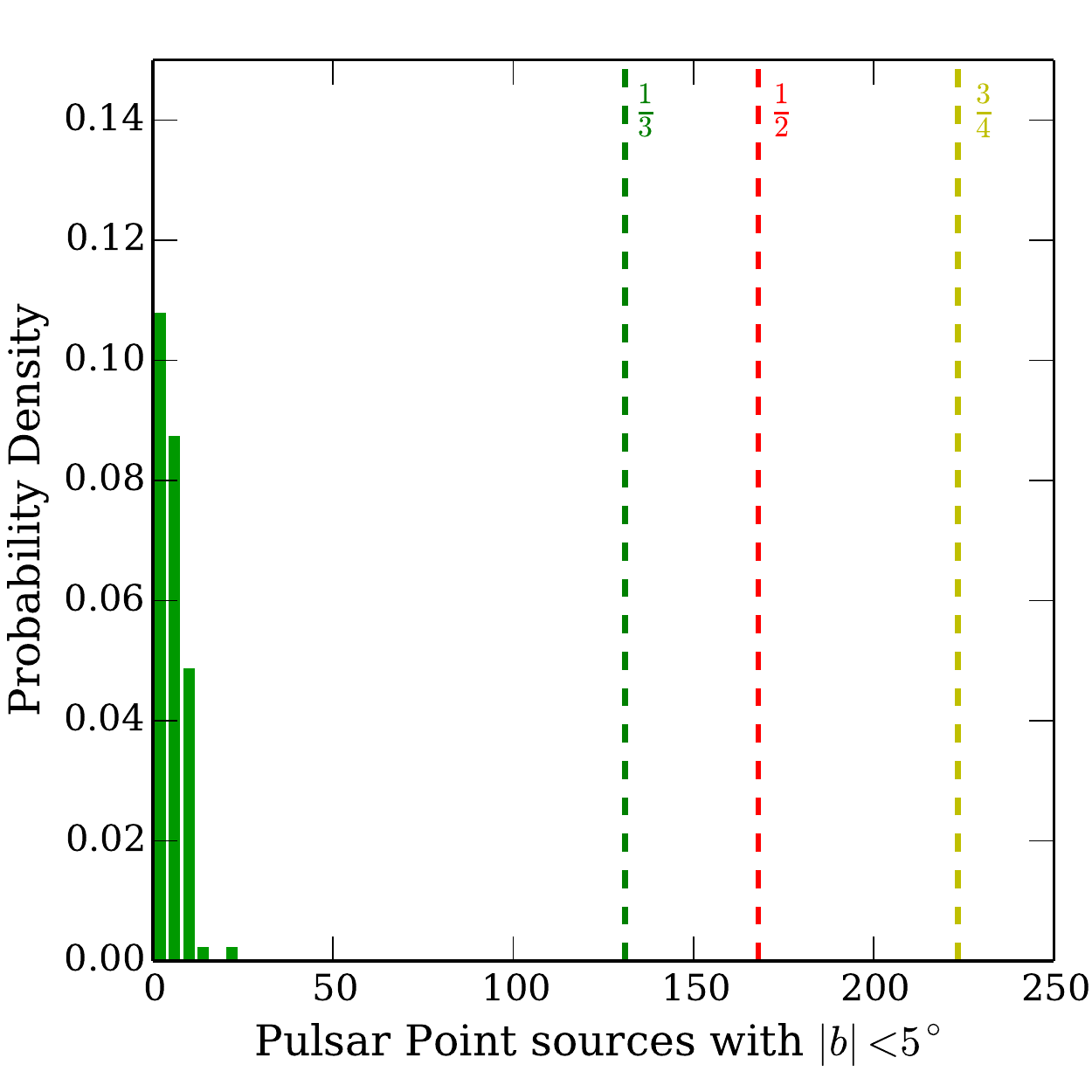}
\caption{\label{fig:ps} Comparison with {\em Fermi} point sources. Each row shows one of our four models in \S\ref{sec:models}.
Starting from the top: Fiducial Model, Scatter Model, Beamed Model, and $20\times$-enhanced MSP Model.  {\em Left} panels show the probability density of the total number of bright pulsars ($F >\! 10^{-10}\,\ergcms$) as determined from 100 random realizations of each model. These are compared to the number of detected pulsars, 31, denoted by the vertical black dashed line.
{\em Center} panels show the number of detectable point sources within $|l| < 20^{\circ}$ and $|b| < 20^{\circ}$, above the $50\%$ {\em Fermi} completeness limit.  The vertical dashed lines show the number expected if the pulsars constitute the denoted fraction of unidentified point sources in the same region from 3FGL.
{\em Right} panels show the number of detectable point sources within the Galactic plane ($|b| < 5^{\circ}$) above the $50\%$ {\em Fermi} completeness limit. The vertical dashed lines show the number expected if young pulsars constitute the denoted fraction of unidentified point sources in the same region from 3FGL.  {\em Note that the bottom (MSP) row uses a different vertical scale.}}
 \end{figure*}

A requirement of any model of the unresolved pulsar population, is that it, at a minimum, not overproduce the number of observed point sources discovered by {\em Fermi}.  Ideally, it should also reproduce the bright {\em Fermi} pulsar luminosity function.

To determine if a pulsar is detectable as a point source by {\em Fermi}, we developed an ad hoc model for the detection of point sources in the {\em Fermi} 2FGL \cite{2012ApJS..199...31N} and 3FGL \cite{2015arXiv150102003T} point source catalogs near the GC \cite{2015PF}.  The 3FGL catalog is complete to much fainter point sources, and provides a more robust constraint on the number of point sources overall.  Most previous studies of the GC diffuse background have used 2FGL (or an approximation to it) to remove point sources in analyzing residual emission, while the recent study by {\it Fermi} uses a unique point source method \cite{FermiGC}.  In our analysis of the flux from unresolved pulsars we will use the 2FGL completeness limit when comparing to studies assuming 2FGL and 3FGL in comparing to \cite{FermiGC}.

We estimate the 50\% completeness limit of {\it Fermi} derived from the brightness distribution of 2FGL point sources within $|b| < 20^\circ$ and $|l|<20^\circ$ \cite{2012ApJS..199...31N} as
\begin{equation}
F_{\rm ps, 2FGL} = 1 \times 10^{-10} \frac{1^\circ}{(b^2+(3 ^\circ)^2)^{1/2}}\,\ergcms, \label{eq:ps2}
\end{equation}
where $b$ is the Galactic latitude of the point source.  We find a similar relation for 3FGL,
\begin{equation}
F_{\rm ps, 3FGL} = 6 \times 10^{-11} \frac{1^\circ}{(b^2+(3 ^\circ)^2)^{1/2}}\,\ergcms. \label{eq:ps3}
\end{equation}
When comparing directly with the 3FGL point source catalog, we also include an additional constraint that the flux must be greater than $4\times 10^{-12}\,\ergcms$.  This constraint is always in effect for $b \geq 14.7^\circ$. 
We should note that these equations were empirically derived by the distribution of known point sources in the GC region, rather than by directly injecting and recovering artificial point sources\cite[cf.,][]{2013pulsarcatalog}.
It is worth emphasizing that these are the thresholds for being detected as a point source.  Much higher fluxes are required to blindly detect pulsations, which is often required to identify which sources are indeed pulsars.

In Fig.~\ref{fig:ps}, we show probability density distributions of pulsar point sources in our four models.  These are calculated by examining the number of point sources in 100 random realizations of each model.  In the {\em left} panels, we first show the probability density distribution of bright pulsars with flux $F > 10^{-10}\,\ergcms$.  For young pulsars,  this is sufficiently bright to be discovered directly from the gamma-ray flux alone, without radio observations.  {\em Fermi} has detected 31 young pulsars ($P > 15\,$ms) and two MSPs ($P \leq 15\,$ms) above this threshold. The typical number of bright pulsars in our Fiducial Model is $\sim\,$20, or only two-thirds of the number detected by {\em Fermi}.  Our MSP model, on the other hand, does a good job at reproducing the entire luminosity function of the known population.

The {\em middle} panels of Fig.~\ref{fig:ps} show the total number of pulsars above the 3FGL 50$\%$ completeness limit within the GC region ($|l| < 20^\circ$ and $|b| < 20^\circ$).  There are 100 unassociated point sources and 10 known young gamma-ray pulsars within this region that are above the same threshold.  On average the Fiducial Model has 66 pulsars identifiable as point sources within this region.  Assuming that 10 of those are identified as pulsars, on average young pulsars in the Fiducial Model constitute $56\,\%$ of the unassociated point sources.  Examining spectra of all the unassociated point sources in this region, we conclude that approximately half are consistent with those of pulsars.

The {\em right} panels of Fig.~\ref{fig:ps} show the probability density of total number of pulsars above Eq.~\ref{eq:ps3}  within the Galactic Plane ($|b| < 5^\circ$) for our four models.  This same region contains 222 unassociated {\em Fermi} point sources and 57 identified young pulsars that lie above the same threshold.  We find that the Fiducial Model, on average, has approximately 176 pulsars above the threshold, constituting $\approx 54\,\%$ of the unassociated population.

It has been argued that the observed distribution of {\em Fermi} point sources excludes MSPs from contributing more than $\sim 10\%$ of the observed diffuse background \cite{2013PhRvD..88h3009H}.  In the bottom panel of Fig.~\ref{fig:ps}, we show the distribution of observable MSP point sources above the $50\%$ detection threshold of {\em Fermi} in the MSP $\times$20 Model.  We find that even when the Galactic Bulge MSP population is twenty times larger than the field population, the MSPs contribute only a small fraction of the point source population.  Typically, fewer than $20\%$ of the observed point source population would be MSPs.  As such, the point source counts near the GC cannot rule out a MSP origin of the Excess.

Overall the Beamed and MSP Models have the least tension with the number of observed point sources.   At the same time, indications from the luminosity function show that pulsars likely constitute more than a third, and close to half, of the unidentified point sources near the GC and Galactic plane.

In this region, some of the measured source spectra may be contaminated by the diffuse background or confused sources, making the spectra tougher to identify.  \citet{2015arXiv150102003T} notes that the Galactic bulge (within $5^\circ$ of the GC), along with the Vela ($l \sim 268^\circ$) and Cygnus ($l \sim 80^\circ$) regions, are particularly troublesome, with 7 sources within $1^\circ$ of the GC and very bright diffuse emission.

\citet{2015ApJS..217....4S} found that a large number of unassociated {\it Fermi} sources within $\sim\! 10^\circ$ of the Galactic plane did not contain radio sources, and that their latitude distribution is similar to the observed gamma-ray pulsar population.  They also note that the average spectral index is $\sim\! -2.34$, and that this differs from the young pulsar population.  We note that the faint $\dot{E} \sim 6 \times 10^{33}$~erg~s$^{-1}$ pulsar J1705--1906 displays a bump in the gamma-ray energy spectrum, although a power law fit yields an index of $\sim\! -2.3$ \cite{2014A&A...570A..44H}.  So steep power law spectra may not be sufficient to rule out faint young pulsars.

\subsection{The Unresolved Population}
We generate maps of the contribution of pulsars to the gamma-ray sky by convolving the gamma-ray flux of each pulsar with the point spread function (PSF) of {\em Fermi}.  This is done twice, instead of once, to reproduce the {\em reconstructed} gamma-ray maps used in the literature \cite[e.g.][]{2014daylanetal,2014arXiv1409.0042C}, who smooth the observed distribution of gamma rays by the PSF or a Gaussian filter after the image has effectively already been smoothed when measured by {\em Fermi}.  We calculate the PSF using the energy dependent model of \citet{2013fermipsf}, who determined that the {\em Fermi} PSF is best reproduced by using the sum of two King functions (see their Eq.~5).  This model reproduces both the core of the flux distribution as well as the much broader tails.  We refer the reader to \cite{2013fermipsf} for details.  We focus on the contribution of pulsars at 2\,GeV, near the peak of the pulsar spectral contribution. 

\begin{figure*}[h!]
  \includegraphics[height=.869\columnwidth,clip=false]{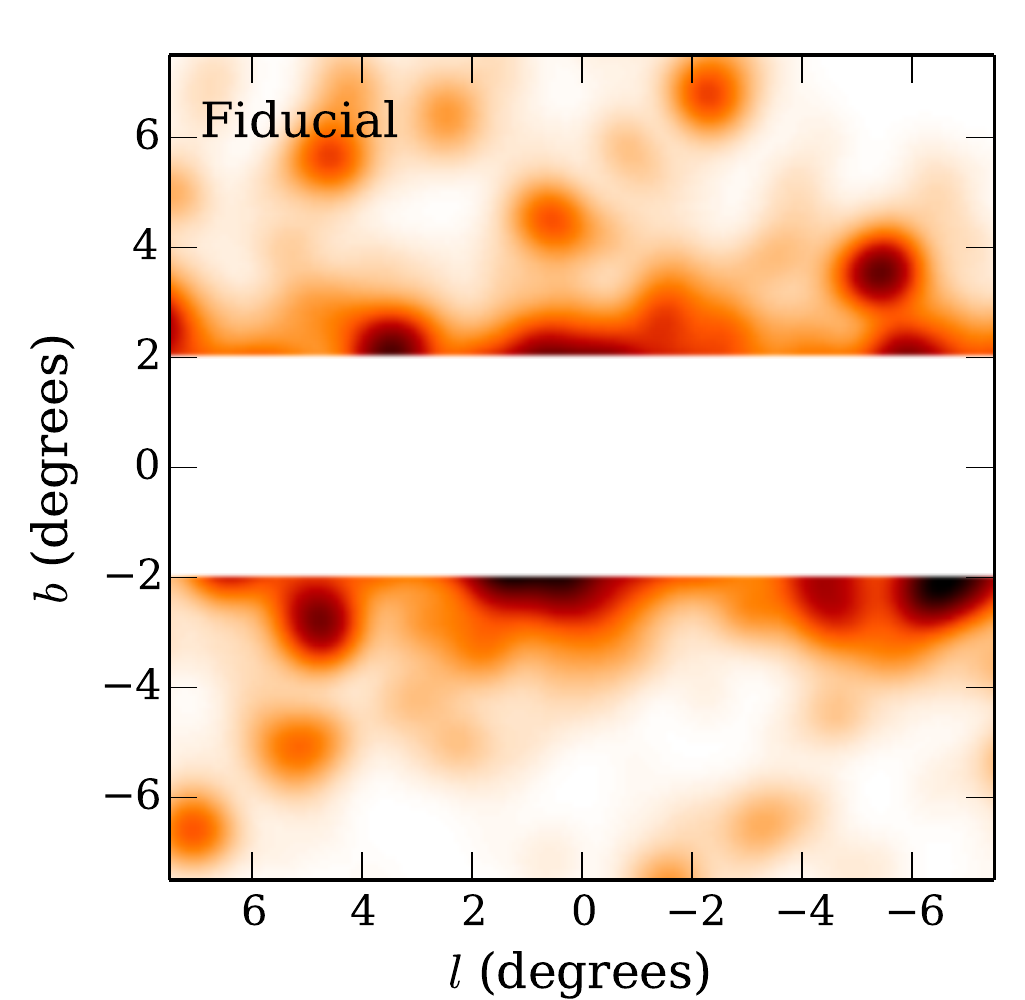}
   \includegraphics[height=.869\columnwidth,clip=true]{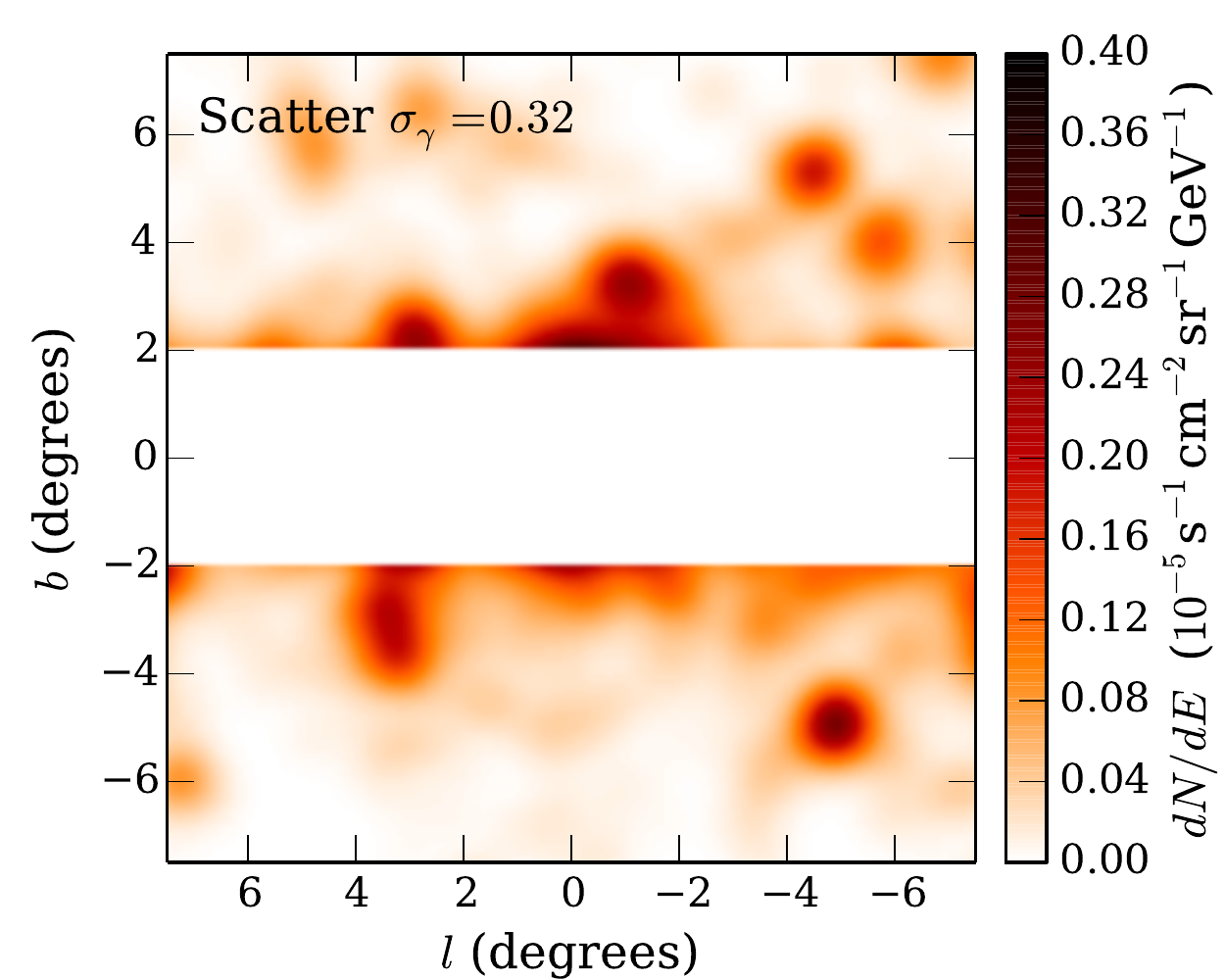}\\
   \includegraphics[height=.869\columnwidth,clip=false]{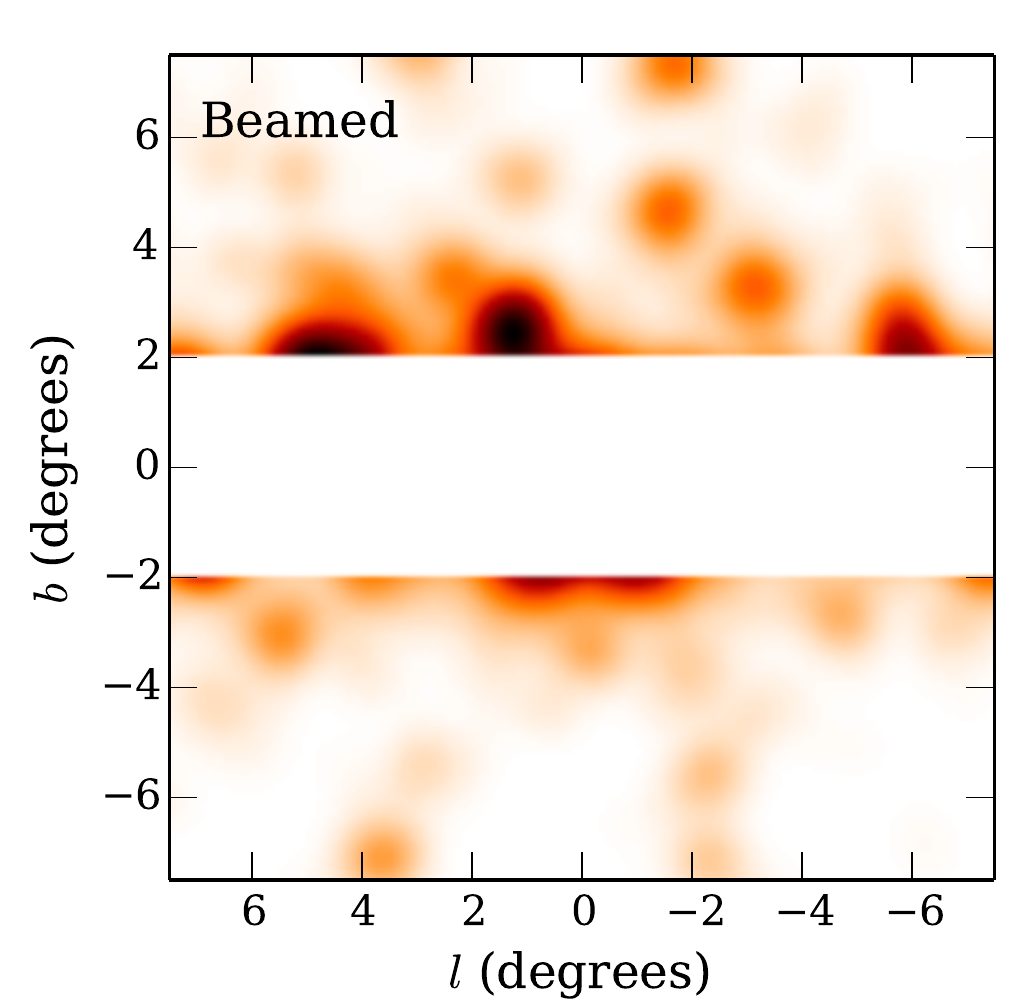}
   \includegraphics[height=.869\columnwidth,clip=true]{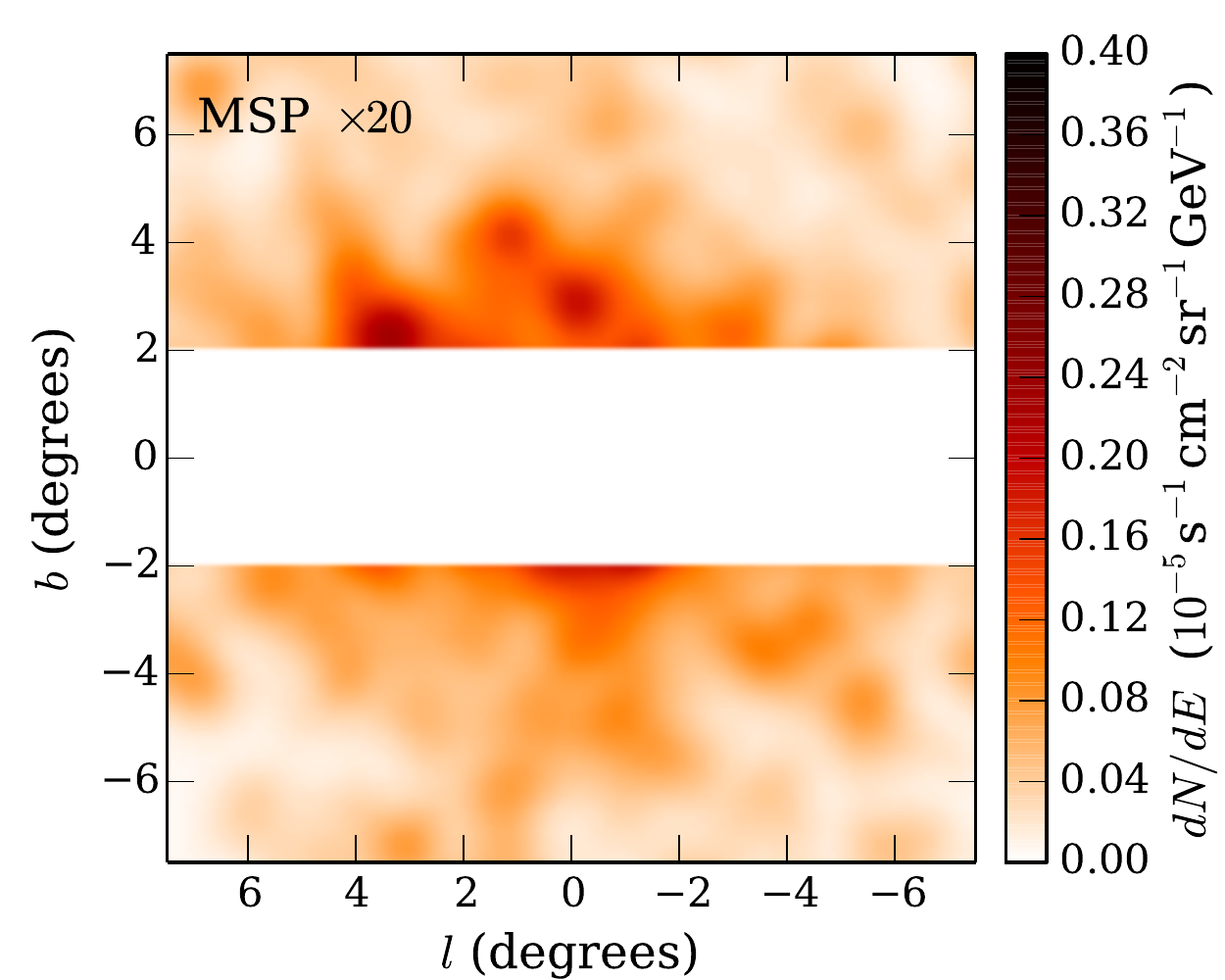}
\caption{The spatial distribution of 2 GeV gamma rays from {\it unresolved} pulsars in one realization for each of our four models, with $2^\circ$ around the GP masked. The panels show, from the top-left, our Fiducial Model, our Scatter Model, our Beamed Model, and the $20\times$-enhanced MSP Model. These do not include any other Galactic emission and the brightest regions are most subject to variation in a single realization. \label{fig:gcmap}}
\end{figure*}

In Fig.~\ref{fig:allsky}, we show a Mollweide projection of gamma-ray flux from young pulsars in the entire MW.  Overall, the pulsar emission follows the Galactic plane, with a peak near the GC due to young pulsars formed in the CMZ, which have not been included in previous analyses of gamma-ray pulsars \cite[cf.][]{2014ApJ...796...14C}.

\begin{figure*}[t!]
\includegraphics[width=\textwidth,clip=true]{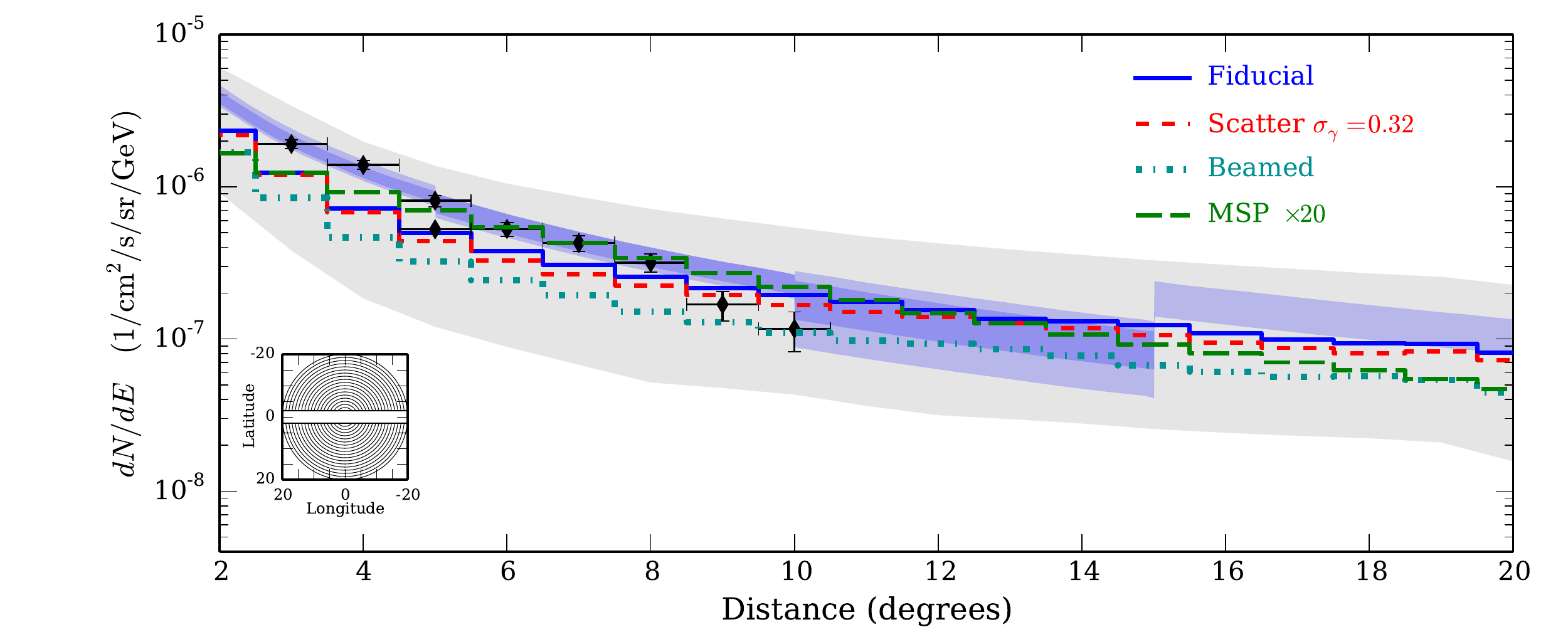}\\
\vspace*{0.0cm}
\includegraphics[width=\textwidth,clip=true]{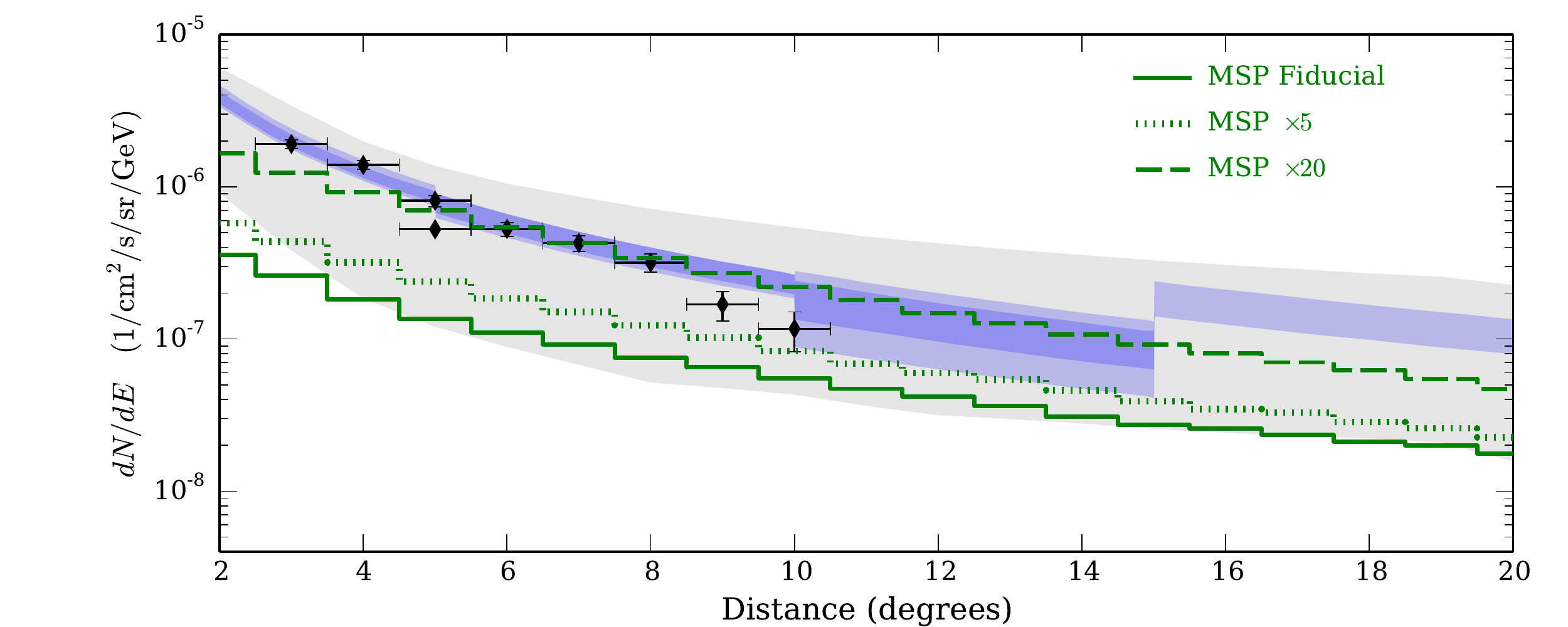}
\caption{\label{fig:daylan} Intensity of the pulsar background at  2\,GeV as a function of angular separation from the Galactic Center, compared to the GC Excess intensity at 2\,GeV inferred in Ref.~\cite{2014daylanetal} ({\it black points and error bars)} and Ref.~\cite{2014caloreetal} (shaded blue regions).
{\it Top:} The {\it solid blue line} shows the flux from our Fiducial Model.  The {\it dashed red line} shows the flux from our Scatter Model with $\sigma_\gamma = 0.32$.  The {\it dashed-dotted cyan line} shows the flux from our YRQ Beamed Model.  And finally the {\it long-dashed green line} shows the flux from our MSP $\times 20$ model.  The {\it inset} shows the rings we  used in our analysis with $|b| > 2^\circ$.
{\it Bottom:} As above, {\it solid green line} from our MSP Fiducial Model.  The {\it dotted line} shows the MSP contribution if the Galactic bulge population is enhanced by 5$\times$ relative to stellar mass than the disk, while the {\it long-dashed line} assumes 20$\times$ bulge enhancement.  In principle, any MSP line can be summed with any young pulsar line to estimate a total pulsar contribution.
  }
\end{figure*}

\subsection{Diffuse GeV Emission near the Galactic Center}
\label{sec:diffuse}
We reconstruct the diffuse emission from the unresolved pulsar population near the GC by first removing all pulsars with fluxes above $F_{\rm ps,2FGL}$.  Although many more point sources are known in the GC region, most current studies mask only those sources discovered in 2FGL \cite[cf.][]{FermiGC}.  At present, we do not mask the brightest pulsars (which usually only covers $95\,\%$ of the flux), but completely remove them, thus our results should be considered a lower bound on the flux in these regions.

In Fig.~\ref{fig:gcmap}, we show our reconstructed maps of the GC region for our four models, with the Galactic plane masked ($|b| < 2^\circ$).  The Fiducial Model also includes an overlay of the different regions of interest.  The High Latitude pulsars are those above the diagonal lines with $|b| \geq |l|$.  Unlike in the rest of this work, this map shows the effect of the finite number of photons collected.  The three young pulsar models show emission in all latitudes, but the flux concentrates around the GC and Galactic plane.   The MSP $\times$20 Model is more spherically symmetric at large angles, as the pulsars are formed using a symmetric Galactic Bulge model.

In Fig.~\ref{fig:daylan}, we plot the angular profile of gamma-ray emission from unresolved pulsars near the GC.  The Fiducial Model is as constructed in \citet{2015PF}.
The lines show the total flux in annular apertures around the GC with width of one degree for each of our four models.  The black points show the results from \citet{2014daylanetal}.  The shaded blue regions show the results from \citet{2014caloreetal}.  Overall, our the models follow the observed Excess of gamma rays from the GC region out to $\approx 20^{\circ}$. The three young pulsar models, we note, are not tuned to match amplitude of the GC excess.  However, the total flux is uncertain by at least a factor of two (shown in the gray region), due to underlying systematic uncertainties, which we explore further in the Appendix.

In the bottom panel of Fig.~\ref{fig:daylan}, we show the unresolved contribution of our Fiducial MSP Model (green solid line), which only reproduces a fraction of the observed diffuse emission near the GC.  We estimate approximately $20\%$ of the diffuse emission comes from MSPs in the Galactic disk.  Fig.~\ref{fig:daylan} also shows the results of enhancing the bulge MSP population by $5\times$ (dotted) and $20\times$ (dashed) the number of MSPs per stellar mass as the Galactic disk. As we discussed previously, the MSP model requires that the bulge have an enhancement of MSPs relative to the Galactic disk.

Overall, we find that a population of young pulsars born near the GC can roughly reproduce the amplitude, spectrum, and angular distribution of the Excess quite directly.  Such a signal has a strong component in the Galactic Plane, consistent with the excess of gamma rays observed in the plane \cite{2012ApJ...750....3A}, as we discuss further in \S~\ref{sec:cmz}.  MSPs can also reproduce the spectrum and angular distribution of the signal.  The spherical morphology imparted by the model appears to be a better match to the observations by \citet{2014arXiv1409.0042C}.  However, as we described in \S~\ref{sec:msp}, this amplitude can only be attained if there is a substantial enhancement of MSPs that extend throughout the bulge, in proportion to the stellar mass.  We explore the latitude dependence of the signal in Appendix~\ref{sec:rois}.  We compare the overall amplitude of the fluxes in the GC region from the Fiducial and $\times$20 MSP models to Excess spectra from \citet{FermiGC} in \S~\ref{sec:extended}.

\section{Inverse Compton from the inner kpc}
\label{sec:extended}
The best fitting models in the detailed recent study for the inner $\sim 1$~kpc of the MW by \citet{FermiGC} include a much higher normalization for the inverse-Compton (IC) component ($6-30 \times$) than their baseline models, possibly due to a larger cosmic-ray electron population or radiation field intensity.  For such an $E_\gamma = 1 - 100$~GeV IC flux, considering photon backgrounds within 1~kpc, requires $e^\pm$ of $E_{e^\pm} = 5 - 500$~GeV.  This energy range corresponds to the positron excess measured by PAMELA \cite{2009Natur.458..607A} and AMS-02 \cite{2013PhRvL.110n1102A,2014PhRvL.113l1102A} most easily ascribed to nearby pulsars (e.g., \cite{yuksel2009,2013ApJ...772...18L,2015ApJ...807..130V}).

While the gamma-ray efficiency of young pulsars increases with age, most of the spin-down power is still carried off by a pulsar wind containing energetic electrons and positrons.  As $\dot{E} \propto t^{-2}$, the youngest pulsars in the population dominate the $e^\pm$ energetics, so this is not sensitive to any low $\dot{E}$ cutoff and is the same for all three of our young pulsar models.


 \begin{figure}[t!]
      \includegraphics[width=0.5\textwidth,clip=true]{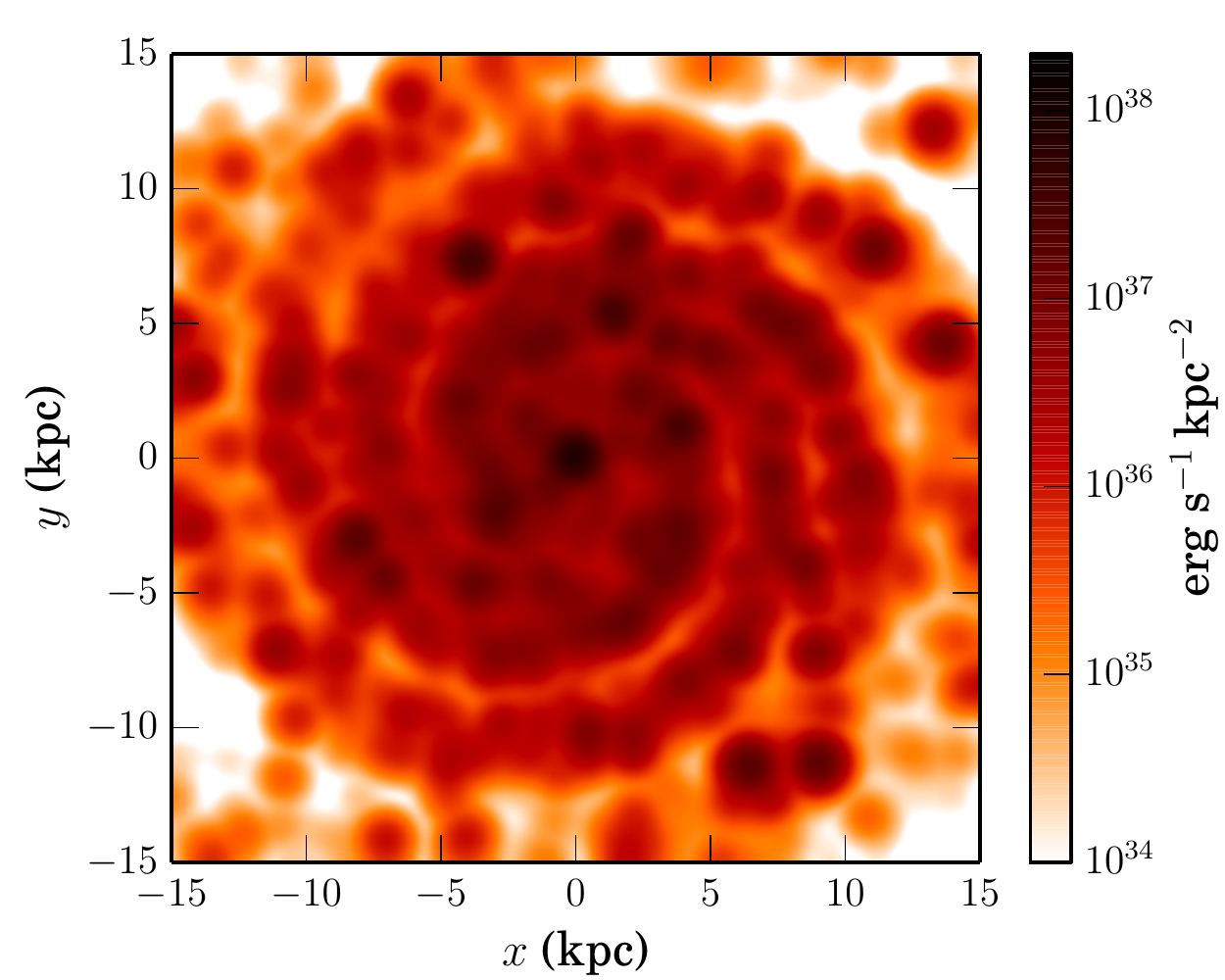}\\
      \includegraphics[width=0.5\textwidth,clip=true]{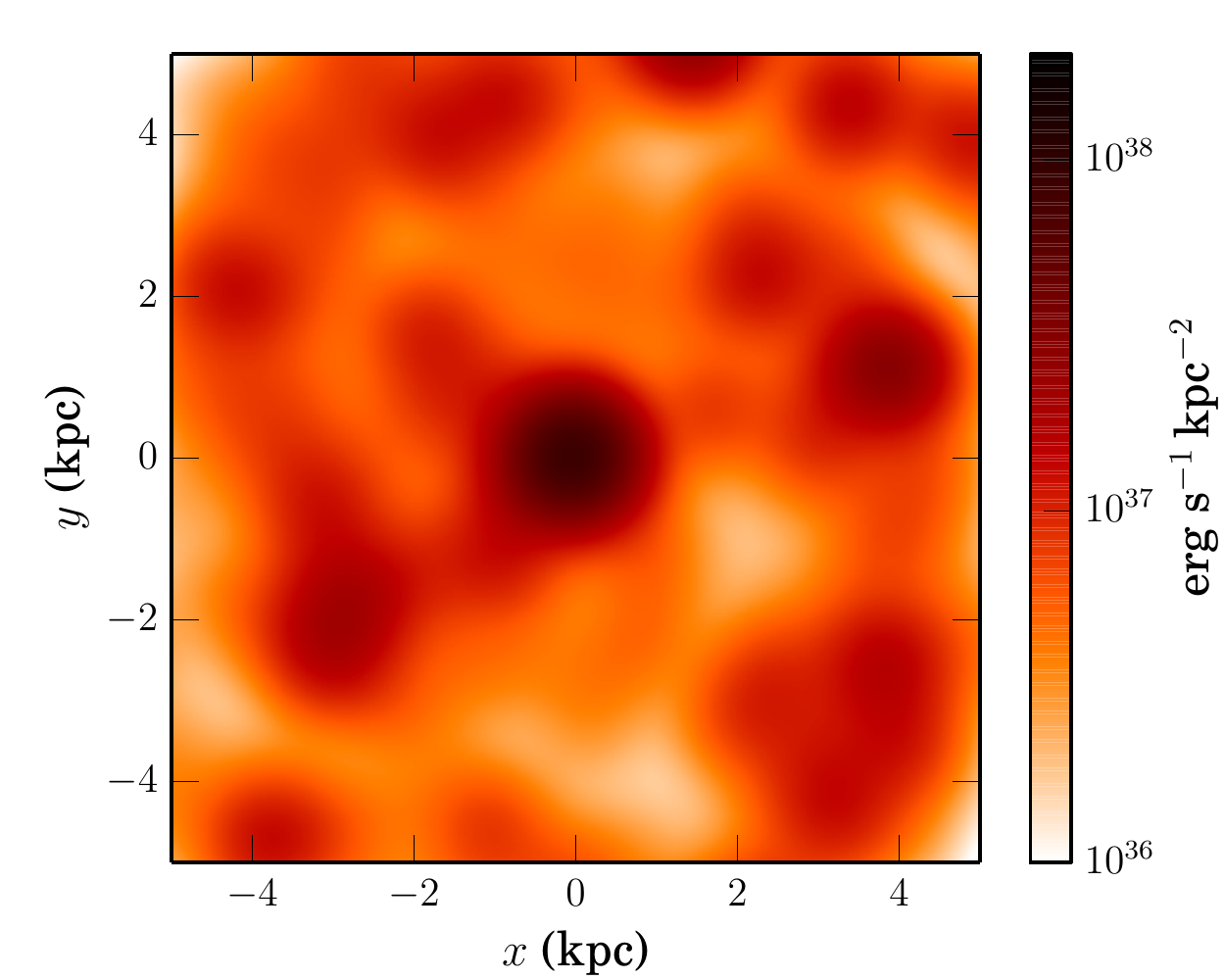}
\caption{\label{fig:edotic}
The distribution of young Galactic pulsar energy output ($\dot{E}$) as viewed wide from above the MW ({\it top}) or zoomed in towards the GC ({\it middle}) and from the side facing the GC ({\it bottom}).
}
 \end{figure}

\begin{figure*}[t!]
\includegraphics[width=1.8\columnwidth,clip=true]{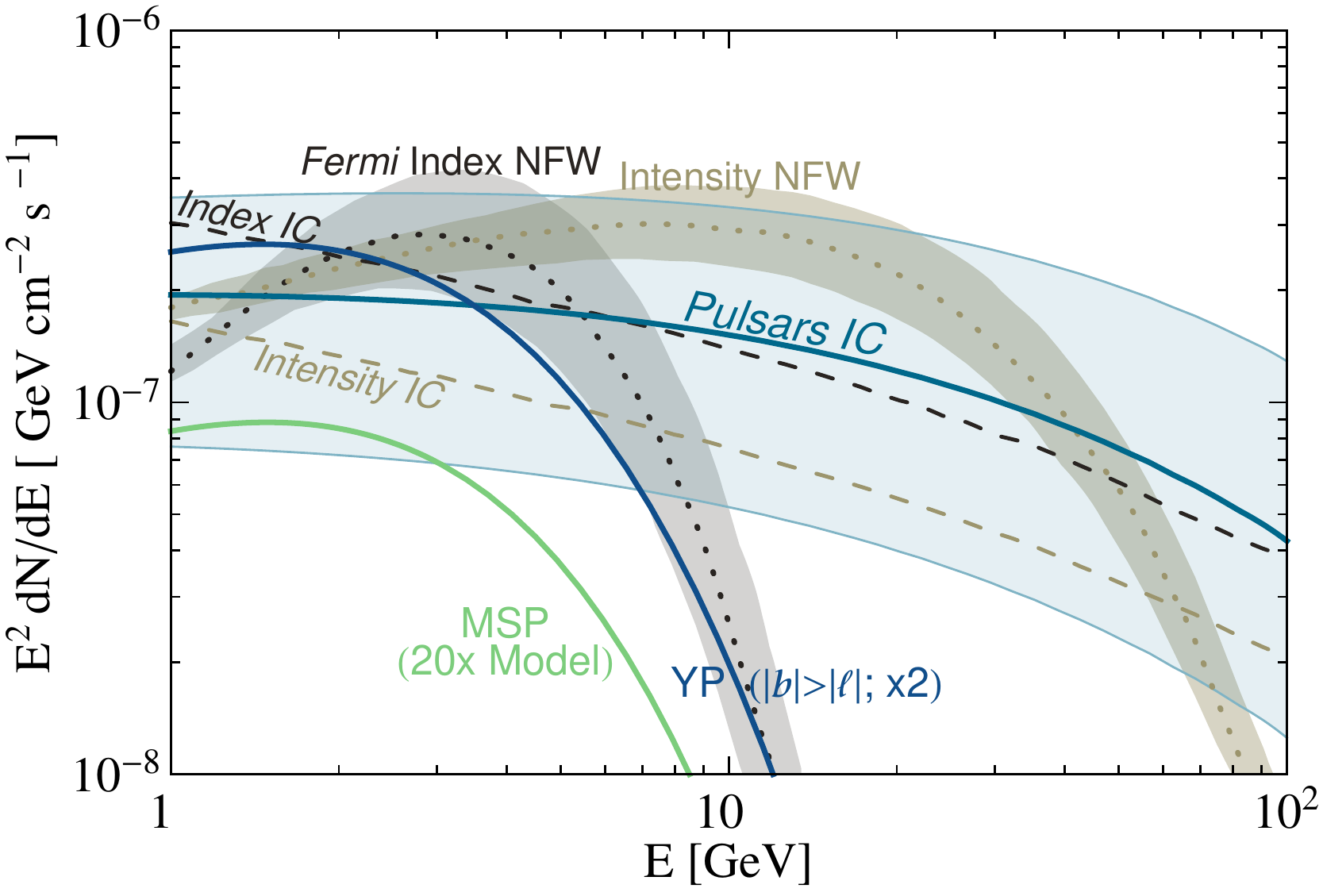}
\caption{
{\it Fermi} NFW ({\it dotted lines}) and IC ({\it dashed lines}) spectra for ``Pulsar Index'' and ``Pulsar Intensity'' IEMs within $15 \!\times\! 15^\circ$ from \cite{FermiGC}.
For young pulsars, we include our composite unresolved High-Latitude YP spectrum (scaled by the effective area to compare with NFW; {\it solid}) and inverse-Compton (IC) flux from the total population with $B = 25\,\mu$G ({\it solid blue}) and $10-50\,\mu$G ({\it blue band}).  We also show the composite unresolved $\times$20 MSP model spectrum ({\it solid green}).    \label{fig:gcic}}
 \end{figure*}

We show the MW distribution of $\dot{E}$ in Fig.~\ref{fig:edotic}, which we see is highly concentrated within 1~kpc of the GC and along the Galactic plane (although propagation should stretch this along the local magnetic field direction \cite{2012arXiv1210.8180K}).  As stated in \cite{2015PF}, we then expect a substantial contribution to the inner Galactic IC background above that usually assumed (see the left panel of Fig.~\ref{fig:dsnr}).  There could also be an increased contribution from SN remnants.  Both of these would be more relevant for the cosmic-ray electrons than for protons, due to the higher rate of $e^\pm$ energy losses making local production more important, especially at high energies due to a harder spectrum (maybe even more at the GC than indicated by PAMELA/AMS, since at Earth the lower density of sources may result in a substantial variation).  This is consistent with the much lower flux attributed to neutral pion decay in the inner $\sim 1$~kpc by \cite{FermiGC}.

We find that a lognormal distribution describes the total summed $\dot{E}$ power within 1~kpc, with $\log_{10} \Sigma_i \dot{E}_i \approx 38.2 \pm 0.1$~erg~s$^{-1}$ for young pulsars, and $\approx 37.5 \pm 0.2$~erg~s$^{-1}$ for the 20$\times$-enhanced MSP scenario.  We note that though we do not include pulsars with age less than $10^4$~yr in this calculation, a single Crab can reach $\dot{E} > 10^{38}$~erg~s$^{-1}$ (at the far right of Fig.~\ref{fig:lumfunc}), though no such object is necessarily present.

While details of cosmic rays will have to be sorted out with more precise modeling, we can simply scale the $e^\pm$ luminosity to $\dot{E}$ to estimate their effect.  Since $\dot{E} \propto t^{-2}$, the youngest pulsars (i.e., those near the disk) are most relevant in this regard, and we content ourselves with an equilibrium, one-zone calculation  \cite{2015arXiv151100723K} using a photon background appropriate for this region \cite{2006ApJ...640L.155M} and a plausible range of magnetic fields within 1~kpc of $B = 10-50\, \mu$G.  For $E_e \gtrsim 10$~GeV, the $e^\pm$ cooling time is then $t_c \lesssim 10^6$~yr, so variations below this scale are smoothed, while this is still a relatively short time for diffusive escape from 1~kpc.  Since $t_c$ decreases roughly as $E_e^{-1}$, higher-energy IC emission may be more localized.

Fig.~\ref{fig:gcic} shows the IC gamma-ray flux using a young pulsar $e^\pm$ source spectrum motivated by positron excess measurements, $dN/dE_{e^\pm} \!\propto\! E_{e^\pm}^{-1.8} {\rm exp}[-E_{e^\pm}/500\,{\rm GeV}]$, and taking $\mathcal{L}_{e^\pm} \!=\! 0.5\, \Sigma_i \dot{E}_i$, for the range of $B$.  These are shown with our Fiducial composite High-Latitude young pulsar spectrum, scaled by effective area for even comparison with the $15 \!\times\! 15^\circ$ {\it Fermi} NFW component flux \cite{FermiGC}, and IC spectra for ``Pulsar Index-Scaled'' and ``Pulsar Intensity-Scaled'' IEMs \cite{FermiGC}.

Although young pulsars do not necessarily account for all of the IC emission, and 100\% of $\dot{E}$ need not go into $e^\pm$, we see they should help account for the enhanced IC without needing to invoke an enhanced photon background or other ad hoc assumptions.  There will also be some GeV--TeV pulsar wind nebula emission (even at old ages as seen around Geminga \cite{2009ApJ...700L.127A} from such a wind \cite{yuksel2009}) and variations owing to the local photon field -- such as near the GC, Arches, and Quintuplet star clusters -- and magnetic field strength \cite{2015arXiv151100723K,2015arXiv151101159K}.  Inverse-Compton emission from the disk, on the other hand, should already mostly be accounted for in models assuming a radial-dependent source density (e.g., \cite{FermiGC}; see again left panel of Fig.~\ref{fig:dsnr}).  While the 20$\times$-enhanced MSP model yields a lower summed $\dot{E}$ within 1~kpc, a less concentrated profile could lead to off-plane IC emission depending on MSP wind properties.

\section{Radio Pulsars}
\label{sec:radio}

\subsection{Young Pulsars in Radio}
\label{sec:ypradio}

\begin{figure*}
  \includegraphics[width=0.49\textwidth,clip=true]{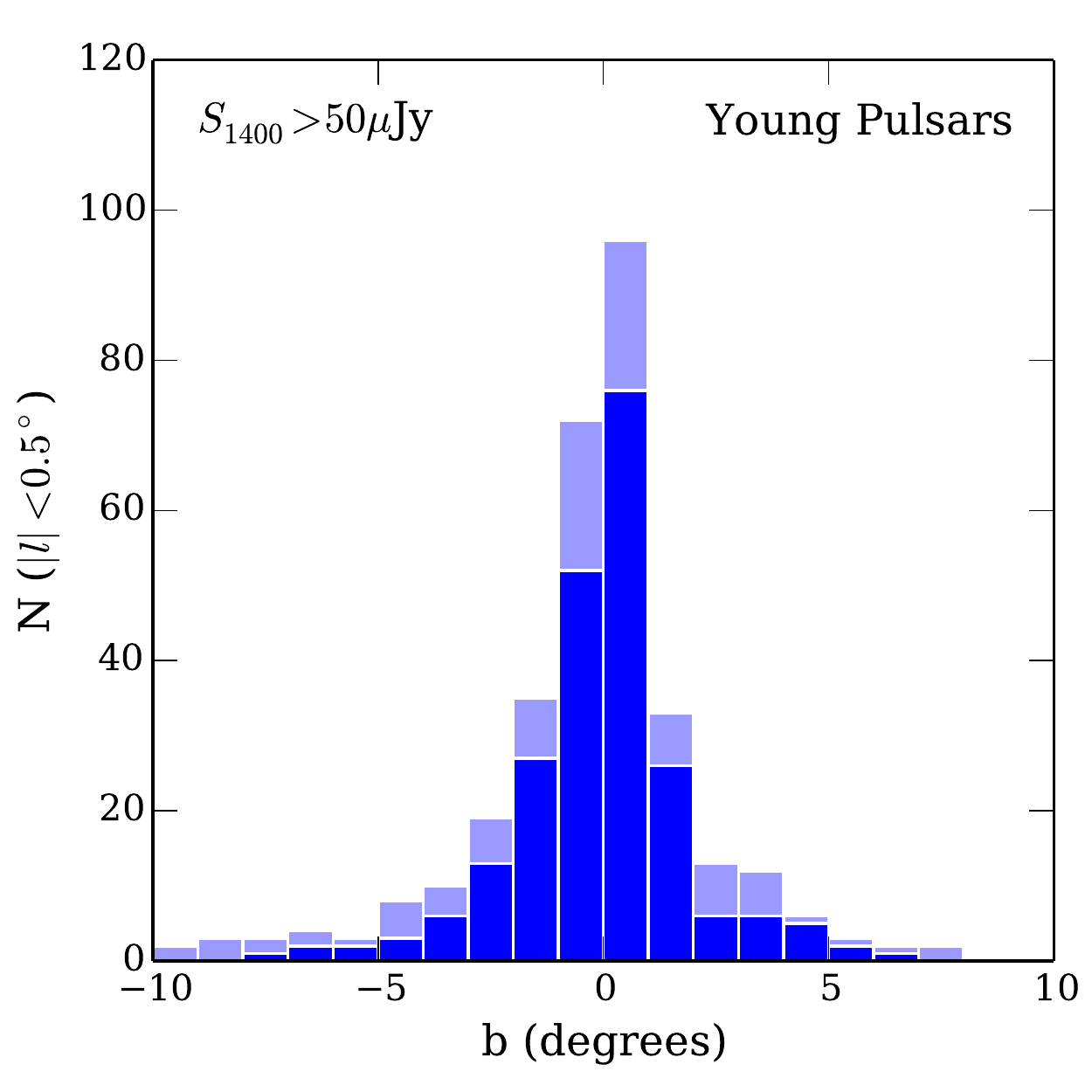}
  \includegraphics[width=0.49\textwidth,clip=true]{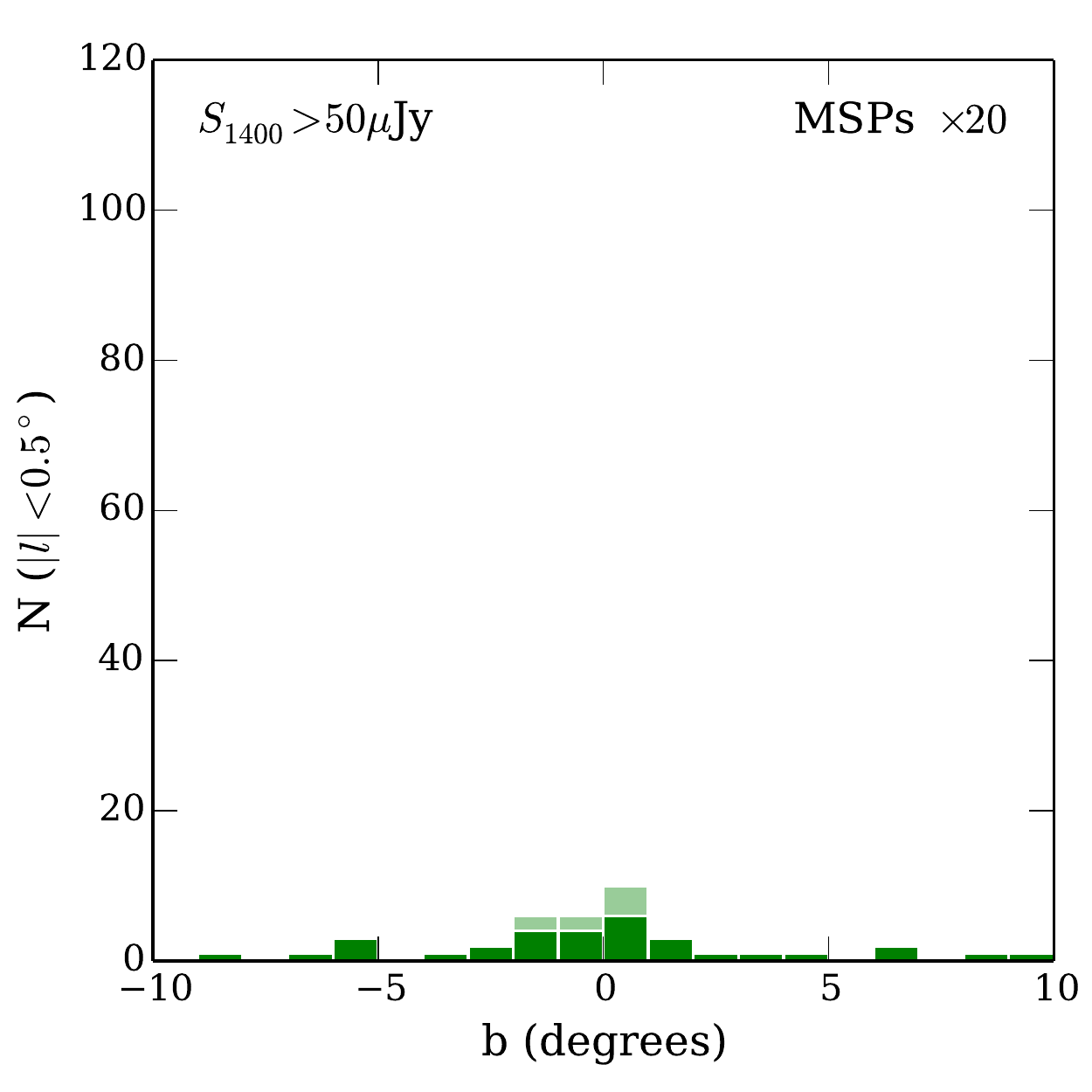}\\
  \includegraphics[width=0.49\textwidth,clip=true]{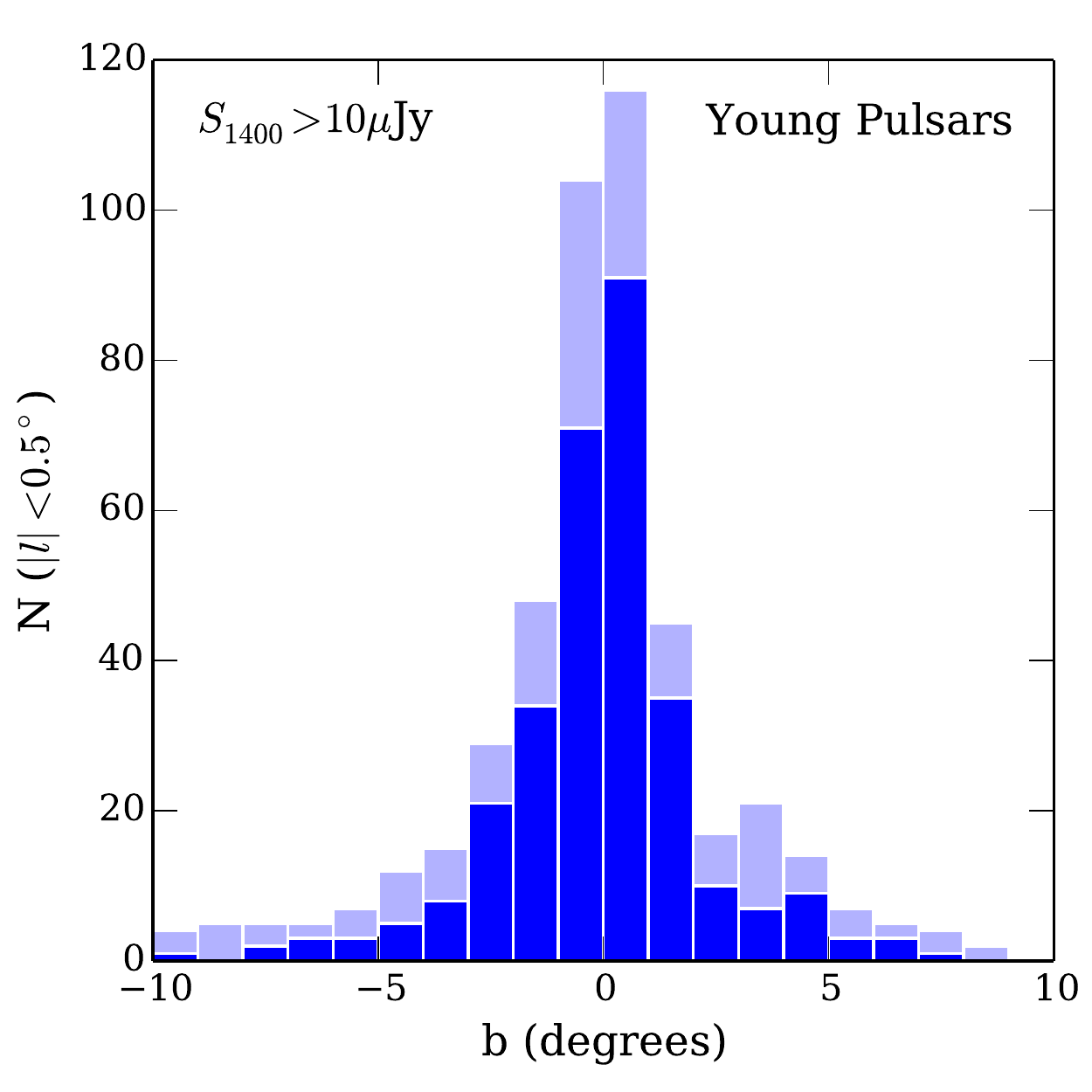}
  \includegraphics[width=0.49\textwidth,clip=true]{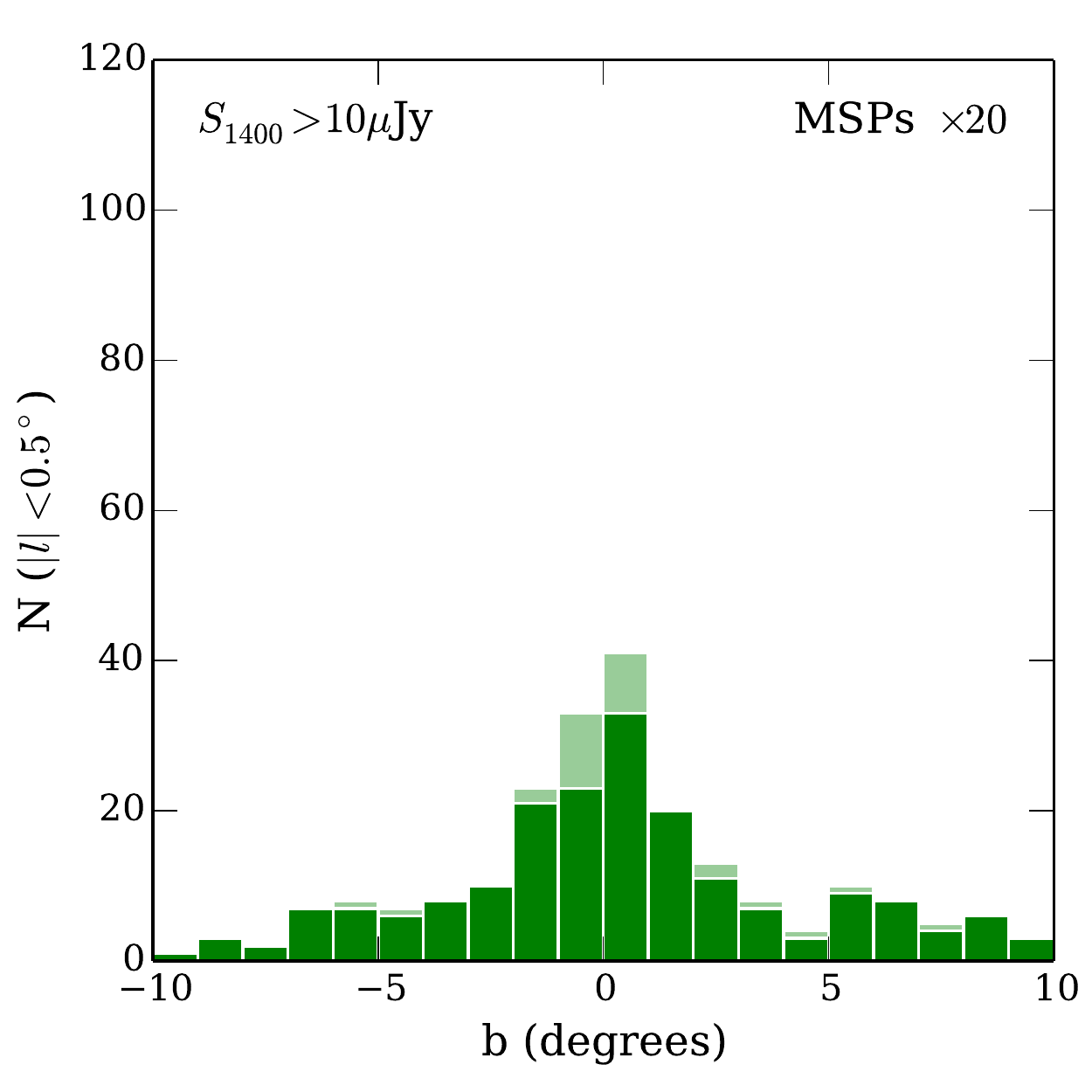}\\
  \caption{\label{fig:radio1} The number of visible radio pulsars in the Galactic Center region. Each figure shows the total number of visible radio pulsars found within $|l| < 0.5^\circ$ as a function of Galactic latitude, $b$. The top row shows the number of radio loud pulsars with flux $S_{1400} > 50\,\mu$Jy for the young pulsar model ({\em left}) and the MSP $\times 20$ model ({\em right}). The bottom row shows the number of radio loud pulsars with $S_{1400} > 10\,\mu$Jy. The dark bins show the pulsars that are associated with the Galactic Center/Bulge. The light shaded show the total number of pulsars visible, including contaminates from the Galactic Disk. Each bin is approximately one square degree.}
\end{figure*}

Radio surveys of the central 50\,pc of the GC probe the underlying distribution of radio pulsars within that region, and hence star formation on a $\sim 10^7\,\yr$ timescale.  To estimate the total number of radio pulsars in this region we use the same simulation parameters as our Fidicual gamma-ray simulations, except we follow all pulsars down to the radio death line \citep{1992A&A...254..198B,fauchergiguerekaspi2006}. 
 We use the empirical pulsar radio luminosity function from \citet[][Eq.~5]{1987A&A...171..152S} to estimate the radio flux from each pulsar.  This luminosity function is nearly identical to the one used in \citet{fauchergiguerekaspi2006} for $\dot{E} \lesssim 10^{33}\,\ergs$.  More energetic pulsars, however, all follow the luminosity distribution of pulsars with $\dot{E} \approx 10^{33}\,\ergs$. The radio emission from pulsars is strongly beamed, and so only a fraction of the pulsars are visible to us. Following \citet{TM98}, we assume that only a fraction of pulsars are beamed in our direction
\begin{equation}
\label{eq:rbeam}
f_{\rm radio} = 0.09 \left [ \log_{10}{(P/10\,{\rm s})}\right]^2 + 0.03.
\end{equation}
With these assumptions we find that there should be $\approx 5 - 15$ radio pulsars beamed towards us that could have been detected by the GC pulsar survey of \citet{denevaetal2009detect}, depending on if we include the inner 7 pointings where the sky is brightest.  This is  consistent with the 5 pulsars observed within the inner $\sim 50\,$pc, given the uncertainty in the spatial distribution of star formation in our model, and variable sky brightness.   Indeed, one could use these radio detections to place a lower limit on the gamma-ray contribution of young pulsars in this region.

Searches for compact radio sources can also limit the total number of pulsars in the GC.  \citet{2008ApJS..174..481L} estimate that no more than $10^4-10^5$ total pulsars may exist within $2^{\circ}$ of the GC. In our simulations $\approx 1\times 10^4$ pulsars are found within the survey area.  The pulsed radio searches in this region from \citet{denevaetal2009detect} place a much tighter constraint on the overall population.

Future radio searches above the Galactic plane would provide an important constraint on the population of young pulsars.  The central $\approx 50\,$pc are partially obscured by a strong scattering screen that smears out signals on short timescales.  Because of this, many deep surveys have searched at high frequencies where the  impact of scattering is less, but the pulsars are intrinsically fainter.  The strongest regions of the scattering screen extends only to $|b| \lesssim 0.2^\circ$ \cite{2014ARep...58..427P}.  Above the Galactic plane, the expected amount of scatter is significantly less, although the pulsar density also decreases with latitude. In Figure~{\ref{fig:radio1}} we show the number density of pulsars in $1$ sq. degree regions above and below the Galactic plane with $|l| \leq 0.5^\circ$.   We find that near $|b| \approx 1^\circ$, the pulsar density is approximately $45\,{\rm deg}^{-2}$ with fluxes above $50\,\mu$Jy at 1400\,MHz. This drops to $25\,{\rm deg}^{-2}$ at $b = 2^\circ$. The majority of these, $\approx 80\,\%$, originate within the CMZ. The remainder are pulsars from the foreground disk.

The successful detection of pulsars in the CMZ is in stark contrast to the non-detections from  deep radio surveys of the central parsec. We estimate, using the same forward modeling and radio luminosity function, that there should be $\approx 10-20$ detectable pulsars in the deepest surveys of the central parsec for a SNR of $10^{-4}\,\yr^{-1}$.  This is slightly larger but consistent with the estimates by \citet{dexter2014} who used purely empirical data and relationships to estimate that $\approx 10$  radio pulsars should have been detected in this region.   Both estimates are at odds with the deepest surveys of the central parsec \cite{kleinetal2004,deneva2010,macquartetal2010,whartonetal2012,2013IAUS..291...57S}, which have found no ordinary, young radio pulsars.  This population of pulsars is not included in any of our gamma-ray models.

\subsection{Millisecond Pulsars in Radio}

Deep pulsed radio searches may be able to constrain the population of MSPs in the Galactic Bulge.  To date, the deepest searches have focused within $\approx 1\,$pc of the GC, where the sky brightness is high and scattering from the interstellar medium (ISM) may smear out the pulsations. Even these deep radio searches are sensitive to only the brightest few MSPs known in the whole galaxy.
Above the plane, the ISM scatters the pulsar signals by orders of magnitudes less.  As such, deep pulsar search near $\sim 1\,$GHz may be able to detect the absolute number of MSPs in the bulge, and constrain the total gamma-ray background produced by this unresolved population.  

 MSPs are typically fainter than the YRL pulsar population, even for pulsars with the same spin-down luminosity.  For this reason, we  use a different radio luminosity distribution for the MSP population. Following \citet{cmsp2015}, we use the best fit model of \citet{2011MNRAS.418..477B} to estimate the flux of MSPs in the Galactic Bulge.
We find that the surface density of radio bright ($S_{1400}> 10\,$mJy) MSPs is approximately 20 deg$^{-2}$ at $2^\circ$ above the Galactic plane (note that this flux limit is five times lower than quoted above for YRL pulsars). Within $2^{\circ}$, we do not reproduce the total flux of the gamma-ray excess.   Detailed estimates of the radio detectability of MSPs have recently been carried out by \citet{cmsp2015} using a complementary approach to model the underlying pulsar population.

As opposed to young pulsars, where there is overwhelming evidence for a substantial radio-quiet population, there is yet no firm evidence for radio-quiet MSPs.  While the predicted values for, say, 47 Tuc are larger than the known radio MSPs, this only arises after assuming an efficiency.  There are still high-latitude point sources that look like MSPs, though ruling out a radio counterpart is very difficult due to the variety of possible signals.

 \begin{figure*}[h!]
      \includegraphics[width=0.49\textwidth,clip=true]{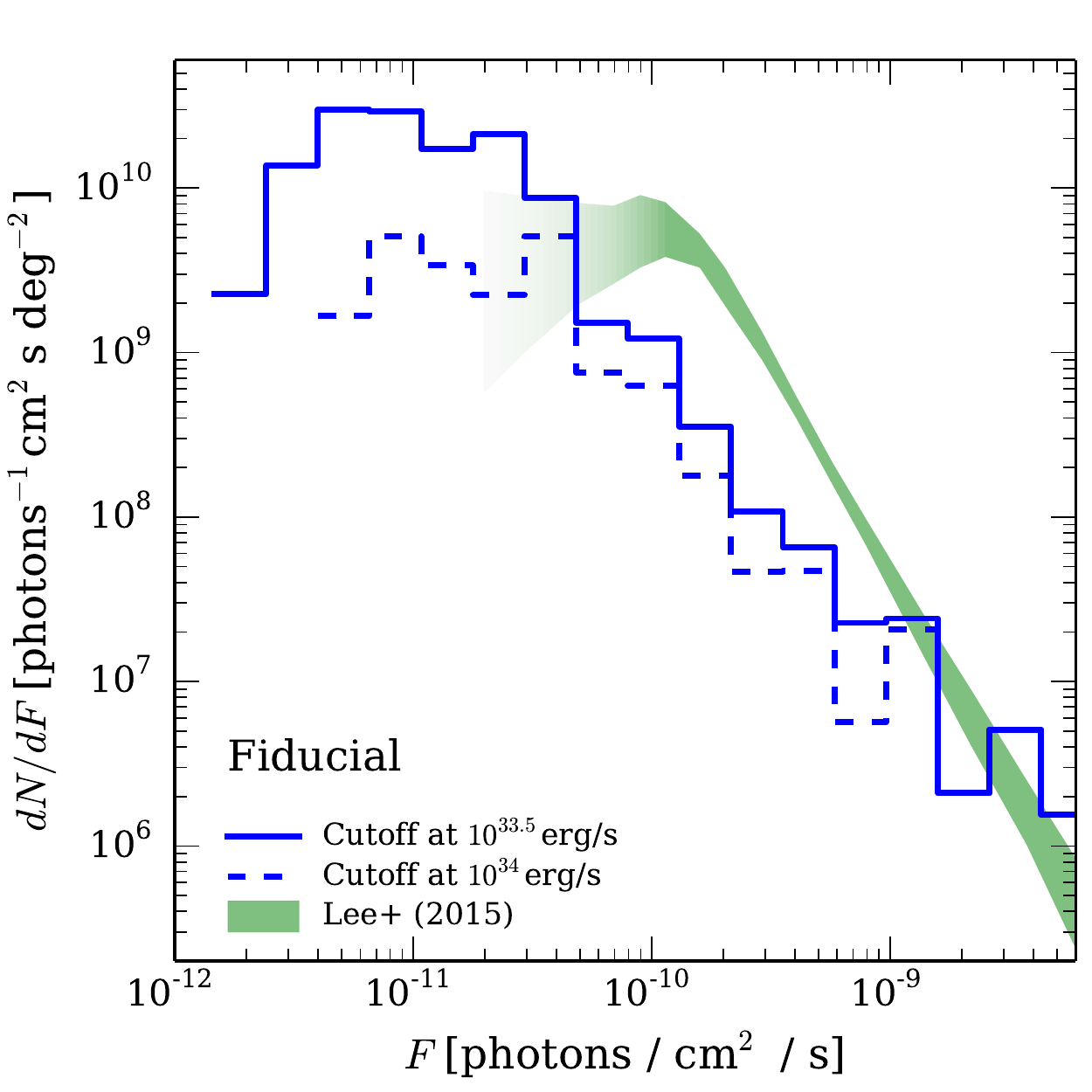}
      \includegraphics[width=0.49\textwidth,clip=true]{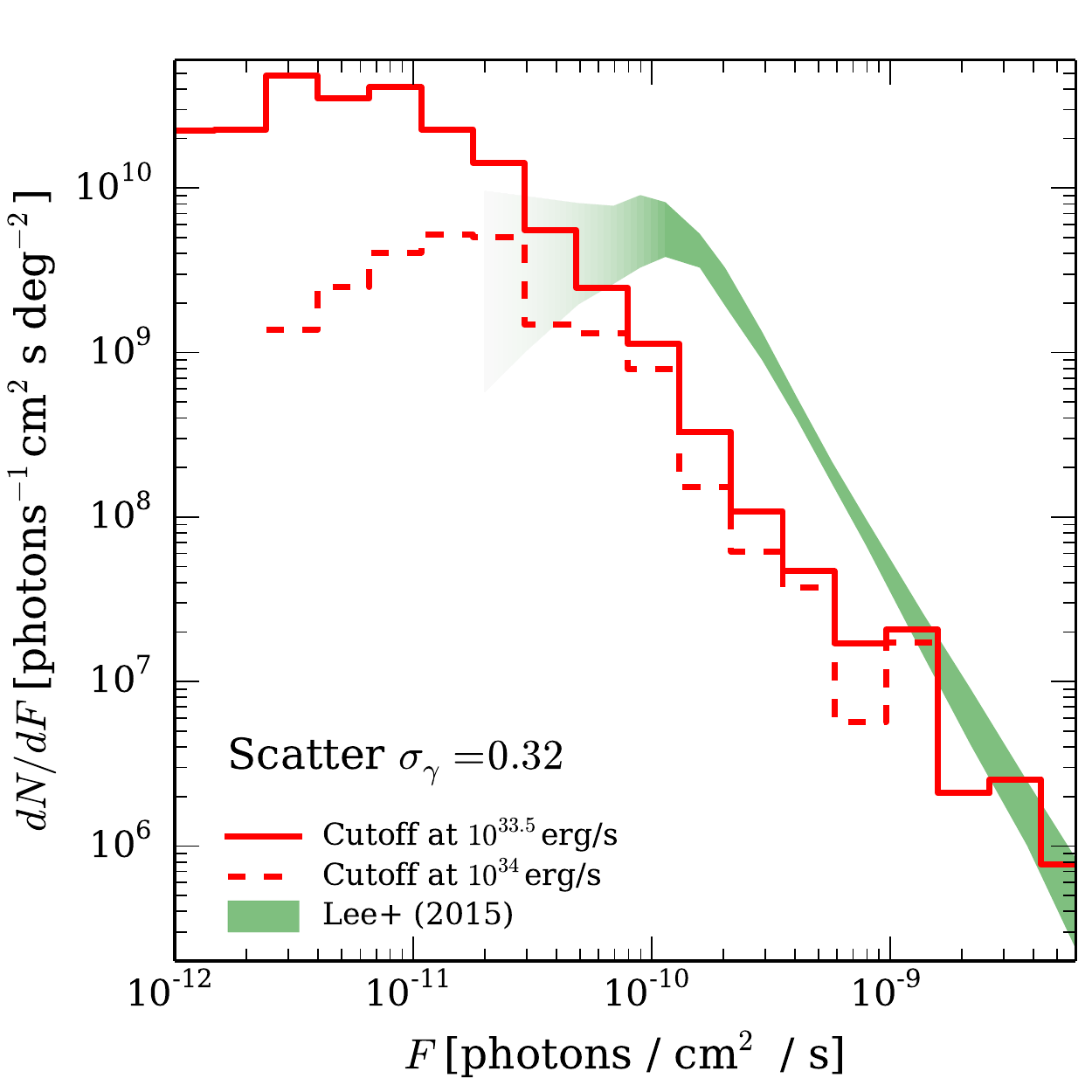}\\
      \includegraphics[width=0.49\textwidth,clip=true]{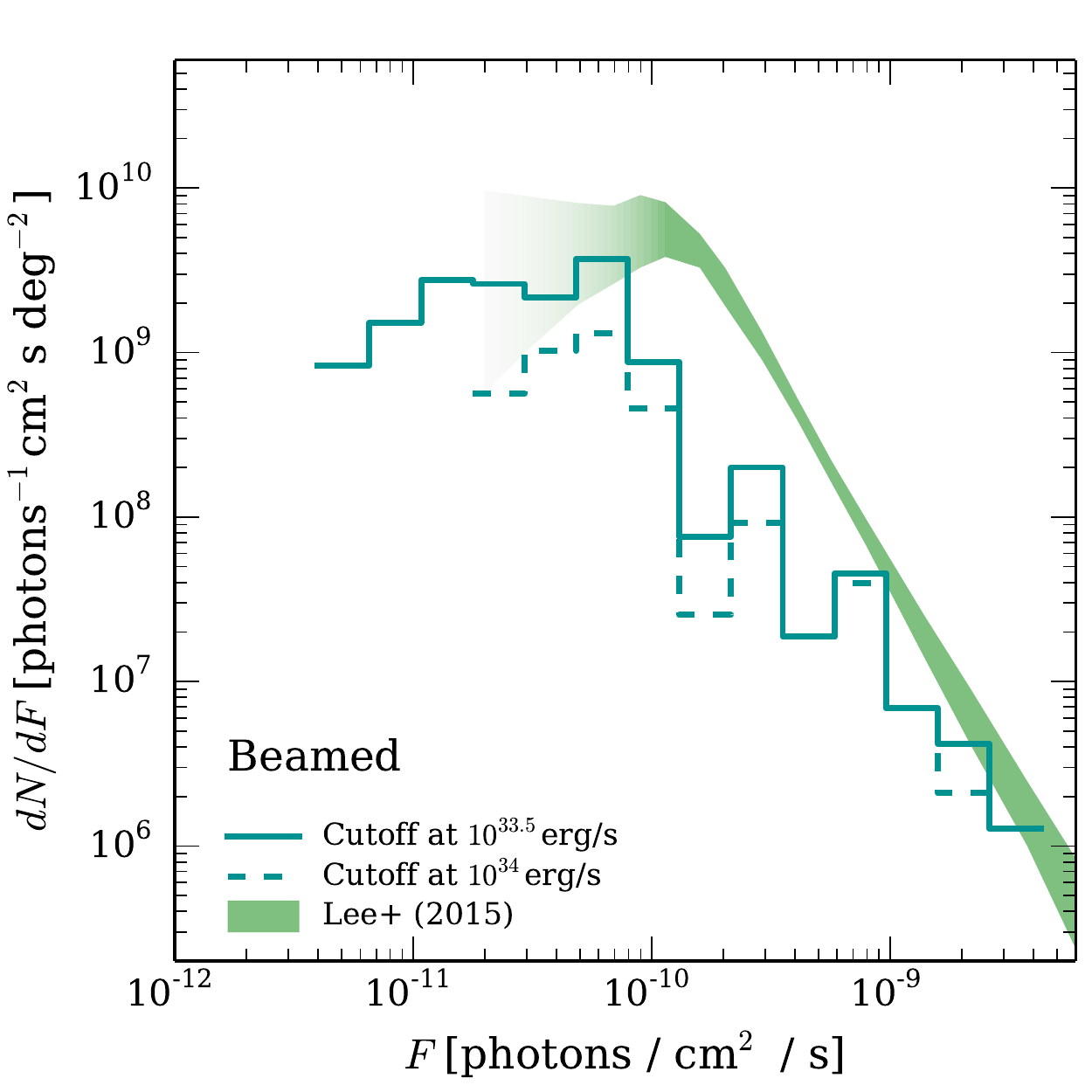}
      \includegraphics[width=0.49\textwidth,clip=true]{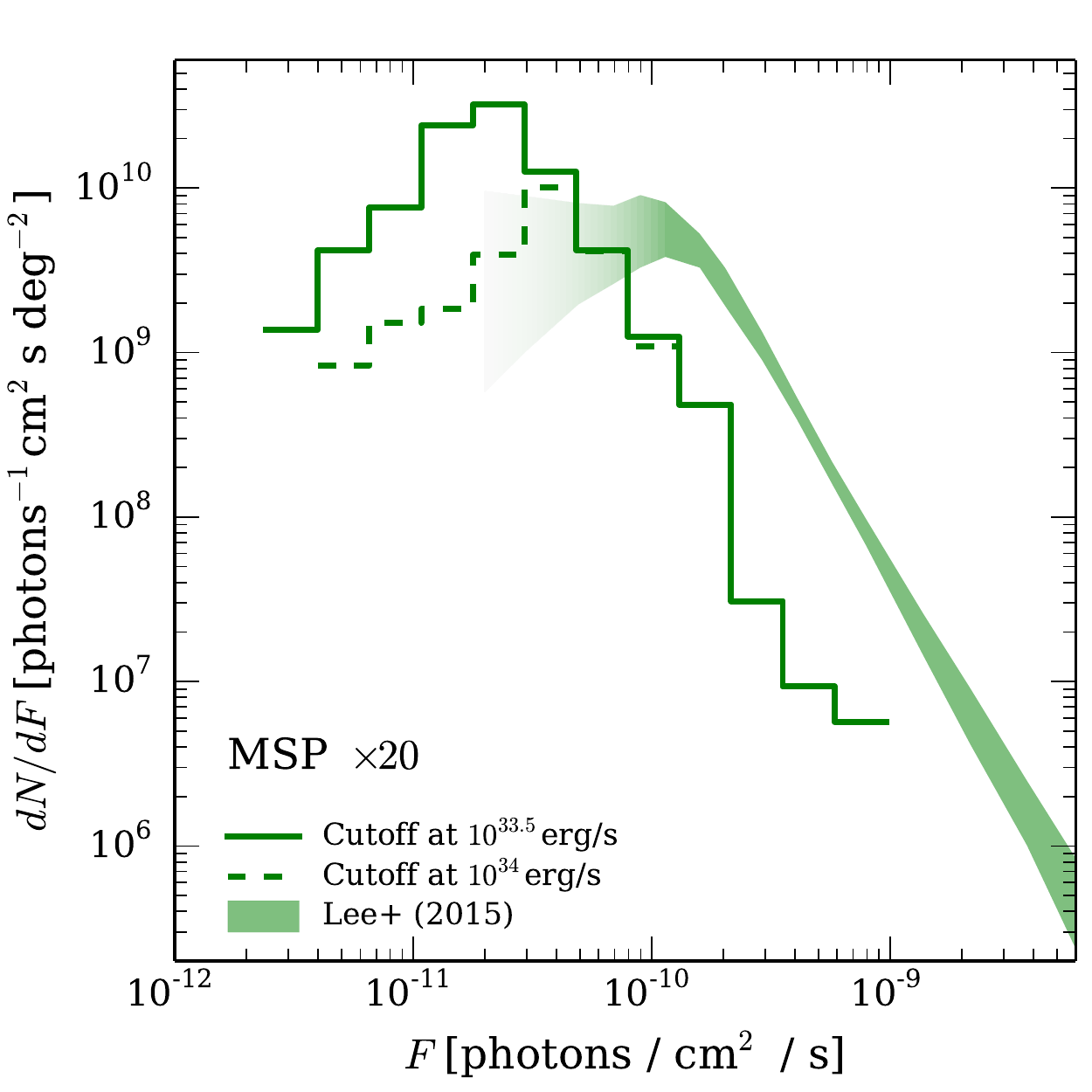}
      \caption{\label{fig:dndf} Flux distribution of  pulsars within $10^\circ$ of the GC ($|b| > 2^\circ$).  Solid lines assume a cutoff of gamma-ray emission at $\dot{E} \!=\! 10^{33.5}\,$erg$\,$s$^{-1}$, dashed lines an $\dot{E} \!=\! 10^{34}\,$erg$\,$s$^{-1}$ cutoff.  The shaded green region shows the model dependent flux distribution of point sources as derived in \citet{2015arXiv150605124L}.  Starting from the {\em Top Left}, we show the results from our Fiducial Model, Scatter Model, Beaming Model, and the MSP $\times 20$ Model.  
}
\end{figure*}

\section{(Ten) Thousand Points of Gamma Rays}
\label{sec:unresolved}
\citet{2015arXiv150605124L} use non-Poissonian photon statistics of pixels to derive the flux distribution of unresolved point sources in the GC region on top of a diffuse emission model.  They focus on the region above the plane ($|b| > 2^\circ$) and within 10$^\circ$ of the GC (though practically the large number of point sources excludes most of the $<\! 5^\circ$ region), using $1.893-11.943\,$GeV gamma rays. The authors found that the entire Excess could be accounted for with a population of unresolved point sources.  In this section, we compare the flux distribution of our four models to see if a population of pulsars is consistent with the unresolved sources inferred in \cite{2015arXiv150605124L}.

In Fig.~2 of \citet{2015arXiv150605124L}, the unresolved background peaks near $2\times 10^{-10}\,$photons$\,$cm$^{-2}$\,s$^{-1}$.  For a typical pulsar spectrum, this corresponds to $3\times10^{-13}\,\ergcms$. Taking a distance of $8.5\,$kpc, this corresponds to a total luminosity of $3\times 10^{34}\,$erg$\,$s$^{-1}$. If we attribute this to a population of unresolved pulsars, this corresponds to $\dot{E}_{\rm peak} = 5\times10^{35}\,$erg$\,$s$^{-1}$ when taking $C=1.3$ ($\dot{E} = 9 \times 10^{35}\,$erg~s$^{-1}$ for $C=1.0$).

Only two out of 40 MSPs have $\dot{E}>\dot{E}_{\rm peak}$.  A much larger fraction of known young gamma-ray pulsars exceed $\dot{E}_{\rm peak}$, 41 of 77 \citep{2013pulsarcatalog}, though only 8 of these are above the plane with $|b|>2^{\circ}$.  Also, the young pulsars have a median characteristic age of only $2 \times 10^4\,$yr.  In our models, this population of pulsars is too rare to reproduce the Excess for the GC SFR we have chosen. Within the region of interest of \cite{2015arXiv150605124L}, we have only {\em two} such pulsars in one realization. Beaming of the gamma-ray emission at low ${\dot E}$  (\S~\ref{sec:beam}) may eliminate this discrepancy. However, the discrepancy with the maximum luminosity of the point sources inferred in \citet{2015arXiv150605104B} is even greater. This is especially the case in the model of \citet{2015ApJ...812...15B}, which is dominated by an old ($\tau \gtrsim 10^9\,\yr$) population of pulsars with low $\dot{E}$.

In Fig.~\ref{fig:dndf}, we directly compare the flux distribution of our four models with the inferred population of unresolved point sources in \cite{2015arXiv150605124L}.  The shaded region in Fig.~\ref{fig:dndf} shows the 68\% contours for the reconstructed flux distribution of resolved  and unresolved point sources in the region of interest. The solid lines show the distribution of point sources in our simulations with a hard cutoff (death-line) at $\dot{E}  = 10^{33.5}\,$erg$\,$s$^{-1}$, as in our Fiducial Model from \cite{2015PF}.  The dashed lines show the same results if we institute a hard cutoff at $10^{34}\,$erg$\,$s$^{-1}$.

In all of our models, the peak of the flux distribution is at lower fluxes, typically corresponding to the flux of pulsars near the gamma-ray death line we imposed.  Increasing the flux of all of our pulsars does not cleanly resolve the discrepancy.  For the $dN/dF$ distributions shown, increasing the flux of all pulsars shifts the lines to the lower right. This can be seen by comparing the Fiducial Model to the Beamed Model, where the observed low $\dot{E}$ pulsars have much larger luminosities (see Fig.~\ref{fig:lumfunc}). The amplitude of $dN/dF$ for the beamed model is in better agreement with the reconstructed flux distribution.

It is a challenge to directly compare the results of our analysis with \cite{2015arXiv150605124L}, since we are reporting the underlying flux distribution from our models, while the derived flux distributions from \cite{2015arXiv150605124L} are created using templates for the point source distribution.  One major assumption of the templates, is that the spectrum of the point sources does not depend on location or brightness.  Our pulsar models violate both assumptions.   

We also note that the $dN/dF$ distribution in \cite{2015arXiv150605124L} does not appear to vary greatly for radial profiles ranging from $r^{-2.2}$ to $r^{-2.8}$, so an even a steeper GC pulsar profile may be consistent with their result, and it may preferentially pick up sources nearer to the plane which are intrinsically brighter.  The break seen in the flux distribution may be caused by a combination of two or more unresolved point source distributions  (e.g., SN remnants, or both YPs and MSPs). Indeed, we estimate that about half of the resolved 3FGL point sources in this region are from pulsars, leaving the other half to other classes of gamma-ray sources.  More detailed modeling may allow us to better recover the properties of the underlying point sources.

The locations of such possible unresolved GC point sources are one place to look for radio pulsars.  However, some fraction may be YRQ pulsars and the ``hot spots'' themselves may not be great proxies for the actual source position.  Systematic uncertainties of the background in the GC region may mean that the effective localization of the point source by {\em Fermi} will be larger than the beam width of a large radio telescope.   A beam width of $\sim\,$few arcmins would require a large area mosaic around such regions, likely requiring a major survey.  A better bet could be continuum imaging with one of the new small-dish wide FoV arrays, e.g., ASKAP, to look for correlations between point sources and {\it Fermi} hot spots.

\section{Discussion and Conclusions}
\label{sec:disc}
Pulsars are ubiquitous in the Milky Way and studies using {\it Fermi} have revealed that pulsed gamma-ray emission is common.  However, other than subtracting the nearest and brightest objects as point sources, young pulsars have been largely neglected in diffuse studies near the Galactic Center, despite a large number expected to result from supernovae in the Central Molecular Zone \cite{2015PF}.  Furthermore, a large population of millisecond pulsars might also be present throughout the bulge from disrupted globular clusters \cite{2015ApJ...812...15B}.  In this work we have explored a variety of emission models of pulsars for both young and millisecond pulsars.  We have used these models to predict both the expected diffuse emission of gamma rays from the unresolved populations of young and millisecond pulsars, as well as their contribution to the unassociated {\em Fermi} point sources in the 3FGL catalog.

We have shown that young pulsars should contribute a significant fraction ($\gtrsim\! 50\,\%$) of the GeV gamma-ray excess observed towards the Galactic Center region using a variety of emission models.  Within $5^\circ$ of the Galactic Center this emission is effectively spherical, with a radial profile that strongly matches that inferred by a number of groups to within the point spread function of {\em Fermi}.
Though a typical luminosity-weighted average for the young pulsar population is dominated by the softer, brightest pulsars confined to the plane, the summed total young pulsar emission becomes harder at larger angles.  Our composite spectra is still somewhat softer than some Excess claims.  However, none of our pulsar models are adequately described by NFW profiles, so we do worry that studies that do not include pulsars self-consistently are oversubtracting fluxes at low energies.

Millisecond pulsars may also be an important component in this region, especially at high latitudes, $|b| \gtrsim 5^\circ$.  We forward modeled MSP evolution throughout the Milky Way in a manner similar to the young pulsars.  We found that the Galactic bulge must have $6 - 20$ times the number of MSPs per stellar mass than the Galactic Disk to explain the GeV excess amplitude outside of $2^{\circ}$.  A few hundred massive, disrupted globular clusters would be sufficient to explain the higher density of pulsars \cite{2015ApJ...812...15B}.  This model, which largely follows the stellar distribution of the bulge, also matched the morphology of the observed excess outside of $2^\circ$ of the Galactic Center.  Within $2^\circ$, we require an even higher concentration of millisecond pulsars per stellar mass than explored here.

We compared predictions of our young and millisecond pulsar models to the observed distribution of {\em Fermi} unassociated point sources in the Galactic Center region, as well as throughout the Galactic Plane.  For the young pulsar population, we expect approximately one-third to one-half of the unassociated point sources near the Galactic Center to be young foreground pulsars depending on the emission model and supernova rate.  Contrary to previous statements \cite[e.g.,][]{2013PhRvD..88h3009H,2015arXiv151204966H}, we find that millisecond pulsars from the Galactic Bulge do not contribute significantly to the point source population inferred in the Galactic Center region.  Old millisecond pulsars are not bright enough to be detected at the distance of the Galactic Center.

Radio observations throughout the Galactic Center region can potentially constrain the contribution of gamma rays from both young and millisecond pulsars. In fact, the young pulsar model discussed here is already consistent with the number of pulsars discovered within the Central Molecular Zone (\S~\ref{sec:radio}).  Within the uncertainties of emission models for gamma-ray-bright pulsars, this suggests young pulsars are a promising source of the gamma-ray emission within this region.  Outside of $\approx 0.4^\circ$, no radio surveys have been sufficiently deep to differentiate between a young pulsar origin and a millisecond pulsar origin of the gamma-ray excess.   Further observations at higher latitudes, whether deep pulsar searches or surveys for point sources, will further constrain the relative contribution from young pulsars and millisecond pulsars.  We found that the young pulsar population can be well characterized by surveys that reach flux limits near $50\,\mu$Jy at 1.4\,GHz. Millisecond pulsars are intrinsically fainter than the young pulsar population, and require surveys reaching $\sim 10\,\mu$Jy to detect more than a few per square degree.

Overall, we find good agreement between our pulsar models and the observed excess distribution. While neither model exactly reproduces the inferred luminosity function of the unresolved point sources near the GC (see ~\S~\ref{sec:unresolved}),  we expect that a significant fraction of flux from unresolved point sources  come from young and/or millisecond pulsars.
One uncertainty is why, in this particular region, the contribution of pulsars is so much more prominent than the rest of the galaxy.  Typically, we find young pulsars contribute $\approx 10\,\%$ of the total GeV flux of the Galaxy.  In the GC region, this is closer to $\approx 30\,\%$.  One possibility is that the strong outflows from this region quickly advect cosmic rays away \cite{2011MNRAS.413..763C,2012MNRAS.423.3512C}, which could explain why the CMZ is also fainter than expected in non-thermal radio emission \cite{2012MNRAS.423.3512C}.

Our estimates suggest that an important implication of YPs, as opposed to dark matter especially, is a large $e^\pm$ luminosity above that typically assumed with the inner kpc of the MW (see left panel of Fig.~\ref{fig:dsnr}).  These will lead to gamma-ray emission through inverse-Compton scattering at a level depending on the amount of spin-down energy converted to energetic $e^\pm$, likely a large fraction with a hard spectrum, and the ambient magnetic field strength relative to the photon field.  We found that an IC flux near to that inferred by {\it Fermi} \cite{FermiGC} can arise.

There may be other contributions to diffuse emission beyond that associated with gas maps or pulsars (possibly including dark matter).
One can also consider primary electron acceleration by supernova remnants, although the escaping flux is uncertain and likely softer.
The much lower gamma-ray flux attributed to neutral pion decay in the inner $\sim 1$~kpc region by \citet{FermiGC} does suggest that secondary $e^\pm$ are far insufficient to account for the IC flux.

We defer more detailed relations to dark matter to \cite{2015toappear}, where we examine a template for dark matter searches (as well as details of the central degrees of the GC).  Obviously we only measure one realization of the pulsar distribution in the MW, which is rather different than for standard assumptions of the smooth dark matter halo (although this too likely depends on peculiarities of Galactic history), although it shows a way forward to extract signals when pulsar backgrounds are relevant.

One could utilize our models with {\it Fermi} data, account for point sources in a consistent manner, and cross correlate observations of dwarf galaxies with the GC to find a signal common to both populations.
Recent dwarf galaxy limits using {\it Fermi} Pass~8 data bite into the parameter space of 30--40~GeV annihilation channel into $b \bar{b}$  \cite{2015arXiv150302641F,2015arXiv150302632T}.  A higher dark matter mass would skirt such present limits (which young pulsar emission automatically satisfy).  A higher-energy component should be least noticeable at the GC, becoming gradually more relevant at larger latitudes (depending on the dark matter density profile).  Signs of this may be evident \cite{2014caloreetal}, although we have also shown that inverse-Compton emission from young pulsars can be at least as luminous than the pulsed emission and likely extends above the disk.

Contrary to a common assumption that the GC GeV excess is already seen with sufficient statistics and that settling the issue will remain entirely in dealing with systematic effects, we suggest that headway can be made with more {\it Fermi} data by just finding more sources and characterizing more pulsars.  Since these are not all from an invisible population, it is realistic to expect improvement by collecting more photons to better understand the signal and the pulsar physics of those objects that cannot be seen individually at the GC.\\

%
We thank John Bally, Casey Law, and Jon Mauerhan  for useful discussions on GC star formation and SN rates; Seth Digel, Troy Porter, and Matt Wood for Fermi discussions.
RMO acknowledges the support provided by NSF grant AST-1313021 and the hospitality of the Aspen Center for Physics.  MDK acknowledges support provided by Department of Energy contract DE-AC02-76SF00515, and the KIPAC Kavli Fellowship made possible by The Kavli Foundation.
 We thank J.\ Bovy for publicly releasing the software package {\sc galpy}, which we used in this work.

\bibliography{pulsars}

\appendix
\section{Systematic Uncertainties}

Throughout our work we have made a number of choices on the properties of gamma-ray pulsars that can have an important impact on their contribution to the diffuse gamma-ray background near the GC.   In this section, we explore a number of these uncertainties, including Galactic properties, the CMZ SNe rate, the magnetic field distribution, and the death line of gamma-ray pulsars, and beaming. 

\subsection{Galactic Uncertainties and Improvements}
\label{sec:cmz}
Although we have simplified the physics of pulsars significantly, i.e., in assuming a constant $B$, $n = 3$, etc., the parameters that we have used were selected to  reproduce the currently observed distribution of pulsars in both the radio \cite{fauchergiguerekaspi2006} and gamma rays \cite{2011ApJ...727..123W}.  Many aspects of these uncertain physical effects are thus practically folded into our  models.

\begin{figure*}
\includegraphics[width=\textwidth,clip=true]{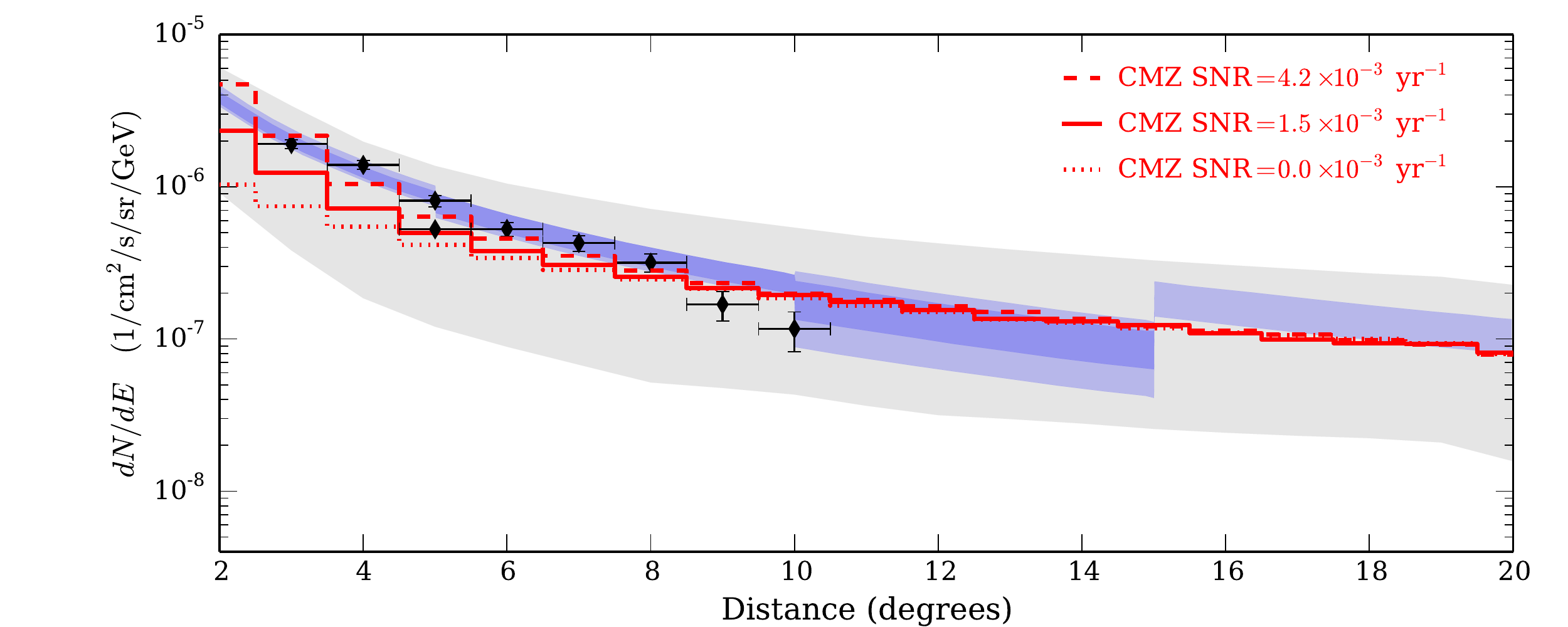}
\caption{\label{fig:snr}
Angular intensity of the GC pulsar background at 2\,GeV for different CMZ SNe rates.
The solid red line shows the pulsar contribution from our Fiducial Model with a CMZ SNR equal to $7\,\%$ of the Galactic rate.
The dotted red line assumes a CMZ SNR proportional to the Galactic fraction of WR stars in the CMZ, $\approx 20\,\%$ \cite{2015MNRAS.447.2322R,2015MNRAS.449.2436R}.
The dotted red line shows the contribution of the Galactic disk alone, which provides the majority of the flux outside of $\sim\! 6^\circ$.
These are compared to the GC Excess intensity at 2\,GeV inferred in Ref.~\cite{2014daylanetal} ({\it black points and error bars)} and Ref.~\cite{2014caloreetal} (shaded blue regions).
}
\end{figure*}

The SN rate at the GC and elsewhere can be better independently estimated, by improved counts of evolved massive stars or other means,  e.g., gamma rays from $^{26}$Al decay ($\tau \sim 1$~Myr) from Wolf-Rayet stars and SNe is also concentrated on the plane within $|l| \lesssim 30^\circ$ \cite{2009A&A...496..713W}.
The Excess may itself prove to be a useful probe of the GC SN rate.  For instance, the Fermi bubbles are natural connected to the history of the GC, whether due to AGN activity or SNe, which could themselves inflate bubbles \cite[see, e.g., ][]{2011MNRAS.413..763C,2014MNRAS.444L..39L}.  Indeed, one possible shortcoming of the young pulsar interpretation, is that the gamma rays appear to be suppressed in the GC region compared to the star formation rate.

As we discussed in detail in \S~\ref{sec:birth}, the birth rate of neutron stars in the CMZ is fairly uncertain, and likely has changed over time.  In our Fiducial Model we assume that pulsars are born at $7\,\%$ of the Galactic pulsar formation rate, as this is consistent with the varied estimates of the SNe rate in the CMZ.  Nonetheless, the actual rate may be as high as $20\,\%$ if we use the WR population as a tracer population for SN.

In Fig.~\ref{fig:snr}, we show the contribution of pulsars to the diffuse background near the GC for three different CMZ SN rates.  The solid red line shows our Fiducial Model with a contribution of $7\,\%$ of the Galactic rate.   The dashed red line shows the contribution from the Galactic Disk only, with no contribution from the CMZ at all. Comparing these two lines, we see that the CMZ contributes only a small fraction of the diffuse flux outside of $\gtrsim 5^\circ$;  foreground pulsars from the Galactic disk contribute the majority of the flux at higher longitudes.   Finally, the dashed line shows the contribution of pulsars if the CMZ SNR is $20\,\%$ .  We see this is sufficient to reproduce the entire observed Excess. Nevertheless, we expect contributions from other sources in this dense region (e.g., MSPs).  In principle, this shows that the CMZ SN rate can not be significantly larger than $\sim 4\times 10^{-3}\,$yr$^{-1}$, otherwise we expect the unresolved pulsar emission from this region to exceed what has been observed.

Here and in \cite{2015PF}, we did not include star formation from the central parsec of the Milky Way, only populating the region within $20 - 200$~pc of Sgr A$^*$.  Some aspects have been discussed in \cite{2015arXiv151100723K,2015arXiv151101159K} and we will devote even more detailed consideration of this interesting region elsewhere, but do focus on their radio characteristics in \S~\ref{sec:radio}.
Also, \citet{2011ApJ...727..123W} included a thick disk component in their models with a 10\% runaway component and 500~pc full-width-half-maximum.  While we have not included such a process in our results, it would shorten travel times for such YP to reach large angles on average. We take our model to be more conservative in estimating the gamma-ray emission at high latitudes.

Relative to MSPs, we have much better prospects for understanding from first principles the young pulsar luminosity density profile.  Further, we also have a much better expectation for the distribution and luminosity of individual pulsars.  Since at later ages beaming may become more relevant, the pulsars that are visible to us will be brighter than the average we have assumed above, improving the prospects for detecting this population with more observations.

In the inner kpc of the MW, there is clearly residual emission along the disk even after the detailed account of foregrounds in \cite{FermiGC}.
There is also an excess of gamma rays observed throughout the Galactic Plane compared to diffuse {\em Fermi} models \cite{2012ApJ...750....3A}.  This is especially evident in the inner galaxy region ($|b| < 8^{\circ}$ and $|l| < 80^\circ$) where the model and the data differ by up to $35\,\%$  \citep[see, e.g., Fig.~17 in][]{2012ApJ...750....3A}.

Overall, we find that the unresolved young pulsar population can contribute of order $\approx\! 5 \!-\! 10\,\%$ of the total diffuse emission in the inner Galactic plane at 2~GeV, depending on the normalization of the disk rate and lower $\dot{E}$ cutoff, consistent with earlier estimates for EGRET \cite{2007Ap&SS.309...35S}.  This may affect inferences of cosmic-ray proton and electron spectra within these regions when included, and thus extrapolations of such spectra to higher energies.


\begin{figure}[t!]\vspace*{-0.55cm}
      \includegraphics[width=0.41\textwidth,clip=true]{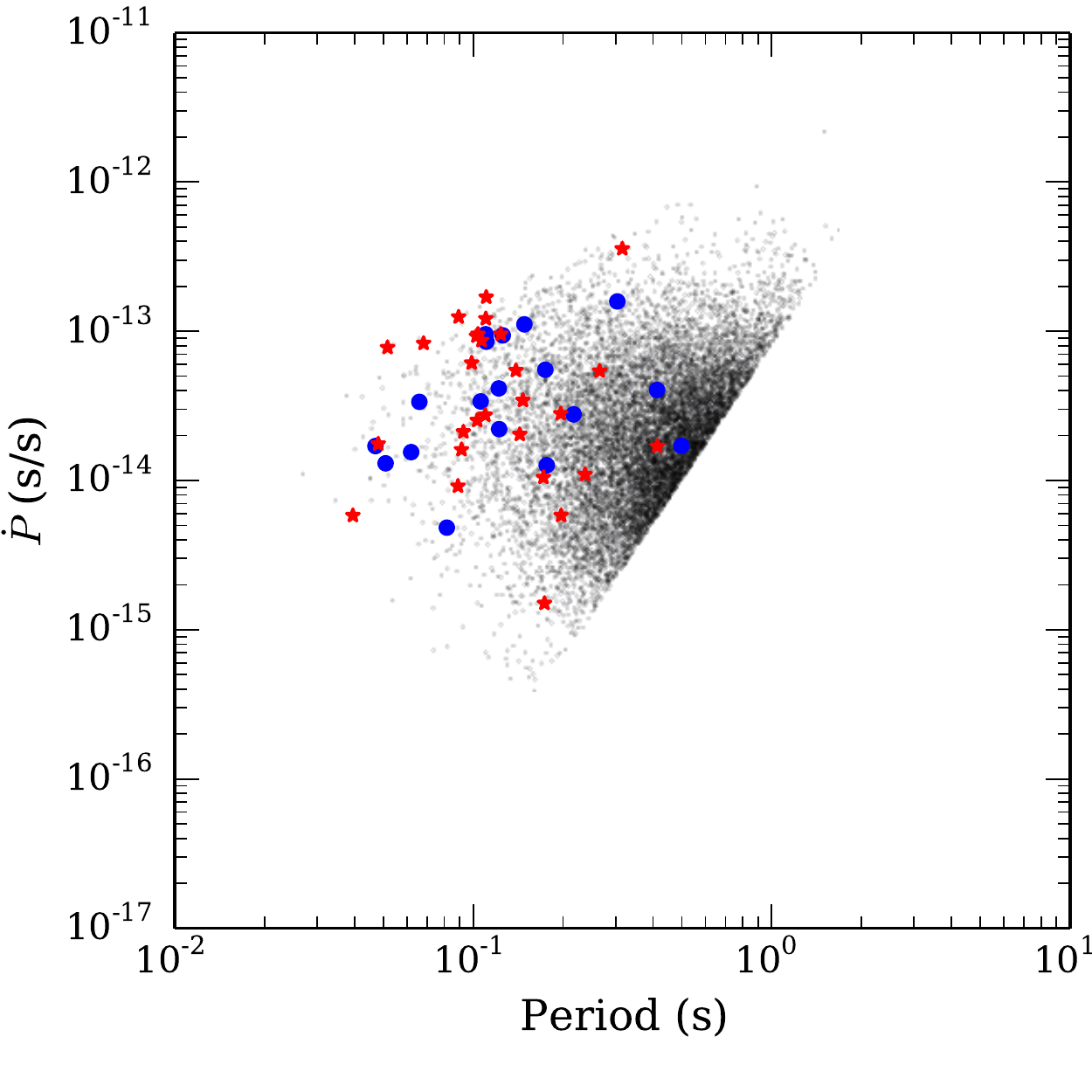}\\
      \vspace*{-0.4cm}
      \includegraphics[width=0.41\textwidth,clip=true]{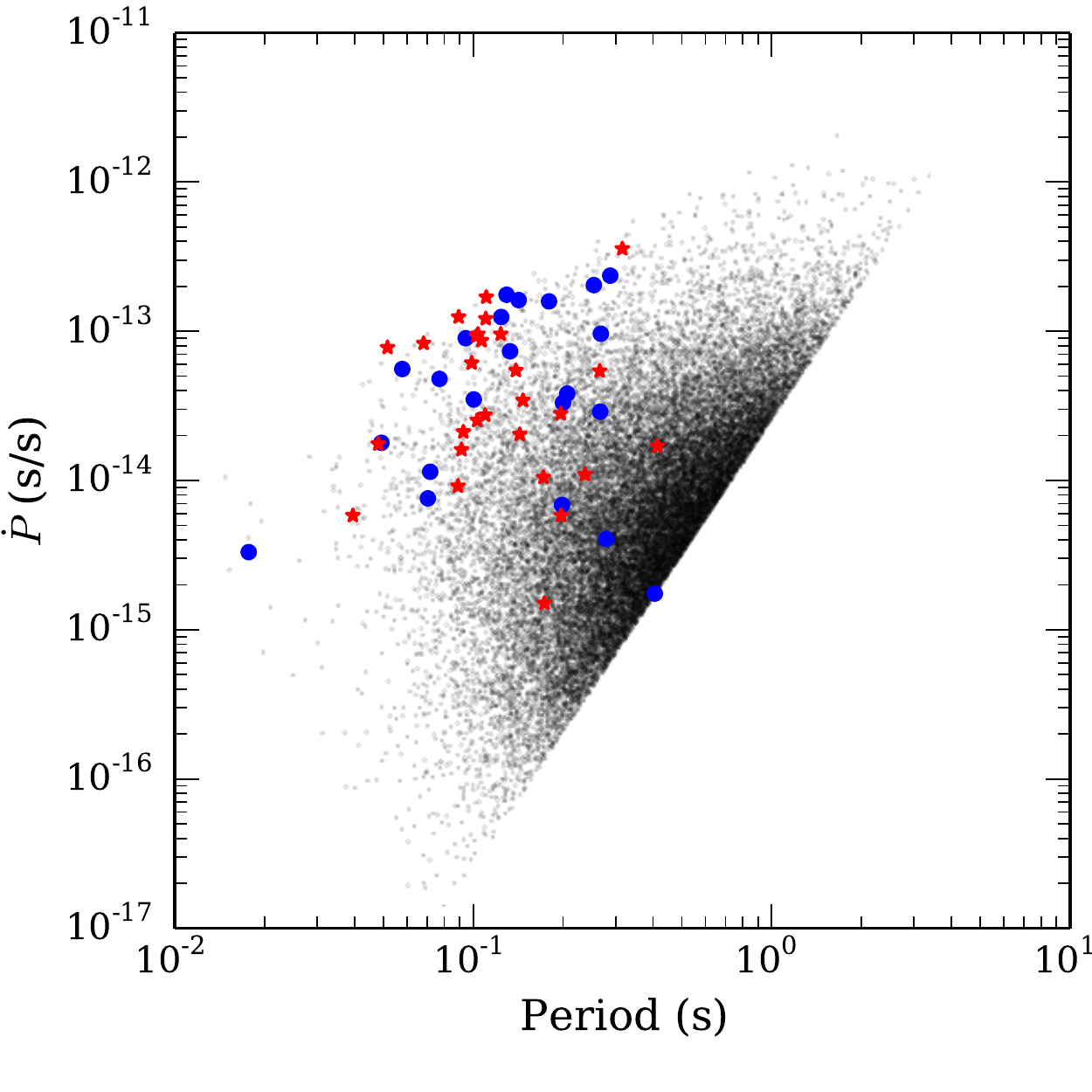}\\
      \vspace*{-0.4cm}
      \includegraphics[width=0.41\textwidth,clip=true]{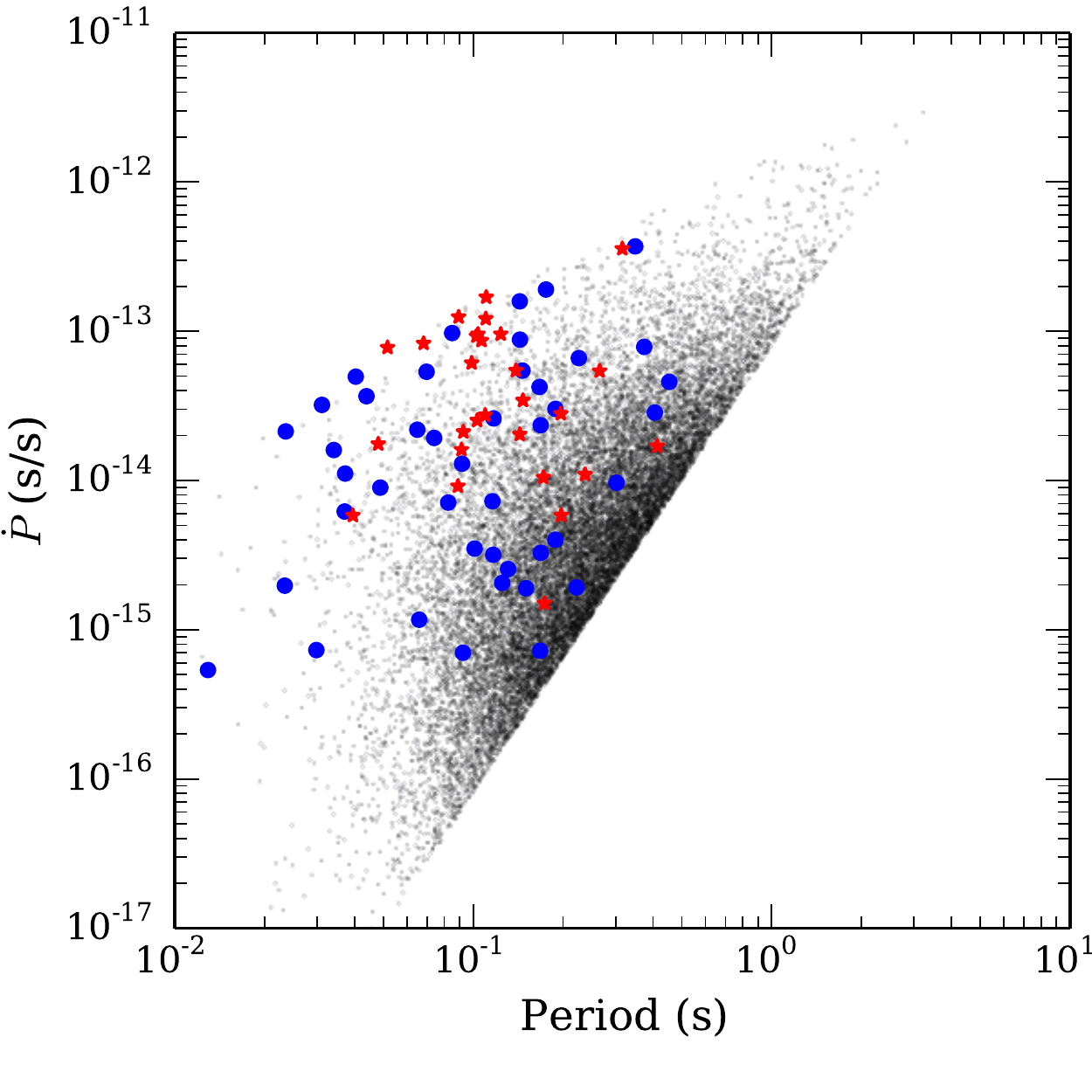}
      \vspace*{-0.6cm}
\caption{\label{fig:ppdot}
Pulsar period-period derivative diagram.  Shown are single models from three different initial magnetic field distributions with $\sigma_{\log{B}} = 0.30$, $0.45$, and $0.55$, from top to bottom.  Red stars show known {\em Fermi} pulsars with flux $> 10^{-10}\,\ergcms$. Blue circles show the model pulsars with the same flux limit. Black points show the entire Galactic model population with $\dot{E} > 10^{33.5}\,$erg$\,$s$^{-1}$.
}
 \end{figure}

\subsection{Magnetic Field Distribution}
\label{sec:B}

There is a general disagreement between the intrinsic properties of pulsars that are assumed in population synthesis studies of radio pulsars and gamma-ray pulsars.  Although both use the same mean magnetic field strength, radio pulsars are typically modeled with a larger variance $\sigmab = 0.55$ \cite{fauchergiguerekaspi2006} than the gamma-ray population $\sigmab = 0.3$ \cite{2011ApJ...727..123W}.  We examine here these two populations in more detail, with a focus on the observational selection effects that impact the gamma-ray pulsar population.

When comparing our population synthesis models to the observed population, we focus on the bright pulsars with flux $\gtrsim 10^{-10}\,\ergcms$.  This population of pulsars should be complete in the {\em Fermi} data.  Using the 2PC, we find that the variance of the {\em observed} population is $\sigmab \approx 0.31$. This is the value used by \citet{2011ApJ...727..123W}.  However we find that observational selection effects cause the {\em observed} variance of the bright pulsars to be smaller than the {\em intrinsic} variance, typically by  $20\,\%$.  This  is a natural consequence of pulsars with large magnetic fields evolving more rapidly to lower luminosities.  This suggests that the intrinsic value of $\sigmab$ is likely closer to $0.4$ than $0.3$.  

In Fig.~\ref{fig:ppdot}, we compare the $P$-$\dot{P}$ distribution of the pulsars with three different initial magnetic field distributions.  With red circles, we show the bright observed gamma-ray pulsars with flux $\gtrsim 10^{-10}\,\ergcms$ that are the basis for most of our comparisons.  The black points show the entire gamma-ray pulsar population for each of our models, and the blue points show the simulated pulsars with flux  $\gtrsim 10^{-10}\,\ergcms$.  Each model shown is selected to be a representative sample with a Kolmogorov-Smirnov (KS) p-value near the mean of all the simulations.  More generally, for each model we can evaluate the observed distribution of $\sigma_{\log B}$ for the brightest pulsars and determine the fraction of realizations that reproduce the observed distribution. For our Fiducial Model with $\sigma_{\log B} = 0.45$, we find $15\,\%$ of our simulations have a simulated variance less than that observed. With $\sigma_{\log B} = 0.40$, we find $30\,\%$ of our simulations have a inferred dispersion less than the observed dispersion.  In contrast, only $2\,\%$ of models with $\sigma_{\log B} = 0.55$ have smaller variance.

\begin{table}[t!]
  \caption{Estimated $\sigma_{\log B}$ of young pulsars in $\dot{E}$ ranges.  We estimate the {\em observed} dispersion of pulsar magnetic fields using a MCMC method that iteratively removes outliers (see text for details).  The first column shows the range of $\dot{E}$ used, in units of $\ergs$, for both the {\em Fermi} gamma-ray pulsars  \cite{2013pulsarcatalog} and the  ATNF catalog \cite{manchesteretal2005}, excluding the pulsars discovered by {\em Fermi}. The error bars show the statistical errors, with the parenthesis showing the difference between the measured $\sigma_{\log B}$ with and without the outliers.  The bright pulsars are {\em Fermi} pulsars with gamma-ray flux above $10^{-10}\,\ergcms$. }
\label{tab:edot}
\begin{ruledtabular}
\begin{tabular}{rcllll}
		&&	& \hspace{1cm}		& {\em Fermi} &    ATNF \\ 
\hline
$5 \times 10^{35}<$&$\dot{E}$&   &     & $0.32^{+0.04(5)}_{-0.03}$   & $0.45^{+0.04}_{-0.04}$ \\
$10^{35}<$&$\dot{E}$&$<5 \times 10^{35}$   &	& $0.29^{+0.06}_{-0.04}$   &  $0.31^{+0.04(17)}_{-0.03}$ \\
$3\times 10^{34}<$&$\dot{E}$&$<10^{35}$   &	& $0.33^{+0.12}_{-0.07}$  & $0.47^{+0.04(17)}_{-0.04}$ \\
$3 \times 10^{33}<$&$\dot{E}$&$<3 \times 10^{34}$&  & $0.43^{+0.15}_{-0.10}$   &  $0.40^{+0.02(22)}_{-0.02}$ \\
$10^{33}<$&$\dot{E}$&$<3 \times 10^{33}$   &	& ---   & $0.42^{+0.02(28)}_{-0.02}$ \\
$10^{32}<$&$\dot{E}$&$<10^{33}$   &	& ---   & $0.46^{+0.02(28)}_{-0.01}$ \\
Bright&&&&$0.32^{+0.05}_{-0.04}$ & --- \\
All&&&&$0.32^{+0.03(5)}_{-0.03}$ &  $0.52^{0.01(20)}_{-.01}$\\
\end{tabular}
\end{ruledtabular}
\end{table}

\begin{figure*}
\includegraphics[width=\textwidth,clip=true]{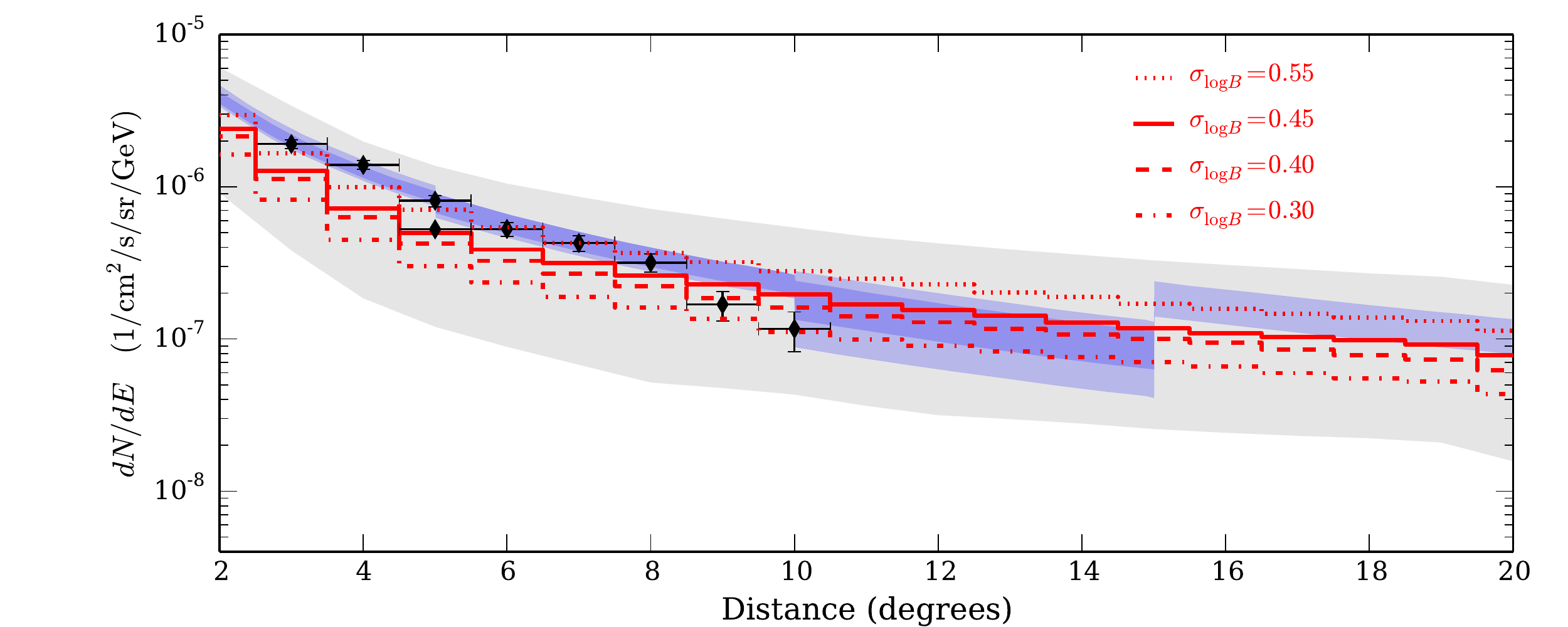}
\caption{\label{fig:sigmab} Intensity of the pulsar background at  2\,GeV as a function of angular separation from the Galactic Center, compared to the GC Excess intensity at 2\,GeV inferred in Ref.~\cite{2014daylanetal} ({\it black points and error bars)} and Ref.~\cite{2014caloreetal} (shaded blue regions).
The solid red line shows the pulsar flux from our Fiducial Model.  The other red lines assume a pulsar $\sigma_{\log{B}} = 0.55$, the value that best describes the radio population of pulsars, and $\sigma_{\log{B}} = 0.30$ as used in \cite{2011ApJ...727..123W}.
}
\end{figure*}

In this work we do not directly model the observational selection effects of the radio discovered pulsars, although we expect the bias to be towards lower values of $\sigmab$.  Nevertheless, our initial conditions should generally be informed by the properties of the radio population. We estimate $\sigma_{\log B}$ using the Markov chain Monte Carlo ensemble sampler {\texttt emcee} \cite{2013PASP..125..306F}, where we iteratively removed $3\!-\!\sigma$ outliers between each estimate until we reached a converged value. We use a uniform prior for $\sigma_{\log B}$.   In Table~\ref{tab:edot}, we report our estimates of $\sigma_{\log B}$ for both the gamma-ray pulsars and radio pulsars in bins of varying $\dot{E}$, as well as the bright population.

Overall, the population of radio pulsars have larger {\em observed} $\sigma_{\log B}$ compared to the gamma-ray pulsars, even when removing the outliers. As we incorporate more pulsars into the radio analysis, most of which do not emit in gamma rays, the distribution begins to approach that inferred by  \citet{fauchergiguerekaspi2006}, but remains smaller. This is likely due to similar observational selection effects. One way to alleviate the discrepancy between the gamma-ray pulsar properties and the radio pulsar properties is to allow the initial spin period to depend on the surface magnetic field of the pulsar.  Alternatively, the differences observed within the radio population may depend on some true physical effects that we do not model here.  For example, some population synthesis models allow the surface field to decay with time.  There is still no compelling evidence from radio surveys \cite{fauchergiguerekaspi2006} or in this work that requires us to include surface field decay.

\begin{figure*}[t]
\includegraphics[width=\textwidth,clip=true]{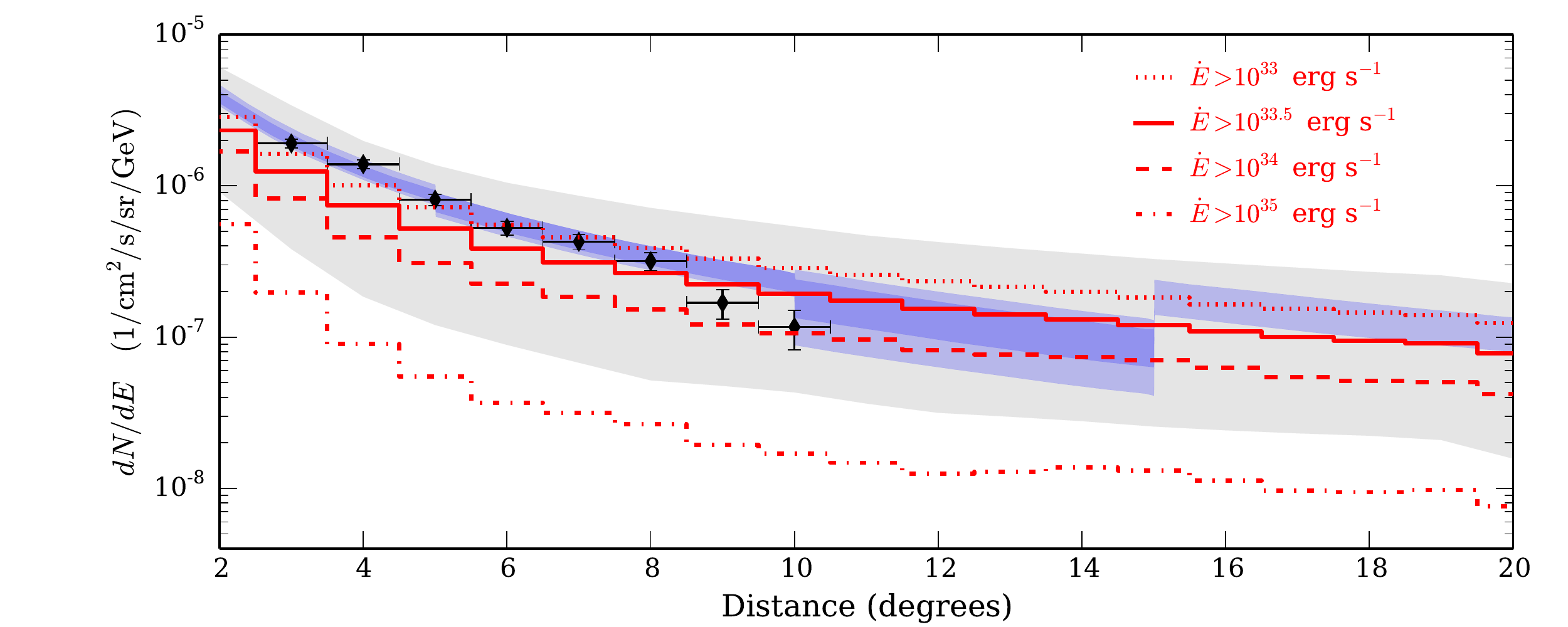}
\caption{\label{fig:edot} Intensity of the pulsar background at 2\,GeV as a function of angular separation from the Galactic Center.
The solid red line shows the pulsar contribution from our Fiducial Model.
The other red lines show the contribution to the Excess for pulsars with ${\dot{E}} > 10^{33}$ (dashed), $10^{34}$ (dotted), and  $10^{35}$\,erg s$^{-1}$ (dash-dotted).  The green   dashed line shows the contribution of pulsars at high Galactic latitude ($|b| > |l|$ and $|b| > 2^\circ$).
These are compared to the GC Excess intensity at 2\,GeV inferred in Ref.~\cite{2014daylanetal} ({\it black points and error bars)} and Ref.~\cite{2014caloreetal} (shaded blue regions).
}
\end{figure*}

In Fig.~\ref{fig:sigmab}, we show the diffuse gamma-ray flux near the GC for four models with $\sigma_{\log B} = 0.3$, 0.40, 0.45, and 0.55 dex.  We see that the estimated flux from pulsars drops by approximately a factor of two when going from $\sigma_{\log B} = 0.55$ to 0.30, which correspond to the preferred values from previous analyses of radio and gamma-ray discovered pulsars, respectively.   Pulsars with large magnetic fields evolve more quickly and emit gamma rays for a shorter time than pulsars with weaker fields.   By having a broad distribution of magnetic fields, such as those implied by radio observations, many more pulsars with weak fields are born. These pulsars can travel farther within their gamma-ray emitting lifetime, where they contribute to the diffuse gamma-ray background.

 Overall, we find that $\sigma_{\log B} = 0.45$ seems to match the properties of the combined population of gamma-ray and radio selected pulsars better than 0.3.  It is only slightly larger than  most likely {\em intrinsic} variance of the gamma-ray population,  $\sigma_{\log B} = 0.37$.

The large systematic difference between the observed gamma-ray pulsars and the radio pulsars in the ATNF catalog \cite{manchesteretal2005}  really drives our use of a larger dispersion.

The pulsar formation rate and the distribution of magnetic fields are both degenerate with how the pulsars evolve with time.  In particular, discrepancies could arise from uncertainties in the moment of inertia in addition to simplifications of the rotating dipole model.  The star formation tracers probably have some systematic uncertainties in their normalization as well.  One way of getting around this is to populate the galaxy with pulsars until one arrives at the number of objects observed (i.e., obtaining an effective pulsar birth rate that may large compared to the SN rate \cite{fauchergiguerekaspi2006,2012A&A...545A..42P}).  In order to keep the number of free parameters to a minimum, we choose not to model these directly for the young pulsar population.

\subsection{Gamma-ray Death Line}
\label{sec:cutoff}

For the heuristic value of $C = 1.0$, the gamma-ray efficiency of pulsars rises to 100\% as $\dot{E}$ approaches $10^{33}\,$erg$\,$s$^{-1}$.   For our Fiducial Model $C=1.3$, the pulsars reach unity efficiency at a higher energy, $\dot{E} \approx 1.7\times 10^{33}\,\ergs$.
We counter this effect in our Fiducial Model by setting the gamma-ray flux to zero for pulsars with $\dot{E} \lesssim 10^{33.5}\,\ergs$, which corresponds to a maximum efficiency of $73\,\%$.  
It is, of course, also implausible that the emission would die just as the efficiency approaches unity.  Nevertheless, we first analyze our results looking at fixed cutoffs for a variety of ${\dot E}$. In Fig.~\ref{fig:edot}, we show the amount of diffuse flux near the GC for a variety of energy cuts from $10^{33}\,\ergs$ to $10^{35}\,\ergs$. Later, lacking the true model of pulsars, we take general, empirical approach that can be scaled for a given magnetospheric emission model.

\begin{figure*}[t]
\includegraphics[width=\textwidth,clip=true]{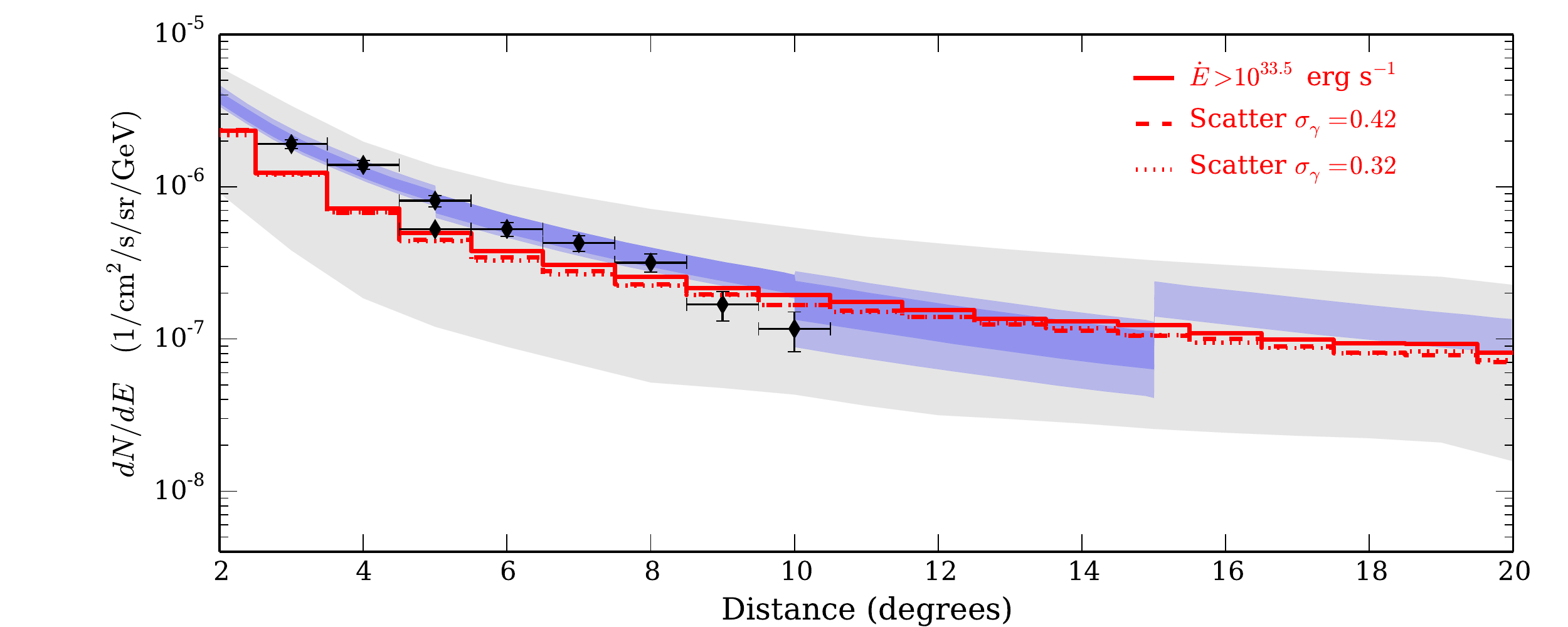}
\caption{\label{fig:scatter} Angular intensity of the GC pulsar background at 2\,GeV for different beaming models.
The solid red line shows the pulsar contribution from our Fiducial Model.
The other lines show the diffuse flux when we include an intrinsic log-normal scatter to the pulsar population, with the noted amplitude, $\sigma_{\log L}$. 
These are compared to the GC Excess intensity at 2\,GeV inferred in Ref.~\cite{2014daylanetal} ({\it black points and error bars)} and Ref.~\cite{2014caloreetal} (shaded blue regions).
}
\end{figure*}

At present, the YRQ gamma-ray pulsar PSR J2139+4716 has lowest inferred $\dot{E}$$\sim 3 \times 10^{33}\,$erg s$^{-1}$ \cite{2012ApJ...744..105P}.  
There is indeed much room at the bottom and the luminosity distribution stands to improve \cite{2011ApJ...738..114R,2013pulsarcatalog,2014A&A...570A..44H}.  This population tends to be YRQ, which means these pulsars must be discovered through dedicated blind searches of the gamma rays.
Presently, we find that the observed distribution of the brightest pulsars makes it hard to exclude  a deathline near $10^{33}\,\ergs$. That is, when looking at pulsars with luminosities between $10^{33}\,\ergs$ and $10^{33.5}\,\ergs$, $44\,\%$ of our simulations have none with fluxes greater than $10^{-10}\,\ergcms$. This is consistent with the observed population.

Allowing the pulsars to approach unity efficiency results in a boost of flux from unresolved point sources near the GC.  
Primarily, magnetic field curvature and spin-down power determine when gamma-ray emission in young pulsars completely shuts off, i.e., when the pair cascades are not able to sustain themselves and can no longer produce enough pairs to short out the accelerating gaps.  This is also when the gap size increases ($w$) so that the emitting surface moves onto less-curved field lines (i.e, a smaller beam.)   Eventually, as the field line curvature and potential become small enough a charge-separated magnetosphere results (like the classical Goldreich-Julian picture).  This is the death line predicted in open gap (OG)  models.  For example, \citet{2011ApJ...736..127W}, estimate the death line to be approximately
\begin{equation}
  \label{eq:wangdl}
\log_{10} {\dot P} = -13.49 + 3.67 \log_{10} P.
\end{equation}
For $P = 0.7\,$s this corresponds to $\dot{E} \approx 1\times 10^{33}\,\ergs$, the lowest value we consider in Fig.~\ref{fig:edot}. 
Other estimates of the outer gap death line are at even lower energies \cite[e.g.][is closer to $\approx 2\times 10^{32}\,\ergs$]{2004ApJ...604..317Z}.

The death line evolves a fair amount depending on the amount of physics included in the calculation.  The most sophisticated vacuum models are those based on 2- and 3-dimensional solutions to the particle transport and Poisson equation in the gaps.  The exact details of gamma-ray emission approaching this regime remain unclear, so showing the results of cutoff at various $\dot{E}$ values can be used to estimate the effect of a more smooth transition to the death line.

\subsection{Intrinsic Scatter}
We also test whether the pulsar population can be more accurately described by including an intrinsic log-normal scatter in the total flux of the pulsars. This is motivated, in part, by the observed scatter in pulsar fluxes, as well as previous models of the radio luminosity of pulsars \cite[e.g.][]{1987A&A...171..152S}.

\begin{figure*}[t]
\includegraphics[width=\textwidth,clip=true]{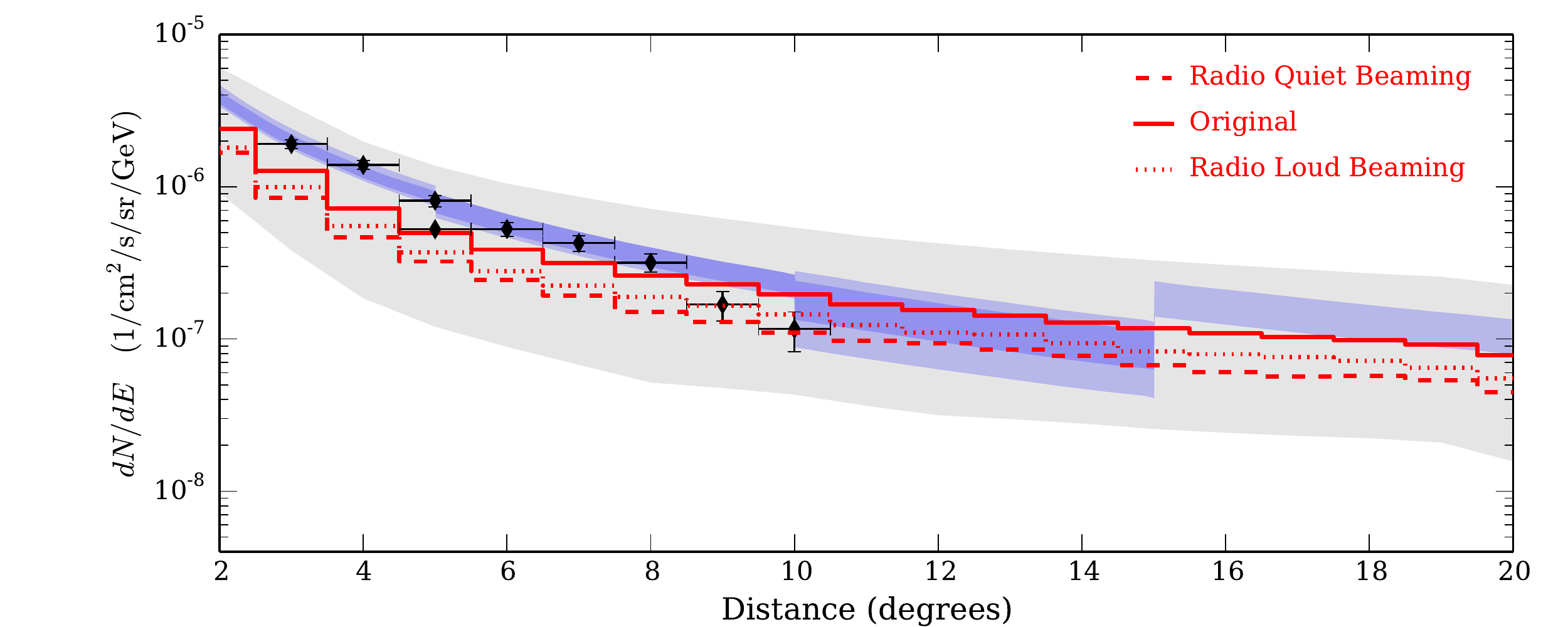}
\caption{\label{fig:beamr} Intensity of the pulsar background at 2\,GeV for different beaming models.
The solid red line shows the   pulsar contribution from our Fiducial Model.
The   dotted red line shows the excess expected if we use the YRL beaming model described in \S~\ref{sec:beam}.
Likewise, the dashed red line shows the excess expected if we use the YRQ beaming model.  
These are compared to the GC Excess intensity at 2\,GeV inferred in Ref.~\cite{2014daylanetal} ({\it black points and error bars)} and Ref.~\cite{2014caloreetal} (shaded blue regions).
}
\end{figure*}

\begin{figure*}
  \includegraphics[width=.45\textwidth,clip=true]{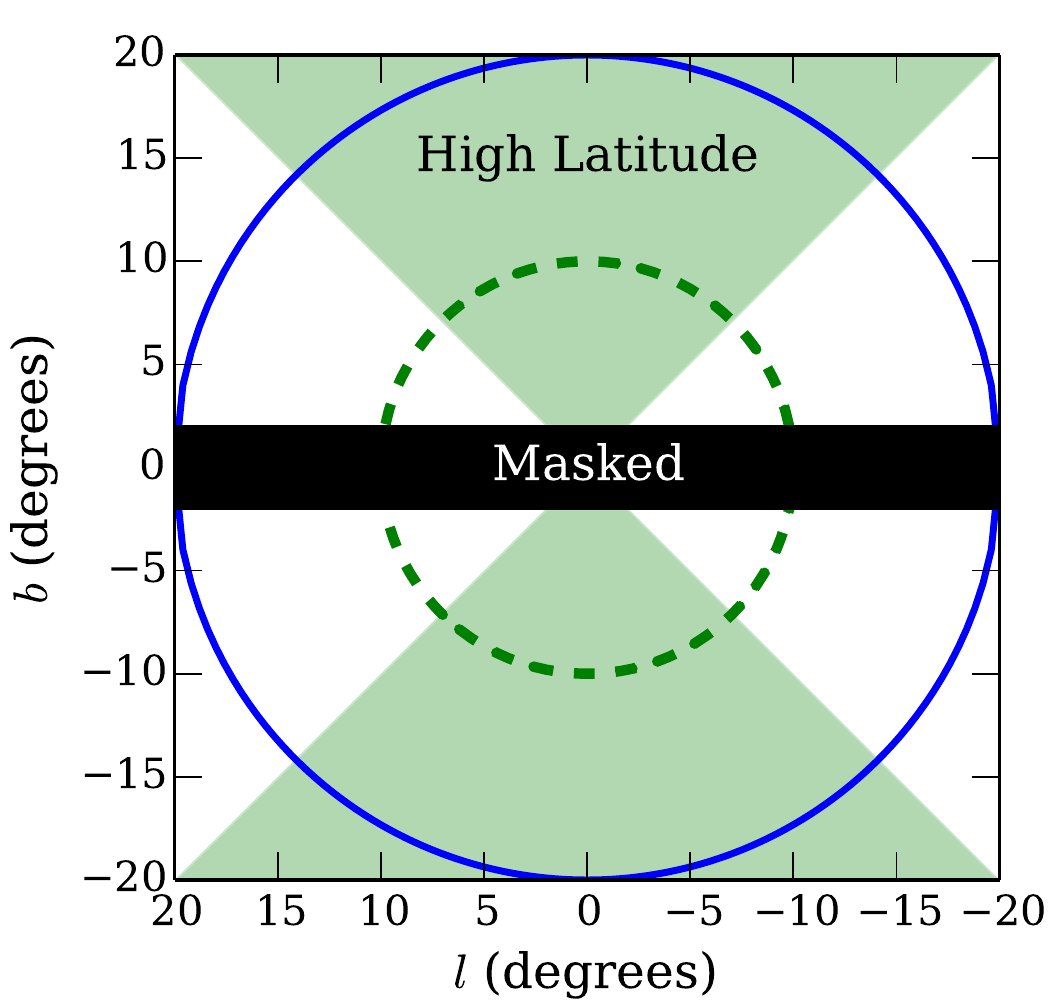}
  \includegraphics[width=.45\textwidth,clip=true]{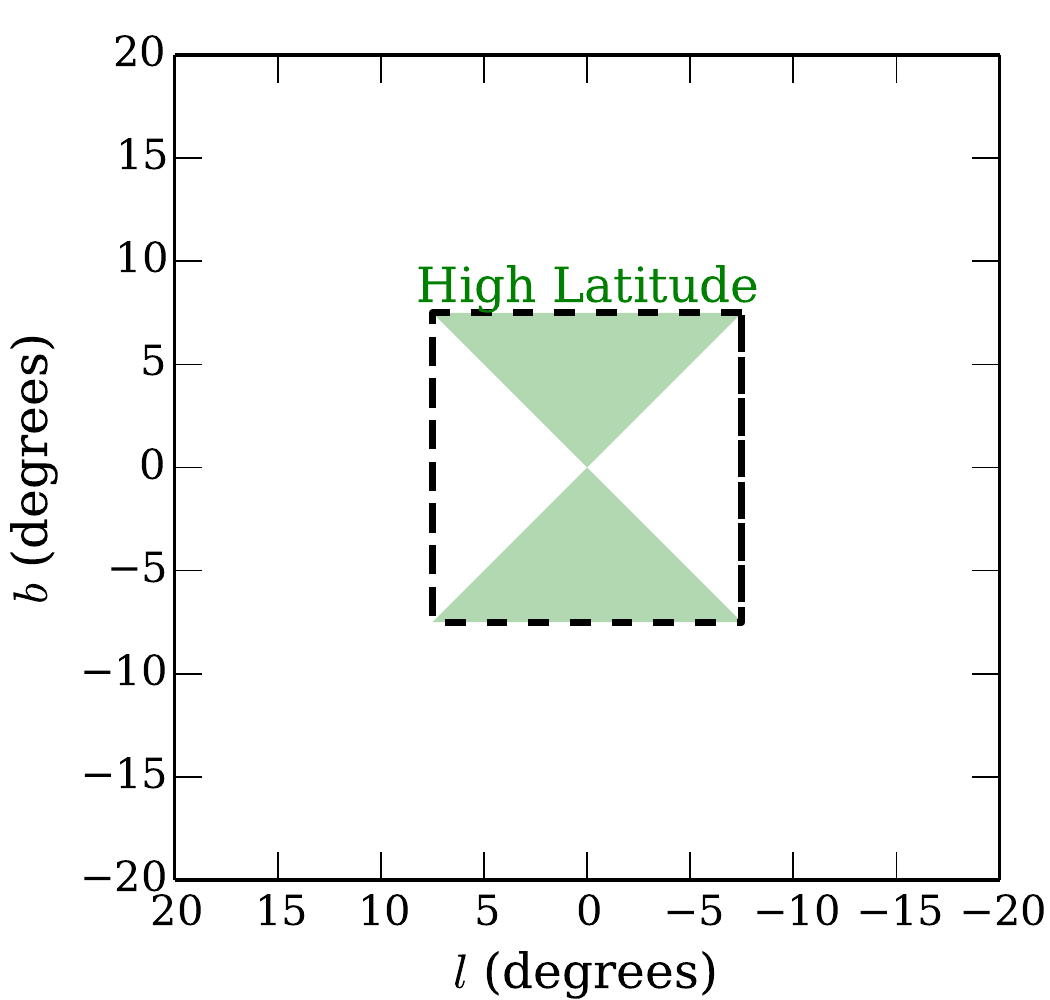}
  
  \caption{\label{fig:rois} Regions of interest.  The {\em left panel} shows the regions used in the analyses of  \citet{2014arXiv1409.0042C} ({\em solid blue line}) and \citet{2015arXiv150605124L} ({\em dashed green line}).  Both analyses mask the Galactic plane ($|b| > 2^\circ$).  
The {\em right panel} shows the region used by  \citet{FermiGC} which does not mask the plane.  The green shaded regions show the high latitude region with $|b| \geq |l|$.  }
  \end{figure*}

\begin{figure*}
  \includegraphics[width=\textwidth,clip=true]{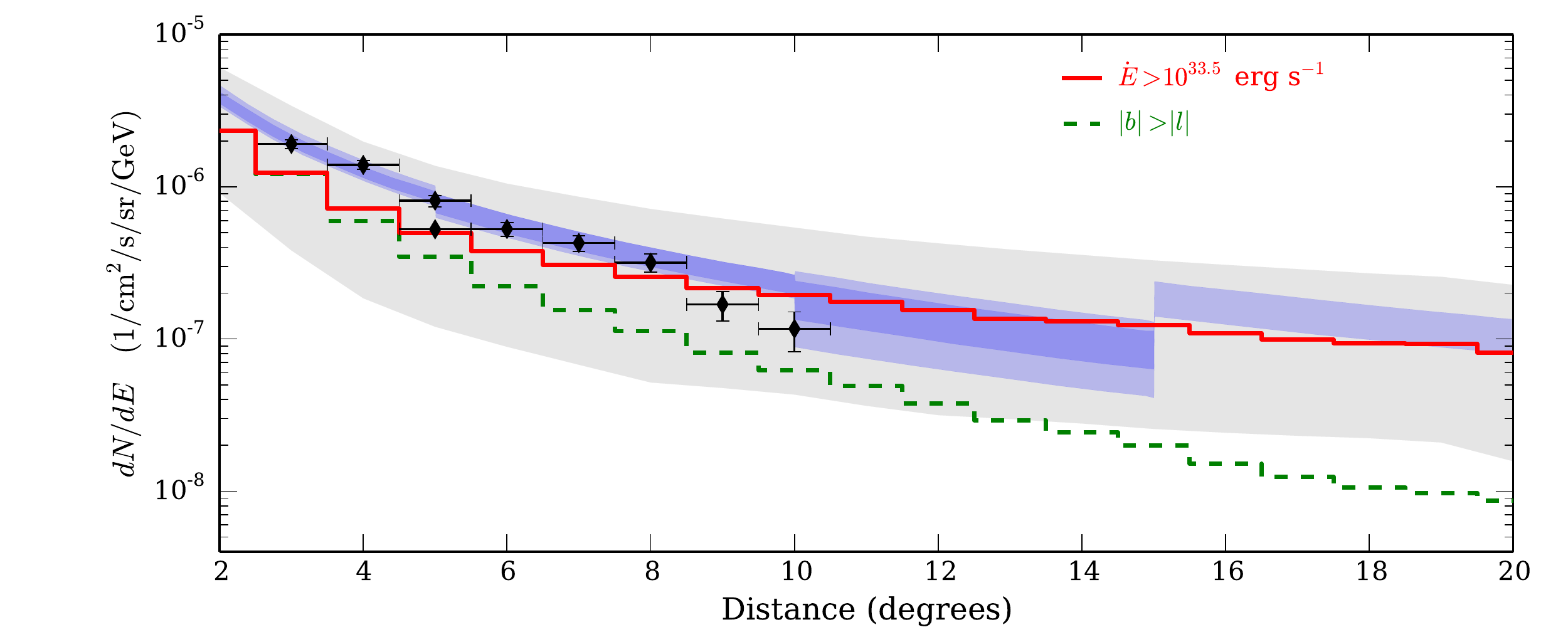}
\caption{\label{fig:highlat}
Angular intensity of the GC pulsar background at 2\,GeV for different regions of interest.
The solid red line shows the pulsar contribution from our Fiducial Model with $|b| > 2^{\circ}$.  
The dashed green line shows the contribution at high latitudes with $|b| > |l|$. Outside of $\sim\! 6^{\circ}$ the majority of flux comes from the Galactic disk which is not spherically symmetric in this region. 
These are compared to the GC Excess intensity at 2\,GeV inferred in Ref.~\cite{2014daylanetal} ({\it black points and error bars)} and Ref.~\cite{2014caloreetal} (shaded blue regions).
}
\end{figure*}

In Fig.~\ref{fig:scatter}, we show how including the intrinsic scatter of the pulsars impacts the diffuse flux towards the GC.  We include our Fiducial Model, and models with $\sigma_{\log{L}} = 0.32$, and $0.42$. In all cases with scatter we set $C=1.0$, yet the scatter effectively increases the mean of the population. For $\sigma_{\log{L}} =$0.32, and 0.42, the mean luminosity of the population is equal to  $1.3$ and $1.6$ respectively, when we impose no limits on the luminosity. In practice,  when $L_\gamma$ is selected to be above the spin down luminosity of the pulsar, we set $L_\gamma = {\dot E}$. 

  Overall, including the scatter only moderately impacts the observed distribution of the diffuse background.  With $\sigma_{\log L} = 0.32$, we get $C \approx 1.3$, and do not markedly change the distribution of point sources, except to increase the total number of bright point sources.  Finally, this scatter automatically tapers the flux of pulsars with $\dot{E} \approx 10^{33}\,\ergs$, because we include a hard ceiling on the flux, but impose no floor.  

We found, in studying the beaming corrected fluxes of pulsars from \citet{2015A&A...575A...3P}, that the intrinsic scatter is actually closer to $\sigma_{\log L} = 0.42$, which would give a mean flux closer to $C = 1.6$.  Using this high of a scatter requires a larger fraction of the unassociated 3FGL point sources to be pulsars that either $\sigma_{\log L} = 0.32$ or our Fiducial Model.

\subsection{Beaming}
\label{app:beam}

In \S~\ref{sec:beam}, we described the beaming models for each the YRQ population and the YRL population.   In Fig.~\ref{fig:beamr}, we show how the GC flux depends on the choice of beaming models (currently assuming C=1.0).  Overall, the YRQ pulsars tend to result in less excess flux for the same model parameters (e.g., SNR, and $\sigma_{\log B}$) as the YRL population or the fidicual model. 

One interesting consequence of this empirically motivated model is that the resulting flux density distribution of the pulsar population is in better agreement with some observations. This is discussed in more detail in \S~\ref{sec:unresolved}.  This is because the observable population of YRQ pulsars has a minimum flux near $ L_\gamma \approx 10^{34}\,\ergs$.

One limitation of the YRQ beaming model is the number of bright pulsars with $\dot{E} \lesssim 10^{33.5}\,\ergs$.  Within this beaming model we typically expect two pulsars with a spin down luminosity below this threshold to be brighter than $10^{-10}\,\ergcms$, where none have been observed.  Only $9\,\%$  of our simulations are consistent with the observations.

\section{Regions of Interest}

\label{sec:rois}

A number of studies focused on the emission from different regions near the GC.  We show the regions of interest from \citet{2014caloreetal} and \citet{FermiGC} in Figure~\ref{fig:rois}. 

The green dashed line shows the flux of high latitude pulsars with $|b| \geq |l|$.  This region has much less contribution from the foreground disk in our YP models.  However, the Fermi bubbles \cite{2010ApJ...724.1044S} are in the same region and are not included in our map.  These have to be carefully subtracted from diffuse maps of this region.  We do not include any systematic residuals that may remain. 

In many analyses of the GC region, spherically symmetric templates are used to fit the diffuse emission \citep[c.f.][]{2014arXiv1409.0042C}.  In order to ascertain the symmetry of the excess towards the GC region, we have divided it into two regions. The main one shown throughout the text is above and below the Galactic Plane with $|b| \geq 2^\circ$.  In Figure~\ref{fig:highlat}, we also show the average emission in the high latitude region with $|b| \geq |l|$.  Near the GC, the young pulsars from the CMZ are spherically symmetric, owing to the large kicks they receive at birth.  However, outside of $5^\circ$, the foreground  YPs begin to dominate the emission.  This population follows the Galactic Disk, and so the high latitude emission is reduced. 

\end{document}